\documentclass[aps,prd,showpacs,notitlepage,nofootinbib,superscriptaddress,floatfix,showkeys,twocolumn]{revtex4-2}
\usepackage{blindtext}
\usepackage{hyperref}
\usepackage{amsmath,amssymb}
\usepackage{float}
\usepackage{microtype}
\usepackage{graphicx}
\usepackage{bm}
\usepackage{latexsym}
\usepackage{epsfig}
\usepackage{psfrag}
\usepackage{color}
\usepackage[dvipsnames]{xcolor}
\usepackage{subfigure}
\def\nn{\nonumber} 
\def\f{\frac}

\def\l{\left}
\def\r{\right}
\def\d{{\mathrm{d}}}
\def\pa{\partial}
\def\Mpl{M_{_{\mathrm{Pl}}}}
\def\cA{{\mathcal{A}}}
\def\ps{\mathcal{P}_{_{\mathrm{S}}}}

\def\pb{\mathcal{P}_{_{\mathrm{B}}}}
\def\pe{\mathcal{P}_{_{\mathrm{E}}}}
\def\mub{\mu_{_{\mathrm{B}}}}
\def\HI{H_{_\mathrm{I}}}
\def\rhoI{\rho_{_\mathrm{I}}}
\def\ee{\eta_{\mathrm{e}}}
\def\e1i{\epsilon_{1\mathrm{i}}}
\def\Ne{N_{\mathrm{e}}}
\def\ke{k_{\mathrm{e}}}
\def\nb{n_{_{\mathrm{B}}}}
\def\ne{n_{_{\mathrm{E}}}}
\def\phii{\phi_{\mathrm{i}}}
\def\phie{\phi_{\mathrm{e}}}

\allowdisplaybreaks[1]

\begin{document}
%%%%%%%%%%%%%%%%%%%%%%%%%%%%%%%%%%%%%%%%%%%%%%%%%%%%%%%%%%%%%%%%%%%%%%%%%%%%%%%
\title{On the challenges in the choice of the non-conformal\\ 
coupling function in inflationary magnetogenesis}

\author{Sagarika Tripathy}
\email{E-mail: sagarika@physics.iitm.ac.in}
\affiliation{Centre for Strings, Gravitation and Cosmology,
Department of Physics, Indian Institute of Technology Madras, 
Chennai~600036, India}
\author{Debika Chowdhury}
\email{E-mail: debika.chowdhury@swansea.ac.uk}
\affiliation{Department of Physics, Swansea University, Swansea, SA2 8PP, U.K.}
\author{Rajeev Kumar Jain}
\email{E-mail: rkjain@iisc.ac.in}
\affiliation{Department of Physics, Indian Institute of Science, 
Bengaluru~560012, India}
\author{L.~Sriramkumar}
\email{E-mail: sriram@physics.iitm.ac.in}
\affiliation{Centre for Strings, Gravitation and Cosmology,
Department of Physics, Indian Institute of Technology Madras, 
Chennai~600036, India}

%%%%%%%%%%%%%%%%%%%%%%%%%%%%%%%%%%%%%%%%%%%%%%%%%%%%%%%%%%%%%%%%%%%%%%%%%%%%%%%

\begin{abstract}
Primordial magnetic fields are generated during inflation by considering
actions that break the conformal invariance of the electromagnetic field. 
To break the conformal invariance, the electromagnetic fields are coupled 
either to the inflaton or to the scalar curvature.
Also, a parity violating term is often added to the action in order to 
enhance the amplitudes of the primordial electromagnetic fields.
In this work, we examine the effects of deviations from slow roll inflation
on the spectra of non-helical as well as helical electromagnetic fields.
We find that, in the case of the coupling to the scalar curvature, there 
arise certain challenges in generating electromagnetic fields of the 
desired shapes and strengths even in slow roll inflation.
When the field is coupled to the inflaton, it is possible to construct 
model-dependent coupling functions which lead to nearly scale invariant 
magnetic fields in slow roll inflation.
However, we show that sharp features in the scalar power spectrum generated 
due to departures from slow roll inflation inevitably lead to strong features 
in the power spectra of the electromagnetic fields.
Moreover, we find that such effects can also considerably suppress the 
strengths of the generated electromagnetic fields over the scales of 
cosmological interest.
We illustrate these aspects with the aid of inflationary models that have 
been considered to produce specific features in the scalar power spectrum.
Further, we find that, in such situations, if the strong features in the 
electromagnetic power spectra are to be undone, the choice of the coupling 
function requires considerable fine tuning.  
We discuss wider implications of the results we obtain.
\end{abstract}
\maketitle

%%%%%%%%%%%%%%%%%%%%%%%%%%%%%%%%%%%%%%%%%%%%%%%%%%%%%%%%%%%%%%%%%%%%%%%%%%%%%%%

\section{Introduction}

Large-scale magnetic fields are observed in galaxies, galaxy clusters and in 
the intergalactic voids (for reviews on magnetic fields, see 
Refs.~\cite{Grasso:2000wj,Giovannini:2003yn,Brandenburg:2004jv,Kulsrud:2007an,
Subramanian:2009fu,Kandus:2010nw,Widrow:2011hs,Durrer:2013pga,Subramanian:2015lua,
Vachaspati:2020blt}). 
The Fermi/LAT and HESS observations of TeV blazars suggest that the strength of 
magnetic fields in the intergalactic medium is of the order of $10^{-15}\,
\mathrm{G}$~\cite{Neronov:1900zz,Tavecchio:2010mk,Dolag:2010ni,Dermer:2010mm,
Vovk:2011aa,Taylor:2011bn, Takahashi:2011ac}.
Also, magnetic fields of strength of the order of $10^{-6}\,\mathrm{G}$ are 
observed within galaxies (for a recent discussion of the various observational 
constraints, see, for instance, Refs.~\cite{Paoletti:2019pdi,Vachaspati:2020blt}).
It seems challenging to explain the presence of magnetic fields of such strengths, 
specifically in the intergalactic voids, on the basis of astrophysical 
phenomena alone~\cite{Brandenburg:2004jv,Kulsrud:2007an}. 
Hence, it is believed that these magnetic fields may have a cosmological origin 
and they could have been generated during the inflationary epoch in the early
universe (for reviews in this context, see Refs.~\cite{Subramanian:2009fu,
Kandus:2010nw,Durrer:2013pga,Subramanian:2015lua,Vachaspati:2020blt}).

Recall that the standard electromagnetic action is conformally invariant.
Therefore, the energy density of the magnetic fields generated in such a 
theory will be rapidly washed away during inflation.
We should clarify that this is strictly true only in the case of the spatially 
flat Friedmann-Lema\^itre-Robertson-Walker~(FLRW) universe, which is conformally 
flat globally. 
The FLRW universes with non-vanishing spatial curvature are conformally 
flat only locally and, as a result, the adiabatic evolution of magnetic 
fields in such scenarios can be affected (see Refs.~\cite{Barrow:2008jp,
Barrow:2011ic}; however, for further discussions in this context, see
Refs.~\cite{Adamek:2011hi,Shtanov:2012pp,Yamauchi:2014saa}).
In this work, we shall focus on the spatially flat FLRW universe.
The spectrum of magnetic fields generated in the conformally invariant theory
will be strongly scale-dependent, inconsistent with the recent constraints from 
the cosmic microwave background~(CMB)~\cite{Planck:2015zrl}.
The simplest way to generate magnetic fields of observable strengths today seems
to break the conformal invariance of the electromagnetic action (in this 
context, see, for example, Refs.~\cite{Turner:1987bw,Ratra:1991bn,Bamba:2003av,
Bamba:2006ga,Martin:2007ue,Bamba:2008ja,Demozzi:2009fu,Bamba:2020qdj,Bamba:2021wyx}).
Often, this is achieved by coupling the electromagnetic field to either the
scalar field that drives inflation~\cite{Bamba:2003av,Martin:2007ue,Watanabe:2009ct,
Kanno:2009ei,Markkanen:2017kmy} 
or to the Ricci scalar describing the background~\cite{Bamba:2006ga,Bamba:2008ja,
Bamba:2020qdj,Bamba:2021wyx}.
In fact, it has also been discovered that the addition of a parity violating term 
in the electromagnetic action can significantly enhance the amplitude of magnetic
fields generated during inflation~\cite{Anber:2006xt,Durrer:2010mq,Caprini:2014mja,
Chowdhury:2018mhj,Sharma:2018kgs,Giovannini:2020zjo,Giovannini:2021thf,Gorbar:2021rlt,
Gorbar:2021zlr}. 
It can be shown that, for certain choices of the coupling function, the spectrum
of magnetic fields generated can be nearly scale invariant consistent with the 
current constraints over a wide range of scales (see, for instance,
Refs.~\cite{Planck:2015zrl,Zucca:2016iur,Sutton:2017jgr,Paoletti:2018uic,
Minoda:2020bod}).

The CMB observations point to a nearly scale invariant primordial scalar power 
spectrum as is generated in models of slow roll inflation~\cite{Planck:2018jri}.
Nevertheless, there has been a constant interest in the literature to examine if 
there exist features in the scalar power spectrum.
During the last decade or two, the possibility of features in the inflationary 
power spectrum has been often examined with the aim of improving the fit to 
the CMB and the large scale structure data (in this context, see, for instance, 
Refs.~\cite{Contaldi:2003zv,Sinha:2005mn,Powell:2006yg,Jain:2008dw,Jain:2009pm,
Hazra:2010ve,Benetti:2013cja,Hazra:2014jka,Hazra:2014goa,Chen:2016zuu,Chen:2016vvw,
Ragavendra:2020old,Antony:2021bgp}).
More recently, with the detection of gravitational waves from merging binary 
black holes~\cite{LIGOScientific:2020kqk}, there has been a tremendous interest 
in investigating whether such black holes could have a primordial 
origin~\cite{DeLuca:2020qqa,Jedamzik:2020ypm,Jedamzik:2020omx,Franciolini:2021tla}.
In this context, a variety of inflationary models generating increased power on 
small scales (compared to the COBE normalized power on the CMB scales) which can lead 
to an enhanced formation of primordial black holes have been investigated (see, for 
instance, Refs.~\cite{Garcia-Bellido:2017mdw,Ballesteros:2017fsr,Germani:2017bcs,
Dalianis:2018frf,Bhaumik:2019tvl,Ashoorioon:2019xqc,Ragavendra:2020sop,Dalianis:2020cla}).
These features in the scalar power spectrum~---~both on the large as well as the 
small scales~---~are usually generated due to deviations from slow roll inflation.
We mentioned above that the spectrum of the magnetic field depends on the choice 
of the function that couples the electromagnetic field to either the inflaton 
or the Ricci scalar.
These coupling functions are often chosen such that the power spectrum of the 
magnetic field is nearly scale invariant in slow roll inflation (actually, the
background is often assumed to be of the de Sitter or power law forms). 
However, if there arise departures from slow roll, the non-trivial dynamics can
influence the behavior of the coupling functions and thereby affect the spectrum 
of the magnetic field.
In other words, the mechanism that generates features in the scalar power spectrum
can also induce features in the spectrum of the magnetic field depending on the 
nature of the coupling that breaks the conformal invariance of the electromagnetic
action or induces violation of parity.

In this work, we shall investigate the effects of deviations from slow roll inflation 
on the power spectra of the electromagnetic fields.
While there have been some earlier attempts to understand the effects of transitions 
during inflation (in this context, see, for instance, Refs.~\cite{Durrer:2010mq,
Byrnes:2011aa,Ferreira:2013sqa,Ferreira:2014hma}; for some recent efforts, see
Refs.~\cite{Shtanov:2019civ,Shtanov:2020gjp}), we find that there does not seem 
to have been any effort to systematically examine the imprints of departures from 
slow roll inflation on the spectra of the electromagnetic fields.
We find that coupling the electromagnetic field to the scalar curvature poses
certain difficulties even in slow roll inflation.
We consider specific inflationary models that lead to features in the scalar power
spectrum. 
We choose functions that are coupled to the inflaton which lead to nearly scale
invariant spectra for the magnetic field either in the absence of departures from 
slow roll or over large scales (which are constrained by the CMB observations) and
examine the effects due to the deviations from slow roll inflation.
We show that, in these cases, unless the non-minimal coupling function is designed 
in a specific manner and is extremely fine-tuned, it is impossible to avoid features 
in the spectra of electromagnetic fields.
Moreover, we notice that, in some cases, the strengths of the magnetic fields can 
be considerably suppressed over large scales.
We believe that exploring the observational signatures of such features can 
help us understand the nature of the non-conformal coupling that is required 
to generate magnetic fields of observable strengths.

This paper is organized as follows.
In the next section, we shall discuss the spectra of electromagnetic fields 
generated during inflation, when the fields are coupled to either the inflaton 
or the scalar curvature.
We shall arrive at the spectra of electromagnetic fields generated in de Sitter 
inflation when the field is coupled to the inflaton.
We shall also evaluate the spectra in the presence of an additional term in the
action that induces the violation of parity.
We shall point out that, even in slow roll inflation, there arise specific 
challenges when considering the coupling of the electromagnetic field to 
the scalar curvature.
In Sec.~\ref{sec:pbe-srm}, we shall construct specific non-minimal coupling 
functions that lead to nearly scale invariant power spectra for the 
magnetic fields in some of the popular models of slow roll inflation.
In Sec.~\ref{sec:im-lf}, we shall introduce a few inflationary models that 
lead to features over large, intermediate and small scales in the scalar 
power spectrum.
In Sec.~\ref{sec:eds-s-emf}, we shall examine the effects of deviations from 
slow roll inflation on the spectra of the electromagnetic fields.
In certain cases, we shall support our numerical computations with analytical
estimates of the amplitude and shape of the electromagnetic power spectra.
In Sec.~\ref{sec:iof}, with the help of an example, we shall illustrate that,
given an inflationary model leading to features in the scalar power spectra, a 
suitably designed non-minimal coupling function can largely undo the sharp 
features generated in the spectra of the electromagnetic fields.
Finally, we shall conclude with a summary in Sec.~\ref{sec:c}.
We shall relegate some of the details to an appendix.

Let us now clarify a few points regarding the conventions and notations that
we shall work with. 
We shall work with natural units such that $\hbar=c=1$, and set the reduced 
Planck mass to be $\Mpl=\l(8\,\pi\, G\r)^{-1/2}$.
We shall adopt the signature of the metric to be~$(-,+,+,+)$.
Note that Latin indices will represent the spatial coordinates, except for~$k$ 
which will be reserved for denoting the wave number. 
As we mentioned, we shall assume the background to be the 
spatially flat FLRW universe described by the following line element: 
\begin{equation}
\d s^2=-\d t^2+a^2(t)\,\d {\bm x}^2
=a^2(\eta)\, \l(-\d \eta^2+\d {\bm x}^2\r),\label{eq:FLRW}
\end{equation}
where $t$ and $\eta$ denote cosmic time and conformal time, while $a$ 
represents the scale factor.
Also, an overdot and an overprime will denote differentiation with respect 
to the cosmic and conformal time coordinates.
Moreover, $N$ shall represent the number of e-folds.
Lastly, $H=\dot{a}/a$ and $\mathcal{H}=a\,H=a'/a$ shall represent the Hubble 
and the conformal Hubble parameters, respectively.

%%%%%%%%%%%%%%%%%%%%%%%%%%%%%%%%%%%%%%%%%%%%%%%%%%%%%%%%%%%%%%%%%%%%%%%%%%%%%%%

\section{Generation of magnetic fields during inflation}\label{sec:s-emf}

In this section, we shall quickly summarize the essential aspects related
to the generation of electromagnetic fields during inflation.
We shall outline the spectra that arise in situations wherein a coupling 
function is introduced to break the conformal invariance of the action 
describing the electromagnetic fields. 

%%%%%%%%%%%%%%%%%%%%%%%%%%%%%%%%%%%%%%%%%%%%%%%%%%%%%%%%%%%%%%%%%%%%%%%%%%%%%%%

\subsection{The non-helical case}

As is often done, we shall first consider a coupling between the electromagnetic 
field and the inflaton to break the conformal invariance of the standard action
describing electromagnetism.
We shall assume that the electromagnetic field is described by the action 
(see, for example, Refs.~\cite{Martin:2007ue,Subramanian:2009fu})
\begin{eqnarray}
S[A^\mu] = -\f{1}{16\,\pi} \int \d^4x \sqrt{-g}\, J^2(\phi)\, 
 F_{\mu\nu}\,F^{\mu\nu},\label{eq:s-em}
\end{eqnarray}
where $J(\phi)$ denotes the coupling function and the field tensor~$F_{\mu\nu}$ 
is expressed in terms of the vector potential $A_\mu$ 
as~$F_{\mu\nu}=( \pa_{\mu}\,A_{\nu}-\pa_{\nu}\,A_{\mu})$.
On working in the Coulomb gauge wherein $A_\eta = 0$ and $\pa_i\,A^i=0$, 
one finds that the Fourier modes, say, $\bar{A}_k$, describing the
vector potential satisfy the differential
equation (see, for example Refs.~\cite{Martin:2007ue,Chowdhury:2018blx}):
\begin{equation}
\bar{A}_k''+2\, \f{J'}{J}\,\bar{A}_k' +k^2 \bar{A}_k = 0. 
\label{eq:de-Abk}
\end{equation}
If we write $\bar{A}_k=\cA_k/J$, then this equation reduces to
\begin{eqnarray}
\cA_k''+ \l(k^2- \f{J''}{J}\r)\, \cA_k = 0.\label{eq:de-cAk}
\end{eqnarray}
The power spectra associated with the magnetic and electric fields are
defined to be~\cite{Martin:2007ue,Subramanian:2009fu}
\begin{subequations}\label{eq:psbe}
\begin{eqnarray}
\pb(k) 
&=& \f{k^5}{2\,\pi^2}\,\f{J^2}{a^4}\, 
\vert \bar{A}_k\vert^2
=\f{k^5}{2\,\pi^2\, a^4}\, \vert \cA_k\vert^2, \label{eq:psb}\\ 
\pe(k) 
&=& \f{k^3}{2\,\pi^2}\,\,\f{J^2}{a^4}\, \vert \bar{A}_k'\vert^2
= \f{k^3}{2\,\pi^2\, a^4}\, \biggl\vert \cA_k'
-\f{J'}{J}\,\cA_k\biggr\vert^2.\qquad 
\end{eqnarray}
\end{subequations}
The initial conditions on the quantity $\cA_k$ can be imposed in 
the domain wherein $k\gg \sqrt{J''/J}$ and the spectra associated
with the electromagnetic fields can be evaluated in the limit
when~$k\ll\sqrt{J''/J}$.

Let us now arrive at the power spectra of the electromagnetic fields in 
de Sitter inflation wherein the scale factor is given by $a(\eta) =
-1/(\HI\,\eta)$, with $\HI$ denoting the constant Hubble parameter.
Typically, the coupling function $J$ is assumed to depend on the scale 
factor as follows (see, for instance, Refs.~\cite{Martin:2007ue,
Subramanian:2009fu}):
\begin{equation}
J(\eta)= \l[\f{a(\eta)}{a(\ee)}\r]^n
=\l(\frac{\eta}{\ee}\r)^{-n},\label{eq:J}
\end{equation}
where $\ee$ denotes the conformal time at the end of inflation.
Note that we have chosen the overall constant so that the coupling 
function reduces to unity at the end of inflation.
We should stress here that the parameter~$n$ is a real number and is 
not necessarily an integer.
In such a case, the Bunch-Davies initial conditions on the electromagnetic
modes $\cA_k$ can be imposed in the limit $k \gg \sqrt{J''/J}$, which, for
the above choice of the coupling function, corresponds to the modes being in 
the sub-Hubble domain at early times.
For the coupling function~\eqref{eq:J}, the solution to 
Eq.~\eqref{eq:de-cAk} that satisfies the Bunch-Davies initial 
conditions is given by
\begin{equation}
\cA_k(\eta) 
= \sqrt{-\f{\pi\,\eta}{4}}\,
\mathrm{e}^{i\,(n+1)\,\pi/2}\,
H^{(1)}_{\nu}(-k\,\eta),\label{eq:nhs}
\end{equation}
where $\nu=n+(1/2)$, and $H_\nu^{(1)}(z)$ denotes the Hankel function of 
the first kind.

The spectra of the electromagnetic fields can be evaluated in the limit 
$k\ll \sqrt{J''/J}$, which corresponds to the super-Hubble limit in de 
Sitter inflation for our choice of the coupling function.
In the limit $(-k\,\ee) \ll 1$, the spectra of the magnetic and electric 
fields $\pb(k)$ and $\pe(k)$ can be obtained 
to be~\cite{Martin:2007ue,Subramanian:2009fu}
\begin{subequations}\label{pb-e}
\begin{eqnarray}
\pb(k) &=& \f{\HI^4}{8\,\pi}\, \mathcal{F}(m)\,(-k\,\ee)^{2\,m+6},\\
\pe(k) &=& \f{\HI^4}{8\,\pi}\, \mathcal{G}(m)\,(-k\,\ee)^{2\,m+4},
\end{eqnarray}
\end{subequations}
where, recall that, $\ee$ denotes the conformal time at the end 
of inflation.
The quantities $\mathcal{F}(m)$ and $\mathcal{G}(m)$ are given by
\begin{subequations}
\begin{eqnarray}
\mathcal{F}(m)
&=&\f{1}{2^{2\,m+1}\,\mathrm{cos}^2(m\,\pi)\,\Gamma^2(m+3/2)},\\
\mathcal{G}(m)
&=&\f{1}{2^{2\,m-1}\,\mathrm{cos}^2(m\,\pi)\,\Gamma^2(m+1/2)},
\end{eqnarray}
\end{subequations}
with 
\begin{equation}
m=\begin{cases}
n, & \text{for $n<-\frac{1}{2}$},\\
-n-1, & \text{for $n>-\frac{1}{2}$}.
 \end{cases}
\end{equation}
in the case of $\pb(k)$, and with 
\begin{equation}
m=\begin{cases}
n, & \text{for $n<\f{1}{2}$},\\
1-n, & \text{for $n>\f{1}{2}$}.
\end{cases}
\end{equation}
in the case of $\pe(k)$.
Note that the spectral indices for the magnetic and electric fields, say,
$\nb$ and $\ne$, can be written as
\begin{equation}
\nb 
=\begin{cases}
2\,n+6, & \text{for $n<-\f{1}{2}$},\\
4-2\,n, & \text{for $n>-\f{1}{2}$},
\end{cases}
\end{equation}
and 
\begin{equation}
\ne
=\begin{cases}
2\,n+4, & \text{for $n<\f{1}{2}$},\\
6-2\,n, & \text{for $n>\f{1}{2}$}.
\end{cases}
\end{equation}
To be consistent with observations, the magnetic field is expected 
to be nearly scale invariant and, evidently, this is possible when 
$n\simeq -3$ or when $n\simeq 2$.
In these cases, it is clear that $\ne \simeq -2$ and $\ne \simeq 2$, 
respectively.
At late times, $\ne \simeq -2$ implies that the energy density in the 
electric field is significant leading to a large backreaction.
In order to avoid such an issue, one often considers the $n=2$ case to
lead to a scale invariant magnetic field with negligible backreaction
due to the electric field.
Note that, in these cases, the power spectra reduce to the following
simple forms
\begin{eqnarray}
\pb(k) = \f{9\,\HI^4}{4\,\pi^2},\quad
\pe(k) = \f{\HI^4}{4\,\pi^2}\,(-k\,\ee)^{2}.\label{eq:ne2}
\end{eqnarray}

%%%%%%%%%%%%%%%%%%%%%%%%%%%%%%%%%%%%%%%%%%%%%%%%%%%%%%%%%%%%%%%%%%%%%%%%%%%%%%%

\subsection{The helical case}

Recall that, we had considered the action~\eqref{eq:s-em} to break the 
conformal invariance of the electromagnetic field.
The action can be extended to include a parity violating term as follows 
(in this context, see, for instance Refs.~\cite{Anber:2006xt,Durrer:2010mq,
Byrnes:2011aa,Caprini:2014mja,Chowdhury:2018mhj,Sharma:2018kgs}):
\begin{eqnarray}
S[A^\mu] & = & -\f{1}{16\,\pi}\, \int \d^4x\, \sqrt{-g}\, \biggl[J^2(\phi)\, 
F_{\mu\nu}\,F^{\mu\nu}\nn\\ 
& & -\, \f{\gamma}{2}\, I^2(\phi)\, F_{\mu\nu}\,\widetilde{F}^{\mu\nu}\biggr], 
\end{eqnarray}
where $\widetilde{F}^{\mu\nu} = (\epsilon^{\mu\nu\alpha\beta}/\sqrt{-g})\, 
F_{\alpha\beta}$, with $\epsilon^{\mu\nu\alpha\beta}$ being the completely
anti-symmetric Levi-Civita tensor, and $\gamma$ is a constant.
In such a case, the modes of the electromagnetic field can be decomposed in 
a suitable helical basis. 
Also, we can work in the Coulomb gauge as we had done in the non-helical case. 
In such a case, it is found that the second term in the above action amplifies
the electromagnetic modes associated with one of the polarizations when 
compared to the other, thereby violating parity or, equivalently, inducing 
helicity~\cite{Caprini:2014mja,Chowdhury:2018mhj,Sharma:2018kgs,Gorbar:2021rlt,
Gorbar:2021zlr}.

When we decompose the electromagnetic field in the helical basis, the Fourier 
modes of the field, say, $\bar{A}^{\sigma}_k$, are found to satisfy the 
differential equation
\begin{equation}
\bar{A}_k^{{\sigma}\prime\prime} 
+ 2\,\f{J'}{J}\, \bar{A}_k^{\sigma\prime}
+\l(k^2+\f{\sigma\, \gamma\, k}{J^2}\,\frac{\d I^2}{\d\eta}\r) \bar{A}^\sigma_k =0,
\end{equation}
where $\sigma=\pm 1$ represents positive and negative helicity. 
Let us define ${\bar A}_k^\sigma=\cA_k^\sigma/J$ as we had done in the 
non-helical case.
In terms of the new variable~$\cA_k^\sigma$, the above equation reduces to
\begin{equation}
\cA_k^{\sigma\,\prime\prime} + \l(k^2
+ \f{2\,\sigma\,\gamma\,k\,I\,I^\prime}{J^2}
- \f{J^{\prime\prime}}{J}\r)\,\cA_k^\sigma 
=0.\label{eq:cA-eqn-gen}
\end{equation}
We shall restrict ourselves to the simplest of scenarios wherein $I = J$. 
In such a case, the above equation simplifies to
\begin{equation}
\cA_k^{\sigma\,\prime\prime}
+ \l(k^2 + \f{2\,\sigma\,\gamma\,k\,J^\prime}{J}
- \f{J^{\prime\prime}}{J}\r)\cA_k^\sigma = 0.
\label{eq:cA-h-de}
\end{equation}
The power spectra of the magnetic and electric fields can be expressed in 
terms of the modes ${\bar A}_k^\sigma$ and the coupling function $J$ as 
follows~\cite{Durrer:2010mq,Anber:2006xt,Caprini:2014mja,Sharma:2018kgs}:
\begin{subequations}
\begin{eqnarray}\label{eq:psbe-h}
\pb(k) 
&=& \f{k^5}{4\,\pi^{2}}\,\f{J^2}{a^{4}}\,
\l[\l\vert {\bar A}_k^{+}\r\vert^2 
+ \l\vert {\bar A}_k^{-}\r\vert^2\r]\nn\\
&=& \f{k^{5}}{4\,\pi^2\,a^{4}}\,
\l[\l\vert  \cA_k^{+}\r\vert^2 
+ \l\vert \cA_k^{-}\r\vert^2\r],\label{eq:psb-h}\\
\pe(k) &=& \f{k^3}{4\,\pi^{2}}\,\f{J^2}{a^4}\,
\l[\l\vert {\bar A}_k^{+ \prime}\r\vert^2 
+ \l\vert {\bar A}_k^{- \prime}\r\vert^2\r]\nn\\
&=&\! \f{k^3}{4\, \pi^2\, a^4}\, \l[\l\vert \cA_k^{+\prime}
- \f{J^\prime}{J}\cA_k^+\r\vert^2+\l\vert \cA_k^{-\prime}
- \f{J^\prime}{J}\cA_k^-\r\vert^2\r].\nn\\
\label{eq:pse-h}
\end{eqnarray}
\end{subequations}

For the form of the coupling function given by Eq.~(\ref{eq:J}), the solutions
to the electromagnetic modes satisfying the differential equation~\eqref{eq:cA-h-de}
and the Bunch-Davies initial conditions can be written as follows (for a recent 
discussion, see, for example, Ref.~\cite{Sharma:2018kgs}):
\begin{equation}
\cA_k^\sigma(\eta) 
=\f{1}{\sqrt{2\,k}}\, \mathrm{e}^{\pi\,\sigma\,\xi/2}\,
W_{-i\,\sigma\,\xi,\nu}(2\,i\,k\,\eta),\label{eq:cA-h}
\end{equation}
where $\nu=n+(1/2)$, $\xi = -n\,\gamma$, and $W_{\lambda,\mu}(z)$ 
denotes the Whittaker function.
In the domain $z\ll 1$, the Whittaker function $W_{\lambda,\mu}(z)$ 
behaves as~\cite{Gradshteyn:1702455,Mathematica}
\begin{eqnarray}
W_{\lambda,\mu}(z)
& \to & \f{\Gamma(-2\,\mu)}{\Gamma(\tfrac{1}{2} -\lambda -\mu)}\, z^{(1/2)+\mu}\nn\\
& &+\, \f{\Gamma(2\,\mu)}{\Gamma(\tfrac{1}{2} - \lambda +\mu)}\,
z^{(1/2)-\mu}.\label{eq:Wfn-s}
\end{eqnarray}
Upon using this result and the expression~\eqref{eq:psb-h}, we find that 
the spectrum of the magnetic field evaluated in the limit $(-k\,\ee)\ll 1$ 
is given by~\cite{Caprini:2014mja,Sharma:2018kgs}
\begin{eqnarray}
\pb(k) &=& \frac{\HI^4}{8\,\pi^2}\,
\f{\Gamma^2(\vert 2\,n+1\vert)}{\vert \Gamma(\tfrac{1}{2}
+i\,n\,\gamma+\vert n+\tfrac{1}{2}\vert)\vert^2}\nn\\ 
& &\times\,\f{\mathrm{cosh}\,(n\,\pi\,\gamma)}{2^{\vert 2\,n+1\vert -2}}\,
(-k\,\ee)^{5-\vert 2\,n+1\vert}.
\end{eqnarray}
Let us now turn to the evaluation of the spectrum of the electric field.
In the calculation of the spectrum, the following relation for the derivative 
of the Whittaker function~\cite{Gradshteyn:1702455,Mathematica}:
\begin{equation}
\f{\d W_{\lambda,\mu}(z)}{\d z} = \l(\f{1}{2}-\f{\lambda}{z}\r)\,
W_{\lambda,\mu}(z)-\f{1}{z}\,W_{1+\lambda,\mu}(z)
\end{equation}
and the following recursion relation: 
\begin{equation}
W_{\lambda,\mu}(z)
=\sqrt{z}\,W_{\lambda-\tfrac{1}{2},\mu-\tfrac{1}{2}}(z)
+\l(\f{1}{2}-\lambda+\mu\r)\,W_{\lambda-1,\mu}(z)
\end{equation}
prove to be helpful.
On using the above relations and the behavior~\eqref{eq:Wfn-s} of the Whittaker 
function, we can obtain the spectrum of the electric field in the helical case 
[as defined in Eq.~\eqref{eq:pse-h}] in the limit $(-k\,\ee)\ll 1$ to be
\begin{eqnarray}
\pe(k)&=&\f{\HI^4}{4\, \pi^2}\,
\f{\Gamma^2(2\, \vert n\vert)}{\vert \Gamma(\vert n\vert 
+i\,n\,\gamma)\vert^2}\,\f{\gamma^2}{1+\gamma^2}\nn\\ 
& &\times\,\f{\mathrm{cosh}\,(n\,\pi\,\gamma)}{2^{2\, \vert n\vert-2}}\,
(-k\,\ee)^{4-2\,\vert n\vert}
\end{eqnarray}
with the factor $\gamma^2/(1+\gamma^2)$ arising {\it only}\/ for positive 
values of the index~$n$.
Evidently, the spectral indices for the magnetic and electric 
fields~---~viz. $\nb$ and $\ne$~---~are given by
\begin{equation}
\nb = 5 - \l\vert 2\,n+ 1\r\vert,\quad
\ne = 4 - 2\,\vert n\vert.
\end{equation}
As in the non-helical case, we find that the spectrum of the magnetic 
field is scale invariant when $n=2$ and $n=-3$.
Interestingly, in the helical case, the spectrum of the electric field 
is also scale invariant when $n=2$, whereas, when $n=-3$, the spectrum
has the same tilt (i.e. $\ne=-2$) as in the non-helical case.

In our later discussion, we shall be focusing on the $n=2$ case.
When $n=2$, we find that the spectra of the helical magnetic and electric fields
[evaluated in the limit $(-k\,\ee)\ll 1$] can be written as~\cite{Mathematica}
\begin{subequations}\label{eq:ne2h}
\begin{eqnarray}
\pb(k)
&=&\f{9\,\HI^4}{4\,\pi^2}\,f(\gamma),\label{eq:pb-ne2h}\\
\pe(k)
&=&\f{9\,\HI^4}{4\,\pi^2}\, f(\gamma)\,
\biggl[\gamma^2 -\f{\mathrm{sinh}^2(2\,\pi\,\gamma)}{3\,\pi\,
\l(1+\gamma^2\r)\,f(\gamma)}\,(-k\,\ee)\nn\\
& &+\,\f{1}{9}\,\l(1+23\,\gamma^2+40\,\gamma^4\r)\,\l(-k\,\ee\r)^{2}\biggl],
\end{eqnarray}
\end{subequations}
where the function $f(\gamma)$ is given by
\begin{equation}
f(\gamma)=\f{\mathrm{sinh}\,(4\,\pi\,\gamma)}{4\,\pi\,\gamma\,
\l(1+5\,\gamma^2+4\,\gamma^4\r)}.\label{eq:fg}
\end{equation}
We will soon clarify the reason for retaining the second and third terms 
within the square brackets [despite the fact that we are considering the
$(-k\,\ee)\ll 1$ limit] in the above expression for~$\pe(k)$.
There are two related points that we need to highlight regarding the results 
we have arrived at above.
Firstly, note that, as $\gamma\to 0$, $f(\gamma)\to 1$, and these spectra
reduce to the non-helical results~\eqref{eq:ne2}, as required.
Secondly, in the above spectrum for the electric field, the first two 
terms go to zero in the limit of vanishing helicity (i.e. as $\gamma 
\to 0$).
In other words, even a small amount of helicity modifies the spectrum of 
the electric field considerably, making it scale invariant.
It is only in the case of extremely small helicity~---~to be precise, when 
$\gamma \ll (-k\,\ee) \simeq k/k_\mathrm{e}$, where $k_\mathrm{e}$ is the 
wave number that leaves the Hubble radius at the end of inflation~---~that
the third term becomes dominant leading to the behavior that we had 
encountered in the non-helical case.

%%%%%%%%%%%%%%%%%%%%%%%%%%%%%%%%%%%%%%%%%%%%%%%%%%%%%%%%%%%%%%%%%%%%%%%%%%%%%%%

\subsection{Coupling to the scalar curvature}

Let us now turn to the case of the electromagnetic field that is coupled to
the scalar curvature~$R$ and is described by the following
action~\cite{Turner:1987bw,Bamba:2008ja,Bamba:2020qdj}:
\begin{equation}
S[A^\mu] = -\f{1}{16\,\pi}\, \int \d^4x\, \sqrt{-g}\, J^2(R)\, 
F_{\mu\nu}\,F^{\mu\nu},\label{eq:a-emf-J-R}
\end{equation}
where $F_{\mu\nu}$ is the electromagnetic field tensor defined earlier.
Evidently, in such a case, one can work in the Coulomb gauge and the Fourier
modes of the electromagnetic vector potential $\bar{A}_k$ and the quantity 
$\cA_k =J\,\bar{A}_k$ would continue to be governed by the differential 
equations~\eqref{eq:de-Abk} and~\eqref{eq:de-cAk}.
Therefore, if the coupling function $J(R)$ is chosen so that it depends on the 
conformal time as in Eq.~\eqref{eq:J}, then we can expect scale invariant spectra 
for the magnetic field when $n=-3$ and $n=2$.

Earlier, while considering the coupling function~\eqref{eq:J}, we had assumed the
background to be that of de Sitter.
Note that the scalar curvature $R$ associated with the FLRW line-element~\eqref{eq:FLRW}
can be expressed as
\begin{equation}
R=6\,\f{a^{\prime\prime}}{a^3}=6\,H^2\, (2-\epsilon_1)
\end{equation}
and we should emphasize that this expression is exact.
In a de Sitter universe wherein $H$ is a constant and $\epsilon_1$ vanishes,
the above relation implies that the scalar curvature is time-independent. 
Therefore, we cannot work in the de Sitter limit.
Since we are interested in potentials which typically lead to slow roll inflation,
we can assume the scale factor to be of the slow roll form.
In such a case, it can be shown that the scalar curvature behaves in terms 
of the conformal time as~$R\propto \eta^{2\,\epsilon_1}$.
This suggests that we can possibly work with a coupling function of the form
\begin{equation}
J(R)=\l(\f{R(\eta)}{R(\ee)}\r)^{\alpha},\label{eq:J-R}
\end{equation}
where $R(\ee)$ denotes the scalar curvature at the end of inflation.
In slow roll inflation, such a coupling will behave in terms of the conformal
time coordinate as follows:
\begin{equation}
J(\eta) \simeq \l(\f{\eta}{\ee}\r)^{2\,\epsilon_1\,\alpha},
\end{equation}
which reduces to our original form of the coupling function, as given by 
Eq.~\eqref{eq:J}, if we choose $\alpha=-n/(2\,\epsilon_1)$.
Also, we can expect to arrive at a scale invariant spectrum for the magnetic
field without any backreaction in the case of $n=2$.

\begin{figure}[!t]
\centering
\includegraphics[width=8.50cm]{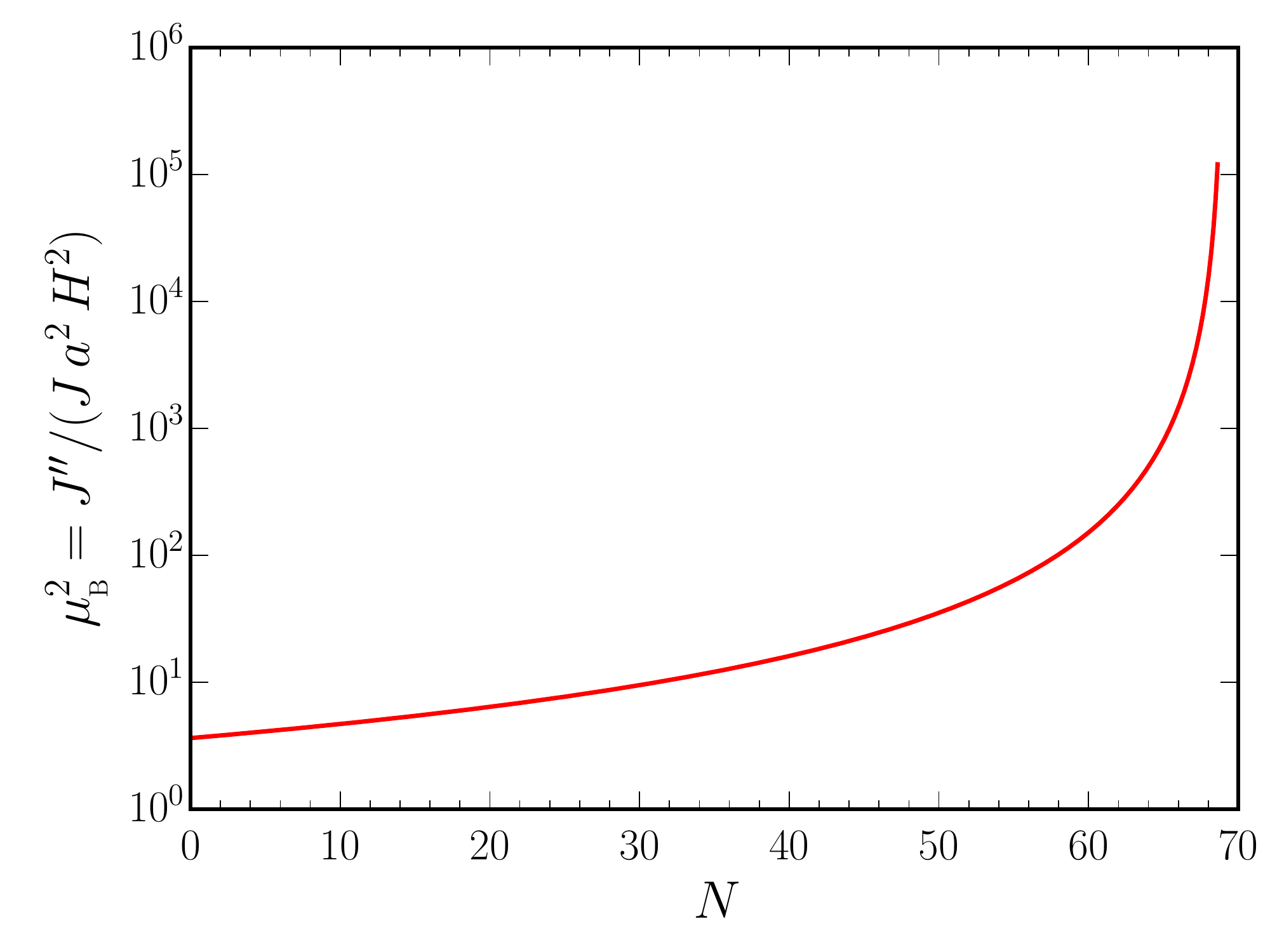}
\caption{The evolution of the quantity~$\mub^2=J^{\prime\prime}/(J\,a^2\,H^2)$, 
with $J$ being given by the coupling function~\eqref{eq:J-R}, as it occurs
in the case of slow roll inflation driven by the quadratic potential (in 
this context, see Sec.~\ref{sec:pbe-srm}), has 
been plotted as a function of e-folds~$N$.
We have set~$\alpha=-1/\epsilon_{1\ast}\simeq -10^2$, where $\epsilon_{1\ast}$ 
is the value of the first slow roll parameter when the pivot scale~$k_\ast$ 
leaves the Hubble radius.
For the value of the parameter~$m$ (describing the quadratic potential) 
and the initial conditions we have worked with, we find that the pivot 
scale~$k_\ast$ leaves the Hubble radius at the e-fold of~$N= 18.63$.
We find that $\mub^2\simeq 6$ near $N\simeq 18$, which is necessary to 
result in a scale invariant spectrum for the magnetic field.
However, since the first slow roll parameter $\epsilon_1$ is not a constant,
$\mub^2$ changes with time and, actually, grows to a large value towards the
end of inflation.
Apart from affecting the shape of the spectra of the electromagnetic fields,
we find that, a large value of $\alpha$ also leads to exceedingly large values 
of the electromagnetic vector potential at either the early or the late stages
of inflation.}
\label{fig:pJR}
\end{figure}

But, there arises a challenge, which, in fact, proves to be a rather 
serious one.
When considering a non-conformal coupling of the form~$J(R)$, we find that, 
in the literature, the scale factor describing the FLRW background is often
assumed to be of a power law form.
Such an assumption works well in power law inflationary scenarios 
wherein the first slow roll parameter~$\epsilon_1$ is strictly a 
constant, but poses difficulties in realistic slow roll models
of inflation wherein $\epsilon_1$ evolves towards unity and 
inflation ends naturally.
Note that, since $\epsilon_1$ is rather small at early times in slow roll 
inflation (in order to be consistent with the constraints on the 
tensor-to-scalar ratio $r$ over the CMB scales; for the latest constraints,
see Refs.~\cite{Planck:2018jri,BICEPKeck:2021gln}), the index 
$\alpha=-1/\epsilon_1$ (for $n=2$) turns out to be large in magnitude, 
typically of the order of~$10^2$ or larger. 
The fact that the index~$\alpha$ has a large magnitude is not surprising and 
can be easily understood.
In slow roll inflation, $R\simeq 12\, H^2$ and hence it hardly changes during 
the initial stages of inflation.
Therefore, one has to raise the scalar curvature to an adequately large power 
to achieve the desired time-dependence of the coupling function.
Moreover, since, in any realistic slow roll model of inflation, $\epsilon_1$ 
is {\it not}\/ a constant, one has to work with an $\alpha$ that is
determined by, say, the value of $\epsilon_1$ when the pivot scale leaves 
the Hubble radius.
However, because $\epsilon_1$ is time-dependent, we are not guaranteed a scale 
invariant spectrum for the magnetic field.
In order to illustrate this point, in Fig.~\ref{fig:pJR}, we have plotted the
quantity $\mub^2=J''/(J\,a^2\,H^2)$ in a slow roll inflationary model described 
by the quadratic potential [which we shall introduce later, see
Eq.~\eqref{eq:qp}].
%%%%%%%%%%%%%%%%%%%%%%%%%%%%%%%%%%%%%%%%%%%%%%%%%%%%%%%%%%%%%%%%%%%%%%%%%%%%%%%
%%%%%%%%%%%%%%%%%%%%%%%%%%%%%%%%%%%%%%%%%%%%%%%%%%%%%%%%%%%%%%%%%%%%%%%%%%%%%%%
We have chosen the parameter $\alpha$ so that~$\mub^2\simeq 6$ when 
the pivot scale leaves the Hubble radius, which is required to lead 
to a nearly scale invariant spectrum for the magnetic field.
But, since $\epsilon_1$ changes with time, the quantity $\mub^2$ grows
to large values at later times.
Such a behavior of $\mub^2$ not only affects the shape of the spectra
of the electromagnetic fields, it influences their amplitude as well.
Importantly, we find that, in general, a large value for $\alpha$ leads 
to rather large values for the electromagnetic vector potential at
either early or late times.

Phenomenologically, the only way out of this difficulty is to choose the
index~$\alpha$ in $J(R)$ [cf. Eq.~\eqref{eq:J-R}] to be dependent on time.
In order to arrive at a scale invariant power spectrum for the magnetic 
field, one may work with a coupling function of the following form:
\begin{equation}
J=\l(\f{R}{6\, H_{\mathrm{e}}^2}\r)^{\alpha(N)}
=\l[\f{H^2\,\l(2-\epsilon_1\r)}{H_{\mathrm{e}}^2}\r]^{\alpha(N)}
\label{eq: CJ(R)}
\end{equation}
and choose $\alpha(N)$ to be
\begin{equation}
\alpha(N)=\f{2\,\l(N-N_{\mathrm{e}}\r)}{{\mathrm{ ln}}\,
\l[H^2\,\l(2-\epsilon_1\r)/H_{\mathrm{e}}^2\r]},
\end{equation}
where $H_\mathrm{e}$ and $N_\mathrm{e}$ denote the Hubble parameter and
the e-fold at the end of inflation.
Such a choice essentially leads to $J(R)\propto a^2$, thereby guaranteeing 
a scale invariant spectrum for the magnetic field.
However, the action~\eqref{eq:a-emf-J-R} of the electromagnetic field described 
by the coupling function~\eqref{eq:J-R} with an $\alpha$ that depends on time 
will not be invariant under general coordinate transformations.
A theory which breaks general covariance seems unattractive and is also quite
likely to be unviable.

%%%%%%%%%%%%%%%%%%%%%%%%%%%%%%%%%%%%%%%%%%%%%%%%%%%%%%%%%%%%%%%%%%%%%%%%%%%%%%%

\subsection{Strength of magnetic fields at the present epoch}

The spectrum of magnetic fields evaluated at the end of inflation allows us 
to arrive at their strengths at the present epoch. 
In the conventional picture, the epoch of reheating is supposed to succeed
inflation.
During reheating, when the energy from the inflaton is being transferred to 
the particles constituting matter, the universe is expected to be filled with
a plasma of charged particles.
The creation of charged particles results in a rapid rise in the conductivity
of the plasma during reheating and, as a result, the electric fields are shorted
out, i.e. they decay exponentially. 
Thereafter, the magnetic fields are supposed to evolve
adiabatically with the expansion of the universe due to the fact that the fluxes
freeze in the highly conducting plasma (for a discussion on these points, see, 
for instance, Refs.~\cite{Subramanian:2009fu,Durrer:2013pga}).

Let us consider the simple scenario wherein reheating occurs instantaneously 
at the termination of inflation.
In such a case, the spectrum of the magnetic field today, say, $\pb^0(k)$, 
can be related to the spectrum $\pb(k)$ at the end of inflation as follows: 
\begin{equation}
\pb^0(k) \simeq \pb(k)\, \l(\f{a_\mathrm{e}}{a_{0}}\r)^4,\label{eq:pbe-pb0}
\end{equation}
where $a_\mathrm{e}$ is the scale factor at the end of inflation, while $a_0$
denotes the scale factor today.
The ratio $a_\mathrm{e}/{a_{0}}$ can be determined from the conservation of
entropy, i.e. the constancy of the quantity $g_\mathrm{s}\,T^3\,a^3$ from the 
end of inflation until today, where  $T$ is the temperature of radiation at 
a given epoch and $g_\mathrm{s}$ represents the effective relativistic degrees 
of freedom that contribute to the entropy.
As a result, we can write
\begin{equation}
\f{a_0}{a_\mathrm{e}} 
= \l(\f{g_{\mathrm{s},\mathrm{e}}}{g_{\mathrm{s},0}}\r)^{1/3}\, 
\f{T_\mathrm {e}}{T_0},
\end{equation}
where $(T_\mathrm{e},g_{\mathrm{s},\mathrm{e}})$ and $(T_0,g_{\mathrm{s},0})$ 
denote the temperature  and the effective number of relativistic degrees of 
freedom at the onset of 
the radiation dominated epoch and today, respectively. 
The quantity $T_\mathrm{e}$ can be determined using the fact that, in the case 
of instantaneous reheating, the energy density at the end of inflation equals
that of radiation at the epoch, leading to $\rhoI \simeq 3\, \HI^2\,\Mpl^2 \simeq 
g_{\mathrm{r},\mathrm{e}}\,(\pi^2/30)\,T_\mathrm{e}^4$, where $g_{\mathrm{r}}$
denotes the effective number of relativistic degrees that contribute to the 
energy density of radiation.
For simplicity, if we assume that $g_\mathrm{r} \simeq g_\mathrm{s}$, upon 
using the above relation, we can arrive at
\begin{equation}
\f{a_0}{a_\mathrm{e}} \simeq \l(\f{g_\mathrm{e}}{g_0}\r)^{1/3}\,
\l(\frac{90\,\HI^2\,\Mpl^2}{g_\mathrm{e}\,\pi^2\,T_0^4}\r)^{1/4}.
\end{equation}
If we consider $g_\mathrm{e} = 106.75$, since $g_0 = 3.36$ and $T_0
= 2.725\, \mathrm{K}$, we obtain that 
\begin{equation}
\f{a_0}{a_\mathrm{e}} \simeq  2.8\times 10^{28}\, 
\l(\frac{\HI}{10^{-5}\,\Mpl}\r)^{1/2}.
\end{equation}
Given the scale invariant spectrum~\eqref{eq:pb-ne2h}
for the magnetic field at the end of inflation in the $n=2$, helical 
case, upon substituting the above expression for $a_0/a_\mathrm{e}$ in
Eq.~\eqref{eq:pbe-pb0}, we can estimate the present day strength of the 
magnetic field, say, $B_0$ (at any scale), to be
\begin{equation}
B_{0} \simeq 4.5 \times 10^{-12}\, \l(\f{\HI}{10^{-5}\,\Mpl}\r)\,
f^{1/2}(\gamma)\; \mathrm{G},
\end{equation}
where  the function $f(\gamma)$ is given by Eq.~\eqref{eq:fg}.
Recall that, in the non-helical case, since $\gamma=0$, we have $f(\gamma)=1$.
Therefore, when parity is conserved, if inflation occurs over energy scales
such that $10^{-10} \lesssim \HI/\Mpl \lesssim 10^{-5}$, then inflationary 
magnetogenesis can be expected to lead to magnetic fields of strength in 
the range $10^{-17} \lesssim B_0 \lesssim 10^{-11}\,\mathrm{G}$ today.
As we shall discuss later, to avoid backreaction due to the generated 
electromagnetic fields, the helicity parameter $\gamma$ is constrained
to be less than about~$2.5$.
We find that, when parity is violated, the above-mentioned strengths of 
the magnetic fields today are amplified by a factor of about $34$ when 
$\gamma \simeq 1$ and by a factor of about $4.4\times10^3$ when $\gamma 
\simeq 2$.

%%%%%%%%%%%%%%%%%%%%%%%%%%%%%%%%%%%%%%%%%%%%%%%%%%%%%%%%%%%%%%%%%%%%%%%%%%%%%%%

\section{Coupling function in slow roll inflationary models}\label{sec:pbe-srm}

Before we go on to discuss inflationary models leading to features in the
scalar power spectrum, we shall evaluate the spectra of electromagnetic
fields generated in slow roll inflation.
Specifically, we shall discuss the forms of the coupling function~$J(\phi)$ 
that are required to generate nearly scale invariant magnetic fields in slow 
roll inflation.
This simple exercise proves to be instructive when we later consider 
situations involving departures from slow roll.

Note that, in terms of e-folds, the coupling function~\eqref{eq:J} is given
by $J(N)=\mathrm{exp}\,[n\,(N-\Ne)]$, where $\Ne$ denotes the e-fold at the 
end of inflation.
Since the evolution of the field $\phi(N)$ will depend on the inflationary
potential, it should be evident that a specific function~$J(\phi)$ will not 
lead to the above-mentioned form of~$J(N)$ in all the models.
We shall now construct the coupling functions $J(\phi)$ that result in the 
required $J(N)$ in some of the popular inflationary models that permit 
slow roll inflation.
For these choices of the coupling functions, assuming $n=2$, we shall also 
numerically evaluate the power spectra of the electromagnetic fields in 
these potentials. 
We shall impose the initial conditions on the electromagnetic modes when $k
\simeq 10^2\,\sqrt{J''/J}$, evolve the modes until late times and evaluate 
the spectra at the end of inflation.

We shall consider three forms for the potential~$V(\phi)$.
The first model we shall consider is the popular quadratic potential given by
\begin{equation}
V(\phi)=\f{m^2}{2}\,\phi^2.\label{eq:qp}
\end{equation}
In such a potential, it is well known that, under the slow roll approximation,
the evolution of the field can be expressed as
\begin{equation}
\phi^2(N)\simeq \phie^2+4\,\, (\Ne-N)\,\Mpl^2,\
\end{equation}
where $\phie\simeq \sqrt{2}\,\Mpl$ denotes the value of the field at the end 
of inflation.
Clearly, we can arrive at the form of $J(N)$ that we desire if we choose $J(\phi)$ 
to be (in this context, see Refs.~\cite{Kanno:2009ei,Watanabe:2009ct})
\begin{equation}
J(\phi)= \mathrm{exp}\,\l[-\f{n}{4\,\Mpl^2}\, 
(\phi^2-\phie^2)\r].\label{eq:J-qp}
\end{equation}
Recall that, COBE normalization determines the value of the parameter~$m$, and 
we find that we need to choose $m=7.18 \times 10^{-6}\,\Mpl$ to arrive at 
the observed scalar amplitude at the pivot scale~\cite{Planck:2018jri}.
To evolve the background, we shall choose the initial values of the field 
and the first slow roll parameter to be~$\phii =16.5\, \Mpl$ and $\e1i = 
7.346 \times10^{-3}$, respectively.
In such a case, we find that inflation lasts for~$68.6$ e-folds in the model. 

The second example we shall consider is the small field model described by 
the potential
\begin{equation}
V(\phi)=V_0\,\l[1-\l(\f{\phi}{\mu}\r)^q\r]\label{eq:sfm} 
\end{equation}
and we shall focus on the case wherein $q=2$.
On working in the slow roll approximation, the evolution of the field in such 
a model can be written as 
\begin{equation}
\mu^2\,\mathrm{ln}\l(\f{\phi}{\phie}\r)
-\f{1}{2}\,\l(\phi^2-\phie^2\r) \simeq 2\,(N-\Ne)\,\Mpl^2,
\end{equation}
with $\phie$ again denoting the value of the field at the end of inflation.
Hence, we can arrive at the $J(N)$ of our interest if we choose the coupling 
function $J(\phi)$ to be
\begin{equation}
J(\phi)\simeq \l(\f{\phi}{\phie}\r)^{n\,\mu^2/2\Mpl^2}\,
\mathrm{exp}\,\l[-\f{n}{4\,\Mpl^2}\, 
(\phi^2-\phie^2)\r].\label{eq:J-sfm}
\end{equation}
If we assume that $\mu\gg \Mpl$, then we find that $\phie\simeq \mu$.
We shall choose $\mu=10\,\Mpl$.
We find that COBE normalization leads to $V_0= 5.38 \times 10^{-10}\,\Mpl^4$.
We have set the initial values of the field and the first slow roll
parameter to be~$\phii =1.6\, \Mpl$ and $\e1i = 5.39\times 10^{-4}$, which 
lead to about $68.4$~e-folds of inflation. 

The third case that we shall consider is the Starobinsky model described by 
the potential
\begin{equation}
V(\phi)=V_0\,\l[1-\mathrm{exp}\l(-\sqrt{\f{2}{3}}\,\f{\phi}{\Mpl}\r)\r]^2.
\label{eq:sm1}
\end{equation}
As we shall consider another model due to Starobinsky later, we shall refer 
to this potential as the first Starobinsky model.
In this model, the evolution of the field in the slow roll approximation is 
described by the expression 
\begin{eqnarray}
N-\Ne &\simeq & 
-\f{3}{4}\, \biggl[\mathrm{exp}\l({\sqrt{\f{2}{3}}}\,\f{\phi}{\Mpl}\r) 
- \mathrm{exp}\l({\sqrt{\f{2}{3}}}\,\f{\phie}{\Mpl}\r)\nn\\
& & -\, \sqrt{\f{2}{3}}\,\l(\f{\phi}{\Mpl} - \f{\phie}{\Mpl}\r)\biggr],
\end{eqnarray}
where the value of the field at the end of inflation, viz. $\phie$, is determined 
by the relation $\mathrm{exp}\,[\sqrt{(2/3)}\,\phie/\Mpl] \simeq 1+2/\sqrt{3}$.
Therefore, to achieve the desired dependence of the coupling function on
the scale factor, we can choose $J(\phi)$ in the model to be
\begin{eqnarray}
J(\phi) 
&=& \mathrm{exp}\,
\biggl\{-\f{3\,n}{4}\, \biggl[\mathrm{exp}\l({\sqrt{\f{2}{3}}}\,
\f{\phi}{\Mpl}\r) - \mathrm{exp}\l({\sqrt{\f{2}{3}}}\,\f{\phie}{\Mpl}\r)\nn\\
& &-\, \sqrt{\f{2}{3}}\,\l(\f{\phi}{\Mpl} - \f{\phie}{\Mpl}\r)\biggr]\biggr\}.
\label{eq:J-sm1}
\end{eqnarray}
Again, COBE normalization fixes the overall amplitude of the potential to be 
$V_0 = 1.43 \times 10^{-10}\, \Mpl^4$.
We have chosen the initial values of the field and the first slow roll 
parameter to be $\phi_\mathrm{i} =5.6\, \Mpl$ and $\e1i = 1.453\times10^{-4}$.
We find that, for the above-mentioned value of $V_0$, these initial conditions 
lead to about $69.5$ e-folds before inflation ends.  

\begin{figure*}
\centering
\includegraphics[width=8.50cm]{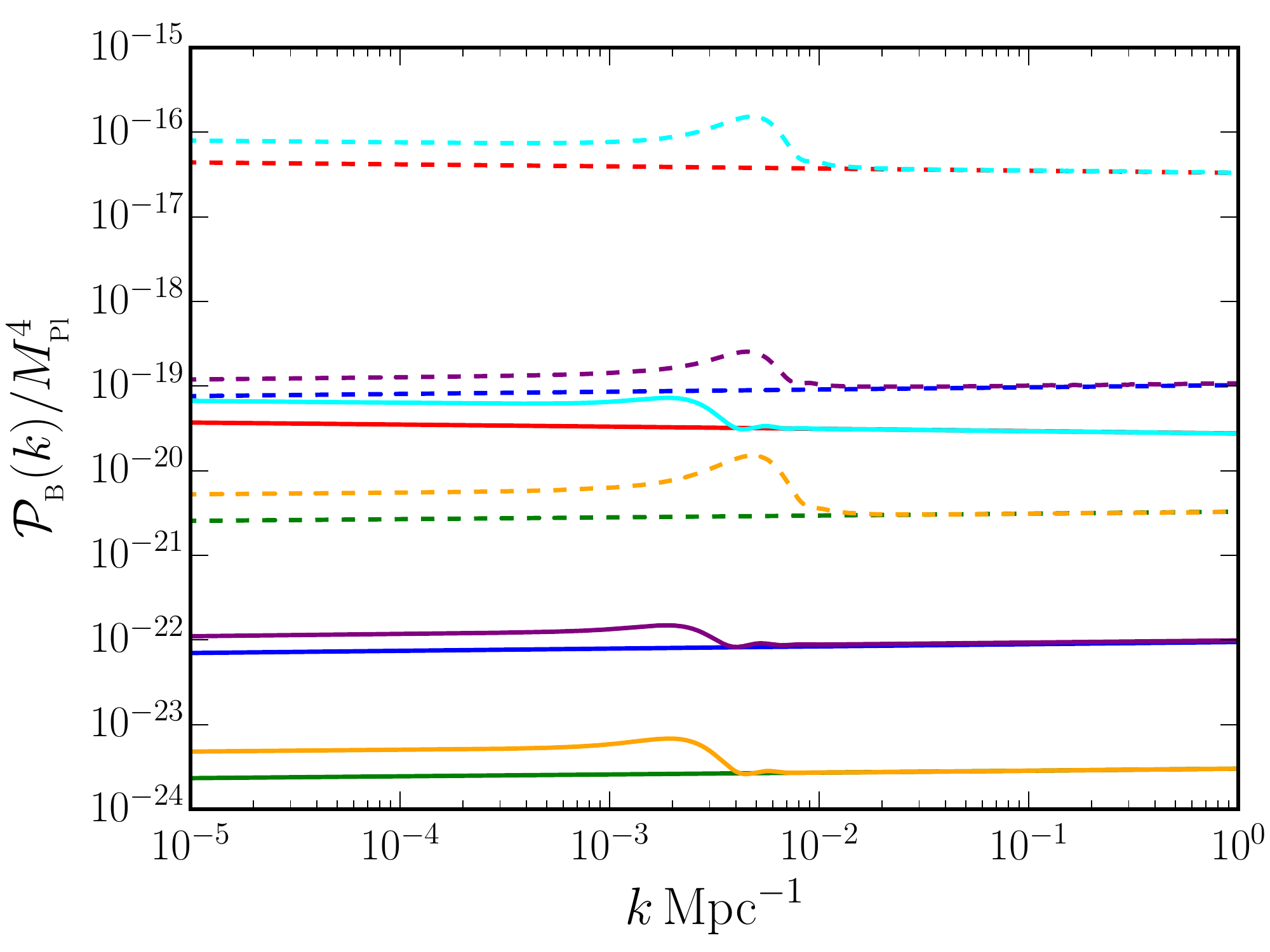}
\hskip 5pt
\includegraphics[width=8.50cm]{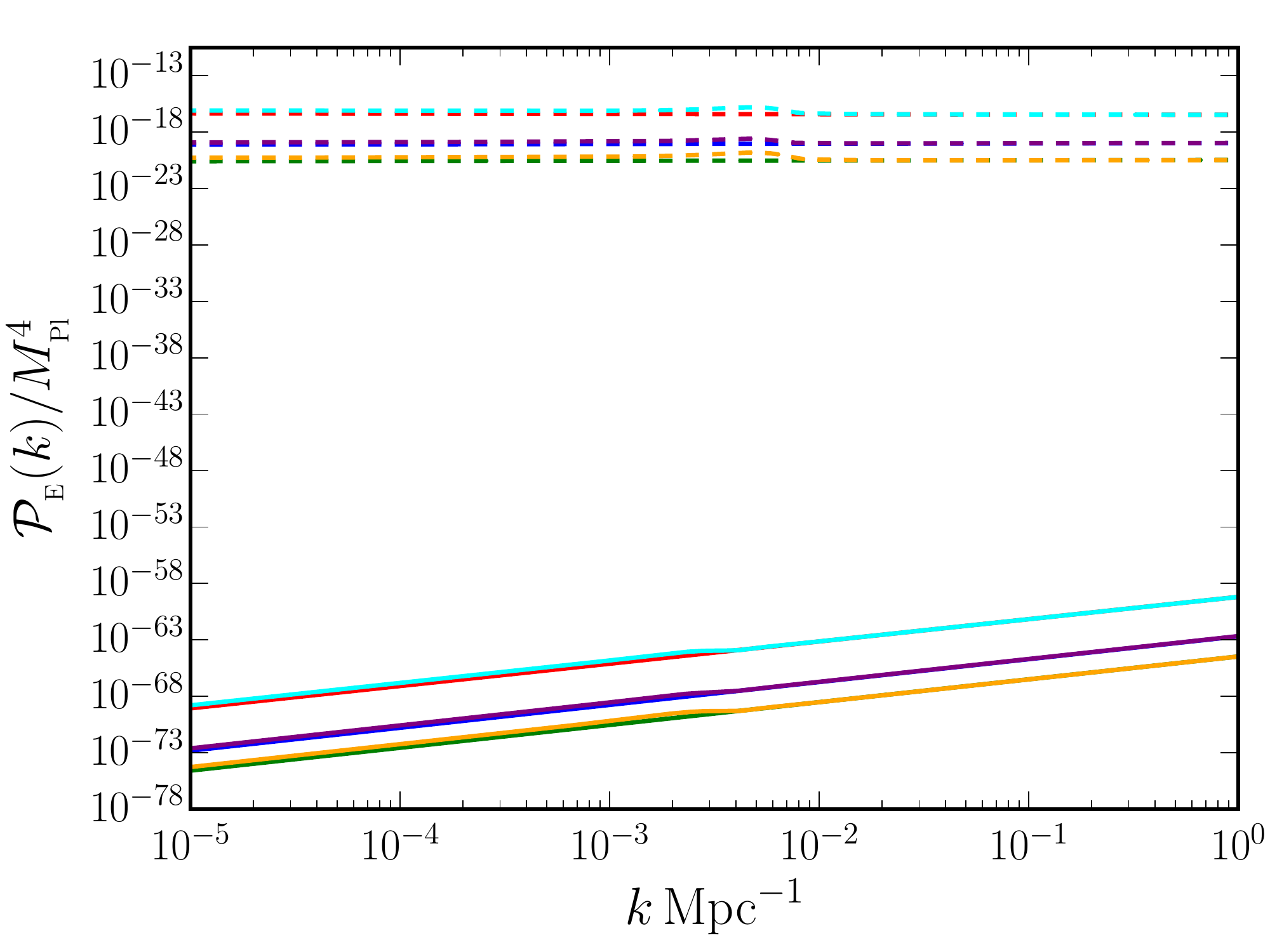}
\caption{The spectra of the magnetic (on the left) and electric (on the 
right) fields arising in the three slow roll inflationary models, viz. 
the quadratic potential (in red), the small field model (in blue) and the 
first Starobinsky model (in green), have been plotted over the CMB scales.
We have also plotted the corresponding spectra when a step has been introduced 
in these potentials (in cyan, purple and orange, respectively), a scenario we
shall discuss later in Subsec.~\ref{subsec:pws-emf}.
Moreover, we have plotted the spectra in both the non-helical (as solid lines) 
and helical (as dashed lines) cases.
We have worked with the parameters mentioned in the text and we have set $n=2$ 
in arriving at the spectra. 
In the helical case, we have set~$\gamma=1$.
We should mention that the shapes and amplitudes of these numerically 
evaluated spectra roughly match the analytical estimates discussed in 
the text.
For instance, the spectrum of the magnetic field is nearly scale invariant 
in all the models (and in both the non-helical and helical cases), modulo 
a small step-like feature that arises when a step is introduced in the potential.
Also, the spectrum of the electric field behaves as $k^2$ in the non-helical
case and it is scale invariant and matches the amplitude of the magnetic field
in the helical case, as we had discussed.
Further, clearly, the amplitude of the spectrum of the helical magnetic field 
is about $10^3$ larger than the amplitude of the non-helical field, as 
expected when~$\gamma=1$.}\label{fig:pbe-srm}
\end{figure*}

Let us now try to understand the amplitude and shape of the spectra of the
electromagnetic fields that arise in these models.
Evidently, to arrive at a nearly scale invariant spectrum for the magnetic
field, we shall choose to work with~$n=2$. 
Since the inflationary models introduced above will lead to a scale factor of 
the slow roll form (rather of the de Sitter type), clearly, we can expect the 
spectrum of the magnetic field in both the non-helical and helical cases to
exhibit a small tilt.
Moreover, in these situations, the spectrum of the electric field can be expected 
to be nearly scale invariant (as the spectrum of the magnetic field) in the helical
case, while it can be expected to behave nearly as~$k^2$ in the non-helical case.
In Fig.~\ref{fig:pbe-srm}, we have plotted the spectra arising in the three slow 
roll models that we discussed above.
%%%%%%%%%%%%%%%%%%%%%%%%%%%%%%%%%%%%%%%%%%%%%%%%%%%%%%%%%%%%%%%%%%%%%%%%%%%%%%%
%%%%%%%%%%%%%%%%%%%%%%%%%%%%%%%%%%%%%%%%%%%%%%%%%%%%%%%%%%%%%%%%%%%%%%%%%%%%%%%
Interestingly, we find that, while the power spectrum for the
non-helical magnetic field arising in the case of the quadratic potential has 
a small red tilt, the spectral tilt happens to be slightly blue in the cases of 
the small field and the Starobinsky models.
One may have naively imagined that, in such situations, it would be possible to 
express the spectral tilts $\nb$ and $\ne$ {\it completely}\/ in terms of the 
slow roll parameters.
This would have indeed been true had we assumed that $J\propto a^n$ and worked 
with the slow roll expression for the scale factor (in this context, see 
App.~\ref{app:si}).
However, our choices for the coupling functions [viz. Eqs.~\eqref{eq:J-qp}, 
\eqref{eq:J-sfm} and~\eqref{eq:J-sm1}] do not {\it exactly}\/ mimic the behavior 
of $J\propto a^n$, but contain small departures from it.
As a result of these deviations, we find that the spectral indices depend on 
the parameters describing the potential apart from the slow roll parameters.
In App.~\ref{app:si}, we show that, a simple analytical estimate of the spectral 
indices indeed match the results we have numerically obtained in all these three 
cases.

Let us now estimate the amplitude of the electromagnetic spectra in the 
slow roll models.
Let us first consider the non-helical case.
It can be easily shown that, when $n=2$, the amplitude of the spectra 
of the magnetic and electric fields at the pivot scale~$k_\ast$ can be 
expressed as [cf. Eqs.~\eqref{eq:ne2}] 
\begin{subequations}
\begin{eqnarray}
\f{\pb(k)}{\Mpl^4} 
& \simeq & \f{9\,\pi^2}{16}\,\l(r\,A_\mathrm{s}\r)^2,\\
\f{\pe(k)}{\Mpl^4} 
& \simeq & \f{\pb(k)}{9\,\Mpl^4}\,\l(\f{k_\ast}{\ke}\r)^2
\simeq  \f{\pb(k)}{9\,\Mpl^4}\,\mathrm{e}^{-100}.
\end{eqnarray}
\end{subequations}
In these expressions, $A_\mathrm{s}=2.1\times 10^{-9}$ denotes the observed 
amplitude of the scalar power spectrum at the pivot scale and $r$~represents 
the tensor-to-scalar ratio~\cite{Planck:2018jri,BICEPKeck:2021gln}.
Note that, we have set $\ke \simeq -1/\ee$, where, as we have indicated earlier, 
$\ke$ is the wave number that leaves the Hubble radius at the end of inflation.
Also, in arriving at the final equality in the above expression for $\pe(k)$, 
we have assumed that the pivot scale leaves the Hubble radius $50$ e-folds 
before the end of inflation, as we have done in the numerical evaluation
of the electromagnetic spectra plotted in Fig.~\ref{fig:pbe-srm}.
In the three slow roll inflationary models of our interest, viz. the quadratic
potential, the small field model and the Starobinsky model,  the tensor-to-scalar
ratio can be easily estimated to be $r\simeq (1.6\times10^{-1},5.79\times10^{-2},
4.8\times10^{-3})$.
The above expressions then suggest that these models will generate non-helical
magnetic fields of amplitudes $\pb(k) \simeq (6.27\times 10^{-19},8.21\times10^{-20},
5.64\times 10^{-22})\,\Mpl^4$.
Moreover, according to expressions above, $\pb(k) \simeq 10^{-20}\,\Mpl^4$ 
implies that $\pe(k)\simeq 10^{-66}\,\Mpl^4$.
These estimates roughly match the results we have arrived at numerically and 
have illustrated in Fig.~\ref{fig:pbe-srm}.
Further, since $\pb(k)\gg \pe(k)$ in the non-helical case, clearly, most of the
energy in the generated electromagnetic fields is in the magnetic field.
Lastly, since $\pb(k)/\Mpl^4 \simeq (\HI/\Mpl)^4 \lesssim 10^{-20}$ in these 
models, we have $\pb(k)/\Mpl^4 \ll \rhoI/\Mpl^4 \sim \HI^2/\Mpl^2$, where,
recall that, $\rhoI$ is the energy density of the inflaton.
This suggests that the energy density in the generated electromagnetic field 
is smaller than the background energy density and hence these scenarios do
not suffer from the backreaction problem (for an early discussion in this 
context, see Ref.~\cite{Kanno:2009ei}, for more recent discussions, see 
Ref.~\cite{Ng:2014lyb,Markkanen:2017kmy}).

Let us now turn to case of the helical electromagnetic fields.
In the helical case, when $n=2$, the amplitude of the spectra of 
the magnetic and electric fields can be expressed as [cf. Eqs.~\eqref{eq:ne2h}] 
\begin{subequations}
\begin{eqnarray}
\f{\pb(k)}{\Mpl^4}
& \simeq & \f{9\,\pi^2}{16}\,\l(r\,A_\mathrm{s}\r)^2\,f(\gamma),\\
\f{\pe(k)}{\Mpl^4}
& \simeq & \f{\pb(k)}{\Mpl^4}\,\gamma^2,
\end{eqnarray}
\end{subequations}
where $f(\gamma)$ is given by Eq.~\eqref{eq:fg}.
Note that, in contrast to the non-helical case, the energy density in the
electric field is now comparable to that of the magnetic field and, in fact,
the contribution due to electric field dominates when $\gamma > 1$.
Therefore, if we need to avoid backreaction due to the helical 
electromagnetic fields which have been generated, we require that $\pb(k)
+\pe(k)\ll \rhoI$.
Since we are considering inflationary models wherein $H/\Mpl\lesssim 10^{-5}$, 
on using the above expressions for the spectra of the electromagnetic fields, 
we find that the condition for avoiding backreaction leads to 
$f(\gamma)\,(1+\gamma^2) \lesssim 10^{10}$.
This limits the value of $\gamma$ to be $\gamma \lesssim 2.5$.
In Fig.~\ref{fig:pbe-srm}, assuming $\gamma=1$, we have also plotted the spectra 
of the helical electromagnetic fields in the three inflationary models discussed
above.
When $\gamma=1$, we find that $f(\gamma)\simeq 10^3$.
As should be evident from the figure, the spectra of the helical magnetic fields 
is indeed amplified by the factor of~$10^3$ when compared to the non-helical case 
in all the models.
Also, it should be clear that, the spectra of the helical electric and magnetic 
fields are comparable, as expected.

%%%%%%%%%%%%%%%%%%%%%%%%%%%%%%%%%%%%%%%%%%%%%%%%%%%%%%%%%%%%%%%%%%%%%%%%%%%%%%%

\section{Inflationary models leading to features in the scalar power 
spectrum}\label{sec:im-lf}

In this section, we shall discuss specific examples wherein deviations from
slow roll inflation lead to features in the scalar power spectrum.
In due course, we shall discuss the effects of such deviations on the spectra 
of the electromagnetic fields.
When departures from slow roll occur, in general, the background and the 
modes describing the scalar perturbations prove to be difficult to evaluate
analytically, and one resorts to numerics. 
We shall begin by recalling a few essential points regarding the evaluation
of the scalar power spectrum.

Let $f_k$ denote the Fourier modes associated with the curvature perturbation.
The modes $f_k$ satisfy the differential equation~(see, for instance, 
the reviews~\cite{Mukhanov:1990me,Martin:2003bt,Martin:2004um,Bassett:2005xm,
Baumann:2008bn,Sriramkumar:2009kg,Kinney:2009vz,Baumann:2009ds,
Sriramkumar:2012mik,Linde:2014nna,Martin:2015dha})
\begin{equation}
f_k''+2\,\f{z'}{z}\,f_k'+k^2\,f_k=0, 
\end{equation}
where the quantity $z$ is given by $z=\sqrt{2\,\epsilon_1}\,\Mpl\,a$, with 
$\epsilon_1=-\dot{H}/H^2$ being the first slow roll parameter. 
In terms of the Mukhanov-Sasaki variable $v_k= f_k\,z$, the above equation 
reduces to  
\begin{eqnarray}
v_k'' + \l(k^2 - \f{z''}{z}\r)\,v_k= 0.\label{eq:de-vk} 
\end{eqnarray}
The standard Bunch-Davies initial conditions are imposed on the variable~$v_k$ 
at very early times when $k \gg \sqrt{z''/z}$, which corresponds to the modes 
being in sub-Hubble regime.
The scalar power spectrum is defined as 
\begin{equation}
\ps(k) = \f{k^3}{2\, \pi^2}\,\vert f_k\vert^2 =
\f{k^3}{2\, \pi^2}\,{\f{\vert v_k\vert^2}{z^2}}. \label{eq:ps-d}
\end{equation}
The modes $f_k$ are evolved from the Bunch-Davies initial conditions and the 
power spectra are evaluated in the super-Hubble regime at late times, i.e. 
when $k \ll \sqrt{z''/z}$. 
Since the modes oscillate in the sub-Hubble domain and the amplitude of the 
scalar modes are known to freeze on super-Hubble scales, numerically, one 
often finds that it is sufficient to evolve the modes from $k\simeq 10^2\, 
\sqrt{z''/z}$ and evaluate the power spectrum when $k \simeq 10^{-5}\,
\sqrt{z/''z}$ (in this context, see, for instance, Ref.~\cite{Hazra:2012yn}).

%%%%%%%%%%%%%%%%%%%%%%%%%%%%%%%%%%%%%%%%%%%%%%%%%%%%%%%%%%%%%%%%%%%%%%%%%%%%%

\subsection{Potentials with a step}\label{subsec:pws}

The first scenario leading to features in the scalar power spectrum that 
we shall consider are inflationary potentials wherein a step has been 
introduced by hand.
Given an inflationary model described by the potential $V(\phi)$, we shall 
introduce a step in the potential as follows (for an early discussion,
see Ref.~\cite{Adams:2001vc}):
\begin{equation}
V_\mathrm{step}(\phi)=V(\phi)\,\l[1
+\alpha\,\mathrm{tanh}\,\l(\f{\phi-\phi_0}{\Delta\phi}\r)\r],
\label{eq:step}
\end{equation}
where, evidently, $\phi_0$, $\alpha$ and $\Delta\phi$ denote the location, 
the height and the width of the step.  
For the original potential $V(\phi)$, we shall consider the three models
admitting slow roll we had discussed in the previous section.
Also, as far as the parameters regarding the original potential is concerned, 
we shall work with the values we had mentioned earlier.
Moreover, we shall work with the following values of the three parameters 
describing the step:~$(\phi_0,\alpha,\Delta\phi)=(14.6616\, \Mpl, 1.55177
\times 10^{-3}, 2.60584\times 10^{-2}\, \Mpl), (2.14\, \Mpl, -0.1153\times 10^{-3},
0.0070\,\Mpl)$ and $(5.3052\, \Mpl, 5.0\times 10^{-5}, 5.0 \times 10^{-3}\, \Mpl)$ 
in the cases of the quadratic potential, the small field model and the first
Starobinsky model, respectively.

\begin{figure*}
\centering
\includegraphics[width=8.50cm]{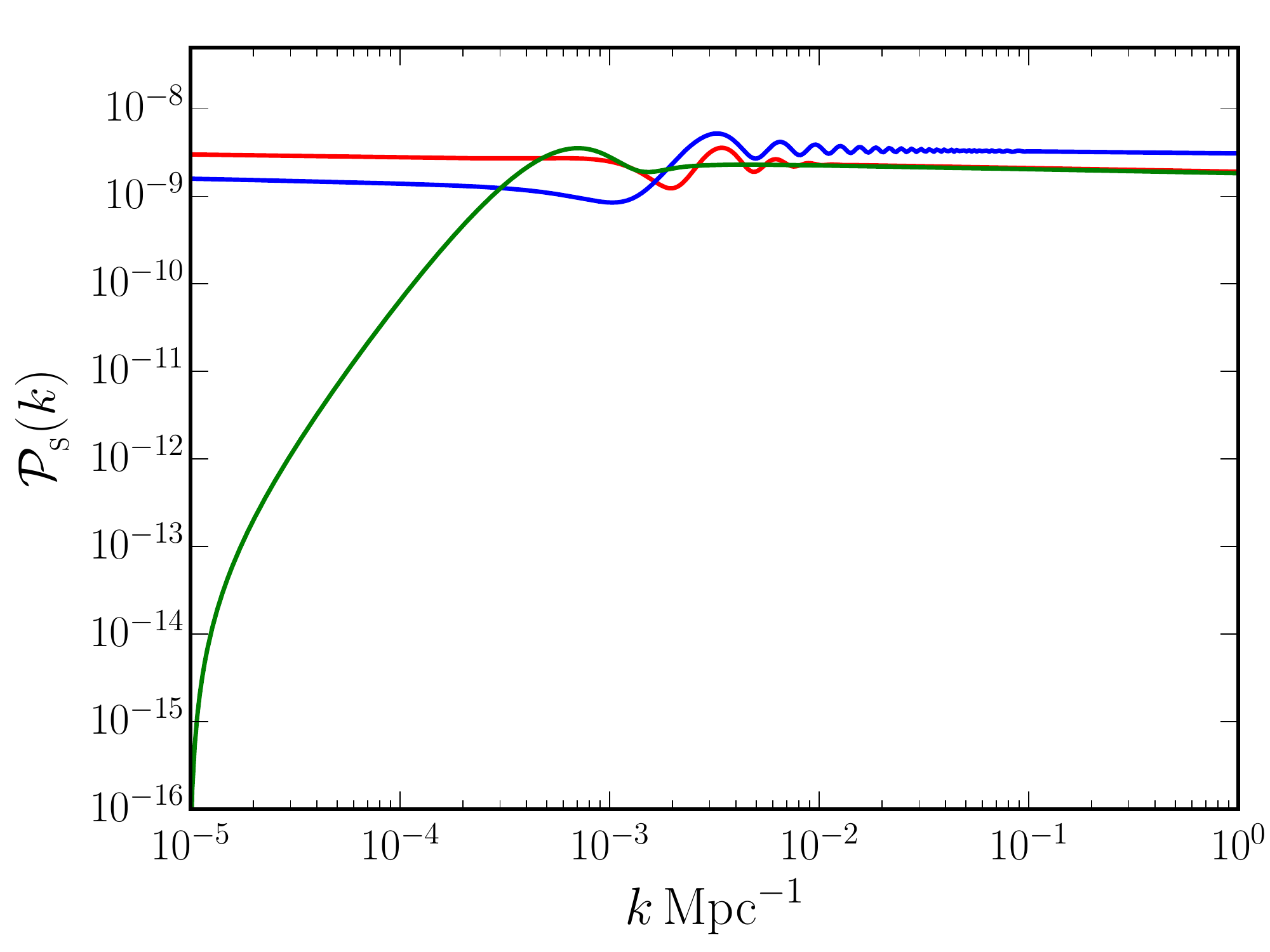}
\hskip 5pt
\includegraphics[width=8.50cm]{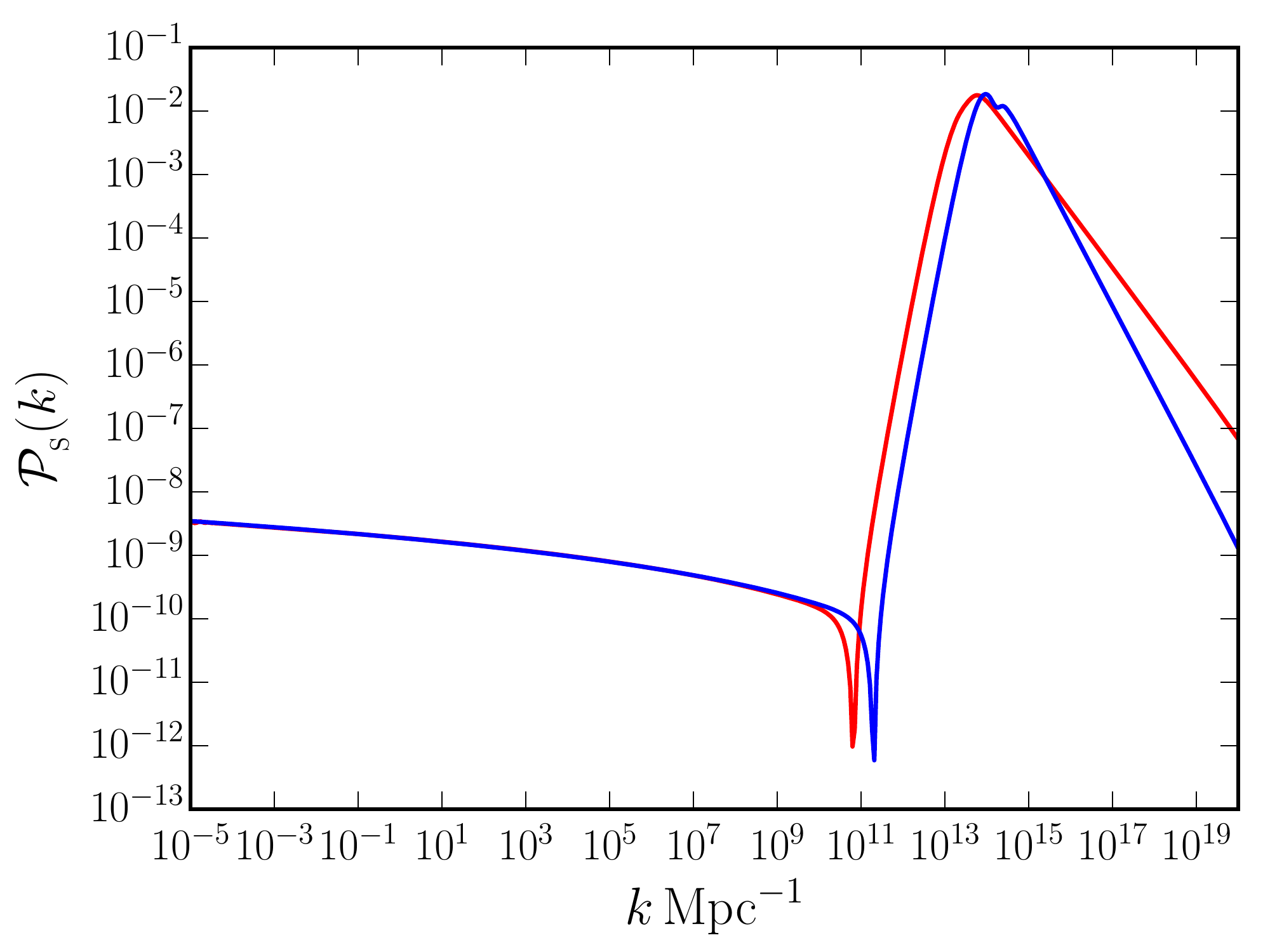}
\caption{The scalar power spectra with features over the CMB and smaller scales 
have been plotted in some of the inflationary models that we have considered.
We have plotted the scalar spectra with features over the CMB scales (on the 
left) in the cases of the quadratic potential with a step (in red), the second 
Starobinsky model described by the linear potential with a sharp change in its
slope (in blue), and the first punctuated inflation model (in green).
We have also plotted the scalar power spectra with a peak in power at small
scales (on the right) that are generated in the  ultra slow roll (in red)
and the second punctuated (in blue) inflation models.
As we shall point out later, the scalar spectra with a sharp rise in power on 
small scales are often considered to produce significant amount of primordial 
black holes.}\label{fig:sps}.
\end{figure*}

As we described above, to arrive at the scalar power spectrum, we impose the 
initial conditions on the modes when $k\simeq 10^2\, \sqrt{z''/z}$ and evaluate 
the power spectrum when $k \simeq 10^{-5}\,\sqrt{z'' /z}$.
Moreover, in these three models, we shall assume that the pivot scale of
$k_\ast=0.05\, \mathrm{Mpc}^{-1}$ leaves the Hubble radius $50$~e-folds 
before the end of inflation.
The scalar power spectrum that arises with the introduction of the step in the 
quadratic potential is illustrated in Fig.~\ref{fig:sps}.
%%%%%%%%%%%%%%%%%%%%%%%%%%%%%%%%%%%%%%%%%%%%%%%%%%%%%%%%%%%%%%%%%%%%%%%%%%%%%%%
%%%%%%%%%%%%%%%%%%%%%%%%%%%%%%%%%%%%%%%%%%%%%%%%%%%%%%%%%%%%%%%%%%%%%%%%%%%%%%%
As one would expect, the introduction of the step in the potential leads to a 
short period of deviation from slow roll as the field crosses the step.
The deviation from slow roll, in turn, generates a short burst of oscillations
in the scalar power spectrum over wave numbers that leave the Hubble radius 
during the period of departure from slow roll.
It is known that such features in the power spectrum can improve the fit to
the CMB data to a certain extent~\cite{Hazra:2010ve,Benetti:2013cja}.

%%%%%%%%%%%%%%%%%%%%%%%%%%%%%%%%%%%%%%%%%%%%%%%%%%%%%%%%%%%%%%%%%%%%%%%%%%%%%

\subsection{Suppressing power on large scales}

Since the advent the WMAP data, it has been known that a suppression 
in power on large scales comparable to the Hubble radius today leads to 
an improvement in the fit to the CMB data (for earlier
discussions, see Refs.~\cite{Contaldi:2003zv,Sinha:2005mn,Powell:2006yg,
Jain:2008dw,Jain:2009pm,Hazra:2014jka,Hazra:2014goa}; for a recent 
discussion, see Ref.~\cite{Ragavendra:2020old}).
In this subsection, we shall discuss two models that have often been 
considered in this context.

The first example that we shall consider is a model due to Starobinsky,
which is governed by the potential~\cite{Starobinsky:1992ts}
\begin{equation}
V(\phi) = \begin{cases}
V_0+A_{+}\, (\phi-\phi_0),&  \text{for $\phi > \phi_0$},\\ 
V_0+A_{-}\, (\phi-\phi_0),&  \text{for $\phi < \phi_0$}.
\end{cases}\label{eq:sm2}    
\end{equation} 
To distinguish from the Starobinsky model~\eqref{eq:sm1} which permits 
slow roll inflation that we had discussed earlier, we shall refer to
the above potential as the second Starobinsky model.
Evidently, the model consists of a linear potential with a sudden change in 
its slope at the point~$\phi_0$.
If we assume that the constant term~$V_0$ in the potential is dominant, then
the first slow roll parameter remains small and the scale factor can be 
described by the de Sitter form.
Under this condition, it is  possible to arrive at analytical solutions for
the evolution of the background~\cite{Starobinsky:1992ts,Martin:2011sn}. 
We shall discuss the evolution of the field later, when we consider the 
coupling between the inflaton and the electromagnetic field.
It is found that, as the field crosses $\phi_0$, while the first slow roll 
parameter remains small, the second and the third slow roll parameters 
turn large leading to a departure from slow roll.
Also, notice that the second derivative of the potential is described by a 
Dirac delta function with its peak at~$\phi_0$.
It is the Dirac delta function that dominates the behavior of the quantity 
$z''/z$ that appears in the Mukhanov-Sasaki equation~\eqref{eq:de-vk}. 
Working in the de Sitter approximation to describe the scale factor as well
as the scalar modes~$f_k$, the deviation from slow roll could be accounted 
for by essentially considering the effects due to the Dirac delta function.
In fact, under these conditions, it is possible to arrive at an analytical 
form for the power spectrum~\cite{Starobinsky:1992ts,Martin:2011sn,
Ragavendra:2020old}.
We shall instead arrive at the scalar power spectrum numerically.
In order to permit numerical analysis, we shall modify the potential 
so that the change in the slope is smooth and not abrupt.  
We shall assume that the potential is given by
\begin{eqnarray}
V(\phi) 
&=& V_0 +\f{1}{2}\, (A_{+}+A_{-})\, (\phi-\phi_0)\nn\\
& &+\,\f{1}{2}\, (A_{+}-A_{-})\, (\phi-\phi_0)\, 
\mathrm{tanh}\l(\f{\phi-\phi_0}{\Delta\phi}\r),\qquad\label{eq:ssm}
\end{eqnarray}
and work with the following values of the parameters involved:~$V_0= 2.98 
\times 10^{-9}\, \Mpl^4$, $A_+= 4.35881 \times 10^{-10}\,\Mpl^3$, 
$A_-=2.499 \times 10^{-10}\, \Mpl^3$, $\phi_0=5.628\,\Mpl$ and 
$\Delta\phi = 10^{-4}\,\phi_0$. 
We shall choose the initial value of the field and the first slow roll 
parameter to be $\phii =8.4348\,\Mpl$ and $\e1i= 10^{-4}$.

The second model that we shall consider is the so-called punctuated 
inflationary model described by the potential (in this context, see
Refs.~\cite{Jain:2008dw,Jain:2009pm,Ragavendra:2020old})
\begin{equation}
V(\phi)=\f{m^2}{2}\, \phi^2-\f{2\,m^2}{3\,\phi_0}\, \phi^3
+\f{m^2}{4\, \phi_0^2}\,\phi^4.\label{eq:pi1}
\end{equation}
It is easy to see that this potential contains a point of inflection 
at~$\phi_0$.
The point of inflection leads to two epochs of slow roll sandwiching
a brief period of departure from inflation, which has led to the name 
of punctuated inflation.
As we shall consider another model of punctuated inflation which leads 
to enhanced power at small scales in the following subsection, we shall 
refer to the above potential as the first model of punctuated inflation.
In this case, we shall work with the following values of the parameters 
involved:~$m=7.17\times10^{-8}\,\Mpl$ and $\phi_0=1.9654\,\Mpl$.
We shall choose the initial values of the field and the first slow roll
parameter to be  $\phii =12.0\,\Mpl$ and $\e1i= 2 \times 10^{-3}$. 

The drawback of these two models is that they lead to much 
longer epochs of inflation than the nominally required~$60$ 
odd e-folds~\cite{Ragavendra:2020old}.
In the Starobinsky model~\eqref{eq:sm2}, we stop the evolution by 
hand after~$72$ e-folds, and assume that the pivot scale leaves the Hubble
radius about $44.5$~e-folds earlier.
In the case of the punctuated inflationary model~\eqref{eq:pi1}, inflation
ends naturally after nearly $110.5$~e-folds and the pivot scale is assumed to 
exit the Hubble radius about $91$~e-folds before the termination of inflation.
The departure from slow roll in these two potentials leads to a step-like 
feature in the scalar power spectrum, as illustrated in Fig.~\ref{fig:sps}.

%%%%%%%%%%%%%%%%%%%%%%%%%%%%%%%%%%%%%%%%%%%%%%%%%%%%%%%%%%%%%%%%%%%%%%%%%%%%%

\subsection{Enhancing power on small scales}

Over the last few years, there has been a considerable interest in examining
models of inflation that lead to enhanced power on scales much smaller than
the CMB scales (in this context, see, for example, 
Refs.~\cite{Garcia-Bellido:2017mdw,Ballesteros:2017fsr,Germani:2017bcs,
Dalianis:2018frf,Bhaumik:2019tvl,Ragavendra:2020sop,Dalianis:2020cla}).
Apart from leading to copious production of primordial black holes, these
models can also generate secondary gravitational waves of considerable 
strengths, which can possibly be detected by the current and forthcoming 
gravitational wave observatories.
Most of these inflationary models contain a point of inflection (just as
the model of punctuated inflation we discussed in the previous subsection), 
which permits a brief period wherein the first slow roll parameter decreases
exponentially.
Such a period of ultra slow roll proves to be responsible for enhancing
the power on small scales in these models.

We shall consider two potentials that lead to enhanced power on small scales.
The first model that we shall consider, which leads to a brief period of
ultra slow roll, is described by the potential~\cite{Dalianis:2018frf}
\begin{eqnarray}
V(\phi) 
&=& V_0\,\biggl\{\mathrm{tanh}\l(\f{\phi}{\sqrt{6}\,\Mpl}\r)\nn\\ 
& &+\, A\,\sin\l[\f{1}{f_\phi}\,
\mathrm{tanh}\l(\f{\phi}{\sqrt{6}\,\Mpl}\r)\r]\biggr\}^2.\label{eq:usr}
\end{eqnarray}
We shall choose to work with the following values of the parameters 
involved:~$V_0 = 2\times10^{-10}\,\Mpl^4$, $A = 0.130383$ and $f_\phi 
= 0.129576$.
For these values of the parameters, the point of inflection in the 
potential is located at $\phi_0 = 1.05\,\Mpl$~\cite{Ragavendra:2020sop}.
Also, if we choose the initial value of the field to be $\phii=6.1\,\Mpl$, 
with $\e1i=10^{-4}$, we obtain about $66$~e-folds of inflation in the model.
Moreover, we shall assume that the pivot scale exits the Hubble radius about
$56.2$~e-folds prior to the termination of inflation.

The second model that we shall consider which permits punctuated inflation 
is described by the potential~\cite{Dalianis:2018frf,Dalianis:2020cla}
\begin{eqnarray}
V(\phi) &=& V_0\,\biggl[c_0 + c_1\,\tanh\, \l(\f{\phi}{\sqrt{6}\,\Mpl}\r)\nn\\ 
& &+\, c_2\,\tanh^2\l(\f{\phi}{\sqrt{6}\,\Mpl}\r)
+ c_3\,\tanh^3{\l(\f{\phi}{\sqrt{6}\,\Mpl}\r)}\biggr]^2.\nn\\
\label{eq:pi2}
\end{eqnarray}
In this case, we shall work with the following values for the parameters 
involved: $V_0=2.1 \times 10^{-10}\,\Mpl^4$, $c_0=0.16401$, $c_1=0.3$, 
$c_2=-1.426$ and $c_3=2.20313$.
As in the previous model, this potential also  contains a point of inflection.
For the above values for the parameters, the point of inflection is located at 
$\phi_0 = 0.53\,\Mpl$.
If we set the initial value of the field to be $\phii = 7.4\,\Mpl$ and 
choose $\e1i=10^{-3}$, for the above choice of parameters, we find that 
inflation is terminated after about $67.8$ e-folds.
Also, we shall assume that the pivot scale leaves the Hubble radius about 
$54.5$~e-folds before the end of inflation.

The scalar power spectra that arise in the above two potentials are illustrated 
in Fig.~\ref{fig:sps}.
Note that the power spectra exhibit a sharp rise in power on small scales 
in these models.
As has been repeatedly emphasized in the literature, it is the period of 
ultra slow roll, with its rather small value for the first slow roll 
parameter~$\epsilon_1$, that turns out to be responsible for the increased 
power in the scalar power spectrum on small scales (in this context, see, 
for instance, Ref.~\cite{Byrnes:2018txb}).

%%%%%%%%%%%%%%%%%%%%%%%%%%%%%%%%%%%%%%%%%%%%%%%%%%%%%%%%%%%%%%%%%%%%%%%%%%%%%%%

\section{Effects of deviations from slow roll on the electromagnetic 
power spectra}\label{sec:eds-s-emf}

Let us now turn to understand the effects of deviations from slow roll 
on the  power spectra of electric and magnetic fields.

%%%%%%%%%%%%%%%%%%%%%%%%%%%%%%%%%%%%%%%%%%%%%%%%%%%%%%%%%%%%%%%%%%%%%%%%%%%%%%%

\subsection{In potentials with a step}\label{subsec:pws-emf}

As we discussed earlier and illustrated in Fig.~\ref{fig:sps}, the 
introduction of the step in a potential which otherwise admits only 
slow roll inflation leads to a short burst of oscillations in the 
scalar power spectrum. 
In Sec.~\ref{sec:pbe-srm}, we had constructed coupling functions $J(\phi)$
[as given by Eqs.~\eqref{eq:J-qp}, \eqref{eq:J-sfm} and \eqref{eq:J-sm1}] in 
the three slow roll models~\eqref{eq:qp}, \eqref{eq:sfm} and~\eqref{eq:sm1}   
so that they lead to nearly scale invariant spectra for the magnetic field
when $n=2$.
Even after the introduction of the step, we have chosen to work with the 
above mentioned coupling functions $J(\phi)$ that we had constructed in the 
slow roll approximation.
In Fig.~\ref{fig:pbe-srm}, we have plotted the resulting spectra of the 
magnetic and electric fields arrived at numerically in both the non-helical
and helical cases. 
As should be clear from the figure, the step in the inflationary potential
only has a small effect on the spectra of the electromagnetic fields.
It essentially generates a small step-like feature in the power spectra.
This is not surprising since, for the choices of the parameters we have worked
with, the step in the potential leads to only a small and brief departure from 
slow roll inflation.

%%%%%%%%%%%%%%%%%%%%%%%%%%%%%%%%%%%%%%%%%%%%%%%%%%%%%%%%%%%%%%%%%%%%%%%%%%%%%%%

\subsection{In models leading to suppression of power on large scales}

In this context, we shall first consider the second Starobinsky model 
described by the potential~\eqref{eq:sm2}.
As we had mentioned earlier, in the model, the field rolls slowly until it
reaches~$\phi_0$ where the slope of the potential changes from $A_+$ to $A-$.
In the slow roll approximation, the evolution of the field prior to it
crossing~$\phi_0$ can be determined to be~\cite{Starobinsky:1992ts,Martin:2011sn}
\begin{eqnarray}
\phi_+(N) 
& \simeq & -\l(\f{V_0}{A_+}-\phi_0\r)\nn\\
& &+\,\biggl[\l(\phi_\mathrm{i}-\phi_0+ \f{V_0}{A_+}\r)^2
-2\,\Mpl^2\,N\biggr]^{1/2},\qquad\label{eq:phip-o} 
\end{eqnarray}
where $\phi_\mathrm{i}$ is the initial value of the field (i.e. at $N=0$).
If we choose to work with a suitably large value of $V_0$ so that it dominates
the potential, then the above expression simplifies to be
\begin{equation}
\phi_+(N) \simeq  \phi_\mathrm{i} -\f{A_+\,\Mpl^2}{V_0}\,N.\label{eq:phip}
\end{equation}
Evidently, once the field has crossed $\phi_0$ and slow roll has been restored, 
the evolution of the field can be expressed as
\begin{eqnarray}
\phi_-(N) 
& \simeq & -\l(\f{V_0}{A_-}-\phi_0\r)\nn\\
& &+\,\biggl[\l(\f{V_0}{A_-}\r)^2-2\,\Mpl^2\,(N-N_0)\biggr]^{1/2},\label{eq:phim-o}
\end{eqnarray}
where $N_0$ denotes the e-fold when the field crosses~$\phi_0$.
If we again assume that $V_0$ is dominant, then the above expression reduces 
to
\begin{equation}
\phi_-(N) \simeq  \phi_0 -\f{A_-\,\Mpl^2}{V_0}\,(N-N_0).\label{eq:phim}
\end{equation}
We should clarify here that, in arriving at the above expressions for the
evolution of the field after it has crossed~$\phi_0$, we have ignored the 
effects that arise due to the change in the slope.
As we had described, the change in the slope causes a brief period of 
departure from slow roll.
If we take into account the effects due to the deviation from slow roll,
the evolution of the field after it has crossed $\phi_0$ can be obtained 
to be~\cite{Starobinsky:1992ts,Martin:2011sn}
\begin{eqnarray}
\phi_-(N) 
& \simeq & \phi_0 
+\f{\Delta A\,\Mpl^2}{3\,V_0}\, \l[1-\mathrm{e}^{-3\,(N-N_0)}\r]\nn\\ 
& &-\,\f{A_-\,\Mpl^2}{V_0}\,(N-N_0),\label{eq:phim-f}  
\end{eqnarray}
where $\Delta A=(A_--A_+)$.
Upon comparing the above two equations, it should be obvious that it is 
the intermediate term that accounts for the departure from slow roll 
which occurs as the field crosses~$\phi_0$.
On using the above expressions describing the behavior of the field, one can 
show that, while the first slow roll parameter remains small, the second and 
the third slow roll parameters turn large as the field crosses~$\phi_0$.

Let us now turn to constructing the coupling function $J(\phi)$ for the second
Starobinsky model.
As we had done in the case of the models discussed in Sec.~\ref{sec:pbe-srm}, 
we can choose to work with the solutions for the field in the slow roll
approximation.
If we choose to do so, we are left with two choices, viz. the slow roll
solutions~\eqref{eq:phip-o} and~\eqref{eq:phim-o} for the field
before and after the transition.
In other words, we can work with either of the following choices for the
coupling function:
\begin{subequations}\label{eq:J-phipm}
\begin{eqnarray}
J_+(\phi) 
&=& J_{0+}\,\mathrm{exp}\,\biggl\{-\f{n}{2\,\Mpl^2}\,\biggl[\l(\phi_+-\phi_0
+\f{V_0}{A_+}\r)^2\nn\\
& &-\,\l(\phi_\mathrm{i}-\phi_0+\f{V_0}{A_+}\r)^2\biggr]\biggr\},\\
J_-(\phi) 
&=& J_{0-}\,\mathrm{exp}\,\biggl\{-\f{n}{2\,\Mpl^2}\,
\biggl[\l(\phi_- - \phi_0 + \f{V_0}{A_-}\r)^2\nn\\
& &-\l(\f{V_0}{A_-}\r)^2-2\,N_0\,\Mpl^2\biggr]\biggr\},
\end{eqnarray}
\end{subequations}
where the constants $J_{0\pm}$ are to be chosen suitably 
so that $J_\pm(\phi_\mathrm{e})=1$, i.e. the value of $J$ is unity at 
the end of inflation.

\begin{figure*}[!t]
\centering
\includegraphics[width=8.50cm]{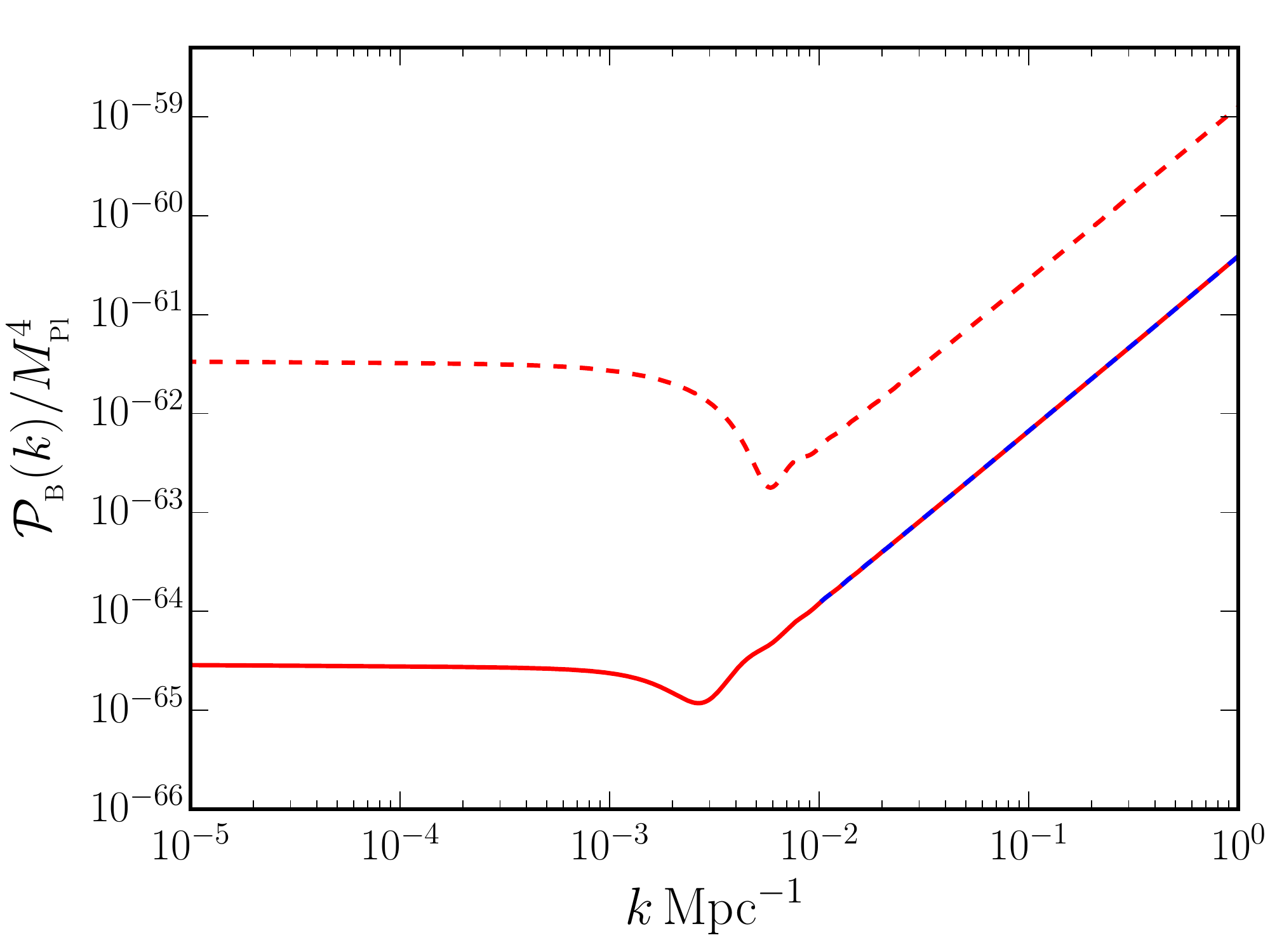} 
\hskip 5pt
\includegraphics[width=8.50cm]{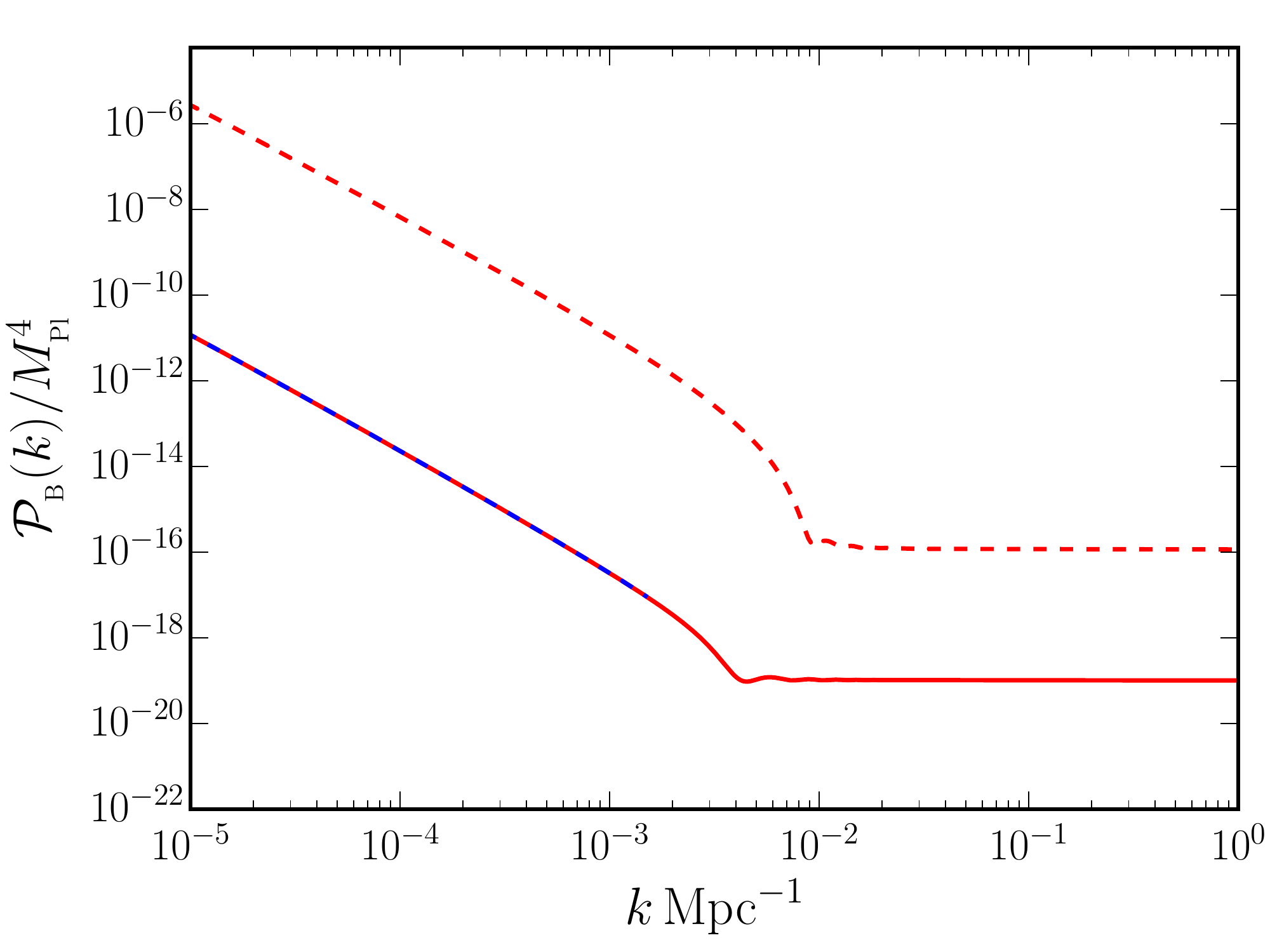} 
\caption{The power spectra of the magnetic field arising in the second 
Starobinsky model for the two choices of coupling functions $J_+(\phi)$ 
(on the left) and $J_-(\phi)$ (on the right) [cf. Eqs.~\eqref{eq:J-phipm}] 
have been plotted for $n=2$ in the non-helical (in solid red) as well as 
the helical (in dashed red) cases.
A linear fit (indicated in dashed blue) to the non-helical power spectra 
over the small and the large scales (on the left and the right) lead to 
the spectral indices $\nb=1.75$ and $\nb=-2.72$, respectively. 
For the values of the parameters we have worked with, the analytical estimates 
for these indices prove to be $\nb=1.71$ and $\nb=-2.98$, which are close to 
the numerically determined values.
As in Fig.~\ref{fig:pbe-srm}, we have set the helicity  parameter~$\gamma$ 
to be unity.
Moreover, note that, for $\gamma=1$, the spectra of the magnetic field over 
the scale invariant domain is about $10^3$ times larger in the helical case 
when compared to the non-helical one, as we had estimated earlier.
Lastly, we should add that, when the coupling function is given by~$J_-(\phi)$,
the strength of the magnetic fields generated is fairly large and hence the 
scenario will lead to a significant backreaction.}
\label{fig:peb-sm2}
\end{figure*}

The power spectra of the magnetic field for the two coupling functions 
$J_{\pm}(\phi)$ for the case of $n=2$ are plotted in Fig.~\ref{fig:peb-sm2}
for both the non-helical and helical cases.
%%%%%%%%%%%%%%%%%%%%%%%%%%%%%%%%%%%%%%%%%%%%%%%%%%%%%%%%%%%%%%%%%%%%%%%%%%%%%%%
%%%%%%%%%%%%%%%%%%%%%%%%%%%%%%%%%%%%%%%%%%%%%%%%%%%%%%%%%%%%%%%%%%%%%%%%%%%%%%%
A few points needs to be emphasized regarding the spectra we have 
obtained.
Firstly, the spectra are scale invariant only over either large or 
small scales.
Let $k_0$ be the mode which leaves the Hubble radius when the field 
crosses~$\phi_0$.
Then, clearly, for the choice of the coupling functions $J_+(\phi)$ 
and $J_-(\phi)$, the magnetic field spectra are scale invariant only 
over $k< k_0$ and $k>k_0$, respectively.
This should not come as a surprise as the coupling functions $J_\pm(\phi)$ 
have been constructed based on the behavior of the field in the slow roll
approximation before and after it crosses $\phi_0$.
Secondly, when $n=2$, for the coupling function $J_+(\phi)$, the spectral 
index of the magnetic field for $k> k_0$ can be estimated to be 
$\nb=-4\,\Delta A/A_+$, while for the function $J_-(\phi)$ the index 
over large scales can be determined to be $\nb=4\,\Delta A/A_-$.
Since $\Delta A=(A_--A_+)<0$, $\nb > 0$ (i.e. the spectrum is blue) in 
the first case and $\nb < 0$ (i.e. the spectrum is red) in the second.
These estimates are indeed corroborated by the numerical results we 
have plotted in Fig.~\ref{fig:peb-sm2}.
Thirdly, while the amplitude of the magnetic field is considerably
suppressed over large scales if we work with the coupling function
$J_+(\phi)$, it is considerably enhanced over these scales in the
case of $J_-(\phi)$.
In fact, for the choice $J_-(\phi)$, the strength of the electromagnetic
fields on large scales are considerable and hence they will lead to a 
significant backreaction.

\begin{figure*}
\centering
\includegraphics[width=8.50cm]{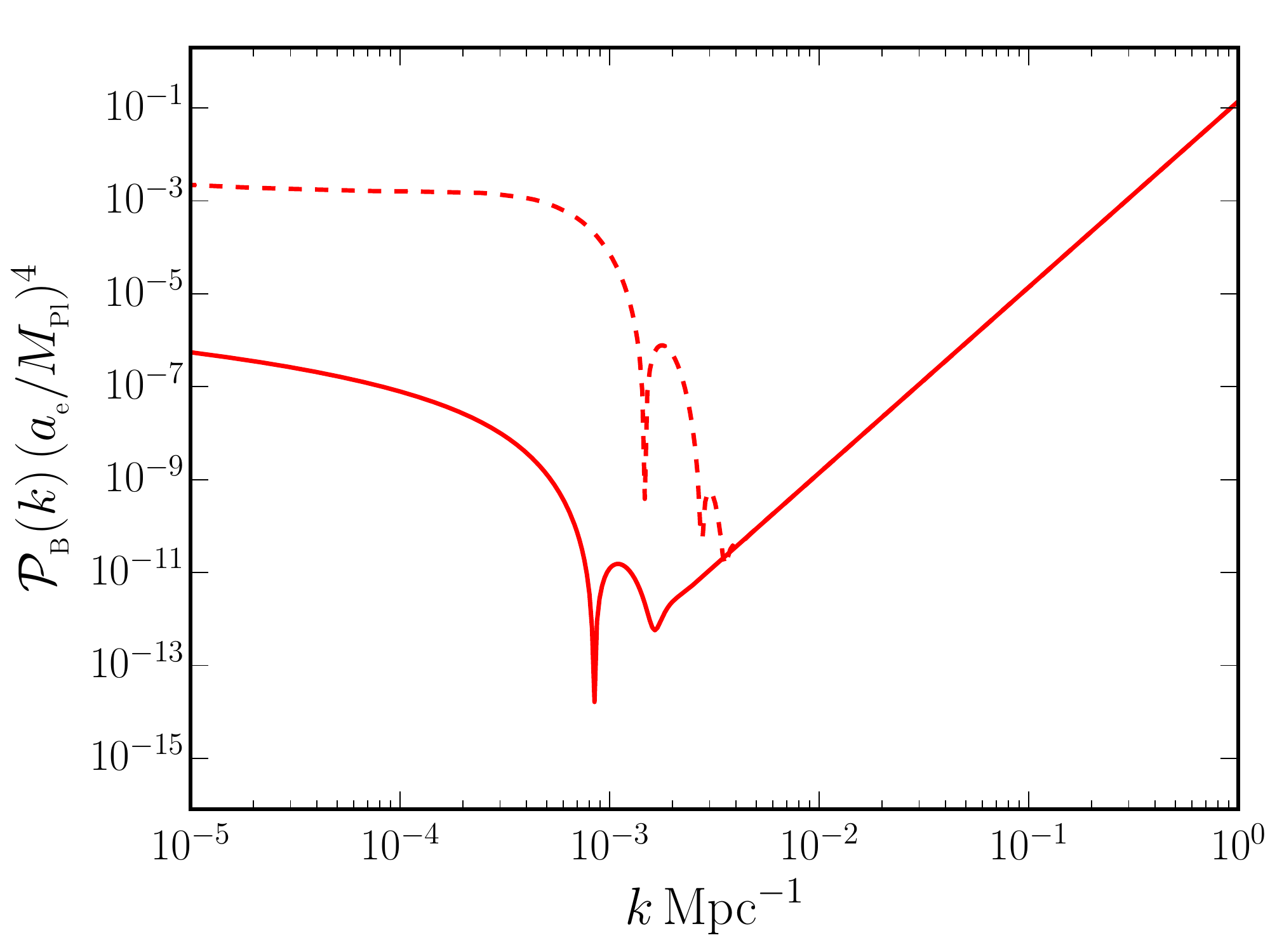}
\hskip 5pt
\includegraphics[width=8.50cm]{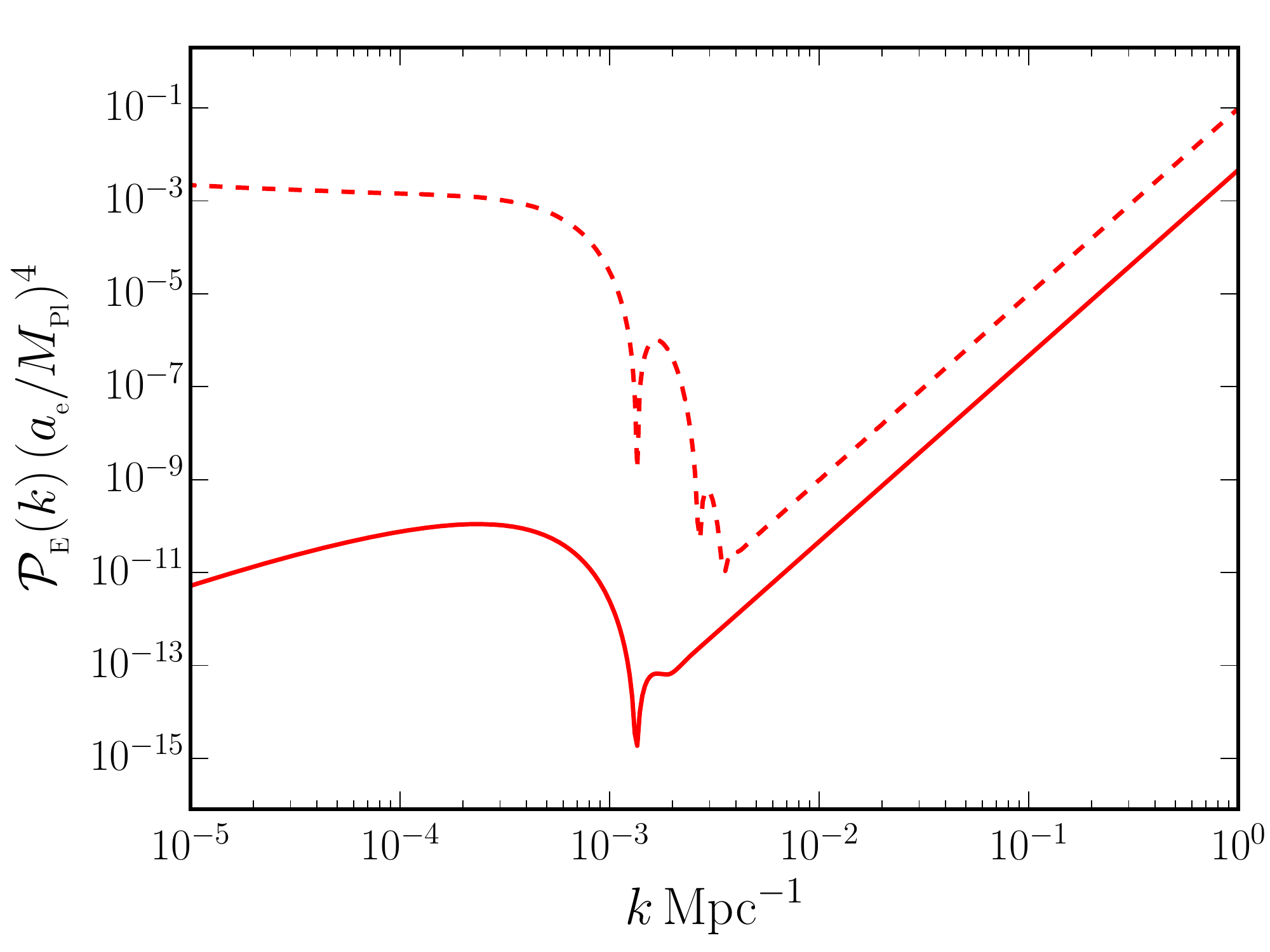}
\caption{The spectra of the magnetic (on the left) and electric (on the right)
fields arising in the case of the first punctuated inflation model~\eqref{eq:pi1} 
have been plotted for both the non-helical (in solid red) and helical (in dashed 
red) cases. 
In arriving at these spectra, we have worked with the coupling function~\eqref{eq:Jcf-pi1}  
and, as earlier, we have set the helicity parameter $\gamma$ to be unity.
As expected, over the large scales, when the modes leave the Hubble radius during
the initial stages of slow roll inflation, the spectra of the magnetic as well as 
the electric fields in the helical case are nearly scale invariant and also have 
roughly the same amplitude.
Moreover, the amplitude of the helical magnetic fields are $10^3$ times greater 
in amplitude than the non-helical fields over the scale invariant domain, as one
may have guessed.
Further, note that the spectra behave as $k^4$ over small scales.
This behavior can be attributed to the fact that, as the background scalar field 
approaches the point the inflection, leading to an epoch of ultra slow roll 
inflation, the non-minimal coupling function $J$ hardly evolves.
We should point out that, in the above plots, we have multiplied the spectra of 
the electromagnetic fields by the factor of by $a_\mathrm{e}^4$ (in contrast to 
the other figures) since their amplitudes turn out to be extremely small otherwise.
As will be evident from the discussion in the following subsection, the rather 
small amplitudes in these cases can be attributed to a very early onset of the 
ultra slow roll epoch required to suppress the scalar power on the largest 
scales.}\label{fig:pbe-pi1} 
\end{figure*}

Let us now turn to the first punctuated inflation model described by the 
potential~\eqref{eq:pi1}.
It proves to be difficult to obtain an analytical solution for the evolution
of the background scalar field in such a potential.
Therefore, we shall solve for the background numerically to first arrive 
at~$\phi(N)$.
We then choose a quadratic function of the form $N(\phi)=a_1\,(\phi^2/\Mpl^2)
 +b_1\,(\phi/\Mpl)+c_1$ to fit the numerical solution we have obtained in 
the initial slow roll regime.
When doing so, for the specific values of the parameters of the potential
and the initial conditions that we have worked with, we obtain the values 
of the three dimensionless fitting parameters to be $(a_1,b_1,c_1)=(-0.104,
-0.0408, 15.949)$.
Finally, to evaluate the spectra of the electromagnetic fields, we shall 
work with a coupling function of the form
\begin{equation}
J(\phi)=\mathrm{exp}\,\l\{n\,\l[a_1\,\l(\f{\phi^2-\phie^2}{\Mpl^2}\r)
+b_1\,\l(\f{\phi-\phie}{\Mpl}\r)\r]\r\}\label{eq:Jcf-pi1}
\end{equation}
and, note that, $J(\phi)$ reduces to unity at $\phie$, as required.
In Fig.~\ref{fig:pbe-pi1}, we have plotted the spectra of the resulting magnetic
and electric fields in both the non-helical and helical cases for~$n=2$. 
%%%%%%%%%%%%%%%%%%%%%%%%%%%%%%%%%%%%%%%%%%%%%%%%%%%%%%%%%%%%%%%%%%%%%%%%%%%%%%%
%%%%%%%%%%%%%%%%%%%%%%%%%%%%%%%%%%%%%%%%%%%%%%%%%%%%%%%%%%%%%%%%%%%%%%%%%%%%%%%
We need to highlight a few points regarding the figure.
The spectra of the electric and magnetic fields in the helical case and the 
spectrum of the magnetic field in the non-helical case are scale invariant 
over large scale modes that leave the Hubble radius during the initial stages
of slow roll.
Also, over the scale invariant domain, the helical amplitudes are~$10^3$ times
larger than the non-helical amplitudes, as expected for~$\gamma=1$. 
For the choice of the coupling function that we have worked with, we find that, 
the spectra of both the magnetic and electric fields behave as $k^4$ (in the 
absence as well as in the presence of helicity) over the small scale modes which 
leave the Hubble radius at later stages.
As we shall discuss in more detail in the following section, when the field 
approaches the point of inflection in the potential and enters a phase of 
ultra slow roll inflation, the coupling function~$J$ hardly changes.
This implies that $J''/J\simeq 0$, which is responsible for the $k^4$ behavior
of the spectra at small scales.
We should also point out that this behavior significantly suppresses the scale 
invariant amplitude of the magnetic field over large scales. 

The two examples discussed in this subsection point to the fact that unless
the coupling function is suitably chosen, strong departures from slow roll
inflation result in spectra of magnetic fields that contain significant 
deviations from scale invariance.

%%%%%%%%%%%%%%%%%%%%%%%%%%%%%%%%%%%%%%%%%%%%%%%%%%%%%%%%%%%%%%%%%%%%%%%%%%%%%%%

\subsection{In models leading to enhanced power on small scales}

Let us now turn to the two models described by the potentials~\eqref{eq:usr}
and~\eqref{eq:pi2} that lead to enhanced scalar power on small scales.
As in the case of the first punctuated inflation model we discussed in the 
previous subsection, these models too lead to an epoch of ultra slow roll
inflation wherein the first slow roll parameter decreases exponentially 
over a short period before it starts rising leading to an end of inflation.
It is the sharp decrease in the first slow roll parameter that is responsible
for the rise in the scalar power in such models (in this context, see
Refs.~\cite{Garcia-Bellido:2017mdw,Ballesteros:2017fsr,Germani:2017bcs,
Dalianis:2018frf,Bhaumik:2019tvl,Ragavendra:2020sop,Dalianis:2020cla}).

In these models, one chooses the parameters of the background potential as
well as the initial conditions such that there occurs an extended period of
slow roll inflation which generates scalar and tensor power spectra that 
are consistent with the CMB observations on large scales.
If we require a nearly scale invariant spectrum of the magnetic field over 
the CMB scales, then, evidently, we need to choose a coupling function $J(\phi)$ 
that is based on the evolution of the field during the long initial epoch of
slow roll inflation.
Since the potentials~\eqref{eq:usr} and~\eqref{eq:pi2} do not seem to admit 
simple analytical solutions, we repeat the exercise we had carried out in 
the case of the first punctuated inflation model.
Utilizing the numerical solution, we arrive at $N(\phi)$ and fit a polynomial
to describe the function. 
We find that we can fit fourth and sixth order polynomials to describe 
the~$N(\phi)$ in the potentials~\eqref{eq:usr} and~\eqref{eq:pi2}.
The coupling functions that we shall work with in these two cases can be 
expressed as 
\begin{subequations}\label{eq:J-pbhm}
\begin{eqnarray}
J(\phi)&=&\mathrm{exp}\,\Biggl\{n\,\biggl[a_2\,\l(\f{\phi^4-\phie^4}{\Mpl^4}\r)
+b_2\,\l(\f{\phi^3-\phie^3}{\Mpl^3}\r)\nn\\
& &+\,c_2\,\l(\f{\phi^2-\phie^2}{\Mpl^2}\r)
+d_2\, \l(\f{\phi-\phie}{\Mpl}\r)\biggr]\Biggr\},\label{eq:J-usr}\\
J(\phi)&=&\mathrm{exp}\,\Biggl\{n\,\biggl[a_3\,\l(\f{\phi^6-\phie^6}{\Mpl^6}\r)
+b_3\,\l(\f{\phi^5-\phie^5}{\Mpl^5}\r)\nn\\
& &+\,c_3\,\l(\f{\phi^4-\phie^4}{\Mpl^4}\r)
+d_3\,\l(\f{\phi^3-\phie^3}{\Mpl^3}\r)\nn\\
& &+\,e_3\,\l(\f{\phi^2-\phie^2}{\Mpl^2}\r)
+f_3\,\l(\f{\phi-\phie}{\Mpl}\r)\biggr]\Biggr\},\qquad\qquad
\end{eqnarray}
\end{subequations}
with the dimensionless fitting parameters being given by $(a_2,b_2,c_2,d_2)$ $=(-0.184, 1.822, -7.040, 10.676)$ and $(a_3,b_3,c_3,d_3,e_3,f_3)$
$=(-1.53\times 10^{-3}, 2.37\times10^{-2}, -0.158, 0.439, -0.459,-0.778)$,
respectively. 

\begin{figure*}
\centering
\includegraphics[width=8.50cm]{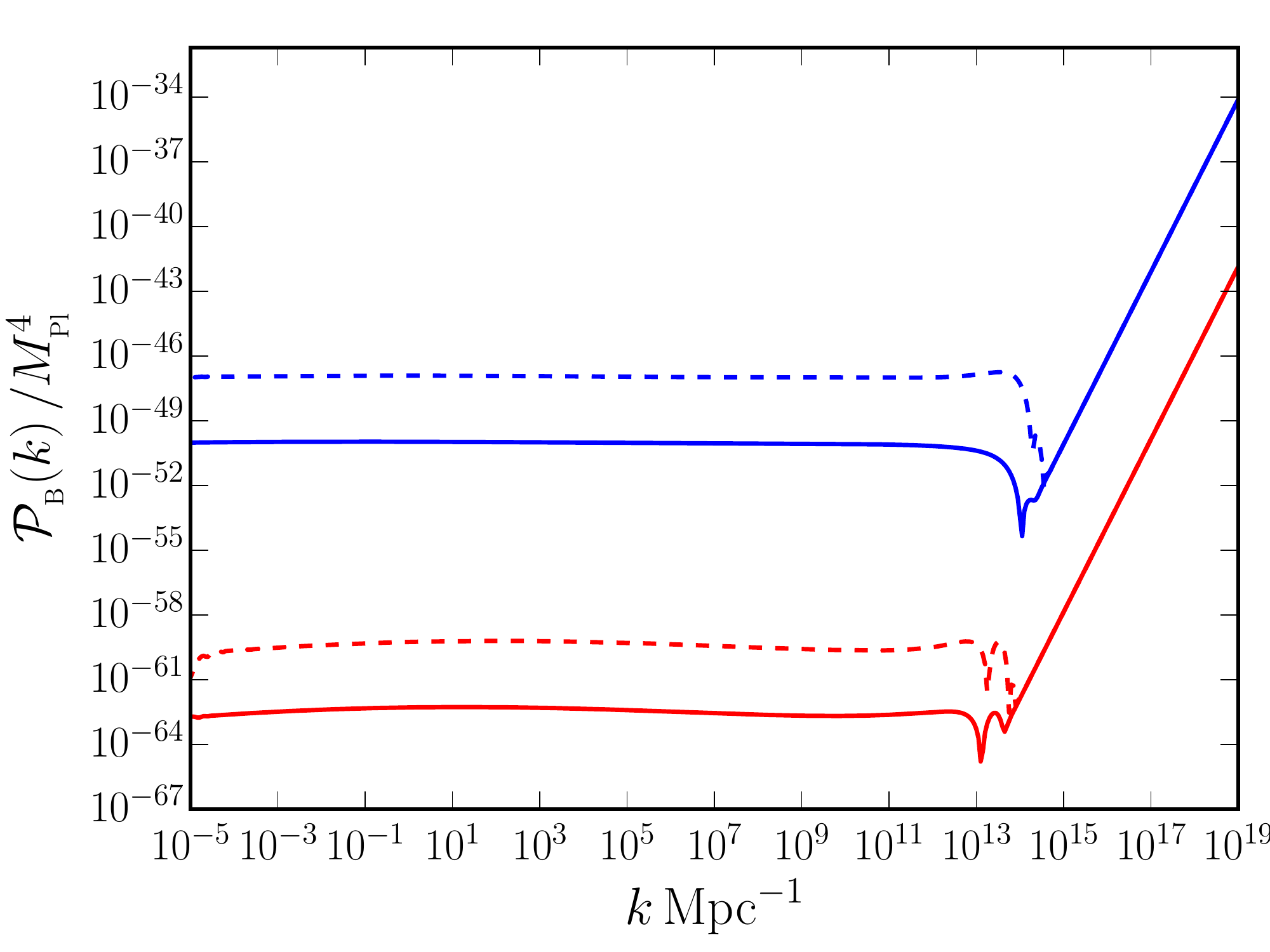}
\hskip 5pt
\includegraphics[width=8.50cm]{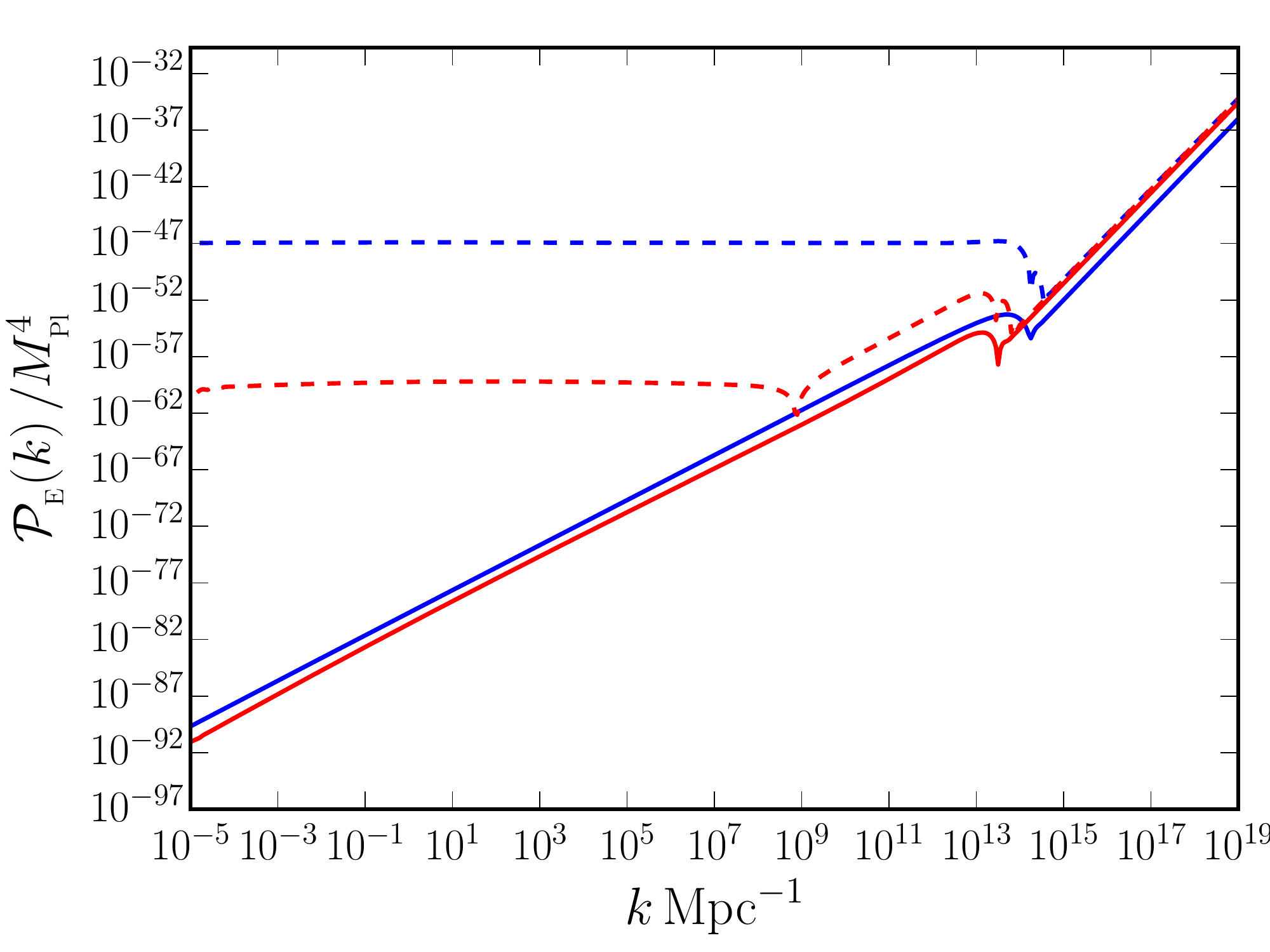}
\caption{The spectra of the magnetic (on the left) and electric (on the right)
fields arising in the ultra slow roll inflationary model~\eqref{eq:usr} 
(in red) and the second punctuated inflationary model~\eqref{eq:pi2} (in blue) 
have been plotted in the non-helical (as solid lines) and helical (as dashed lines) 
cases, respectively.
Note that we have worked with the coupling functions~\eqref{eq:J-pbhm} to arrive
at these spectra.
Also, we have chosen $n=2$ and set $\gamma=1$, as we have done earlier.
Clearly, the spectra of the electromagnetic fields in both the helical 
and non-helical cases are along expected lines, as we have discussed in the 
text.
In particular, we should point out that the spectra in the two models behave as 
$k^4$ at large wave numbers. 
This behavior arises due to the fact the the coupling functions cease to evolve 
as the field approaches the point of inflection in these models.
In such a situation, the electromagnetic modes effectively behave as in the 
conformally invariant case, leading to the $k^4$ behavior.
We should also add that, apart from changing the shape of the spectra at small
scales, the background evolution significantly suppresses the power in the 
spectra on large scales.}\label{fig:pbe-usr-pi2}
\end{figure*}

In Fig.~\ref{fig:pbe-usr-pi2}, we have plotted the spectra of the electromagnetic
fields that arise for the above choices of the coupling functions in the two models
of our interest.
%%%%%%%%%%%%%%%%%%%%%%%%%%%%%%%%%%%%%%%%%%%%%%%%%%%%%%%%%%%%%%%%%%%%%%%%%%%%%%%
%%%%%%%%%%%%%%%%%%%%%%%%%%%%%%%%%%%%%%%%%%%%%%%%%%%%%%%%%%%%%%%%%%%%%%%%%%%%%%%
We should mention that, in arriving at the spectra, we have set $n=2$ and 
$\gamma=1$, as we have done before.
The following points are clear from the figure.
Note that the spectra of the magnetic fields in both the non-helical and 
helical cases are nearly scale invariant over large scales. 
This is because the coupling functions have been determined by the slow roll
behavior of the field. 
Also, as we have seen earlier, the magnitude of the helical magnetic field is 
about $10^3$ larger than the non-helical field over the scale invariant domain.
Moreover, over large scales, as expected, the spectrum of the electric field 
behaves as $k^2$ in the non-helical case and is nearly scale invariant with
an amplitude comparable to the spectrum of the magnetic field in the helical 
case.
Further, at small scales, all the spectra behave as $k^4$ for the same 
reasons as we had encountered in the case of the first punctuated inflation 
model~\eqref{eq:pi1}.
When the background scalar field approaches the point of inflection in these
models, the coupling functions $J$ hardly evolve (in this context, see 
Fig.~\ref{fig:J-usr-pi2}) and the electromagnetic modes effectively behave 
as in the conformally invariant case leading to the $k^4$ behavior.
Lastly, we should mention that such a background behavior not only changes
the shape of the spectra of the electromagnetic fields at small scales, it 
also suppresses the scale invariant amplitudes of the spectra at large scales.

%%%%%%%%%%%%%%%%%%%%%%%%%%%%%%%%%%%%%%%%%%%%%%%%%%%%%%%%%%%%%%%%%%%%%%%%%%%%%%%

\subsection{An analytical estimate}

In this subsection, we shall analytically arrive at the power spectra of
the electromagnetic fields in models which permit ultra slow roll inflation
and lead to enhanced scalar power on small scales.

%%%%%%%%%%%%%%%%%%%%%%%%%%%%%%%%%%%%%%%%%%%%%%%%%%%%%%%%%%%%%%%%%%%%%%%%%%%%%%%

\subsubsection{A simple approximation}

\begin{figure}
\centering
\includegraphics[width=8.50cm]{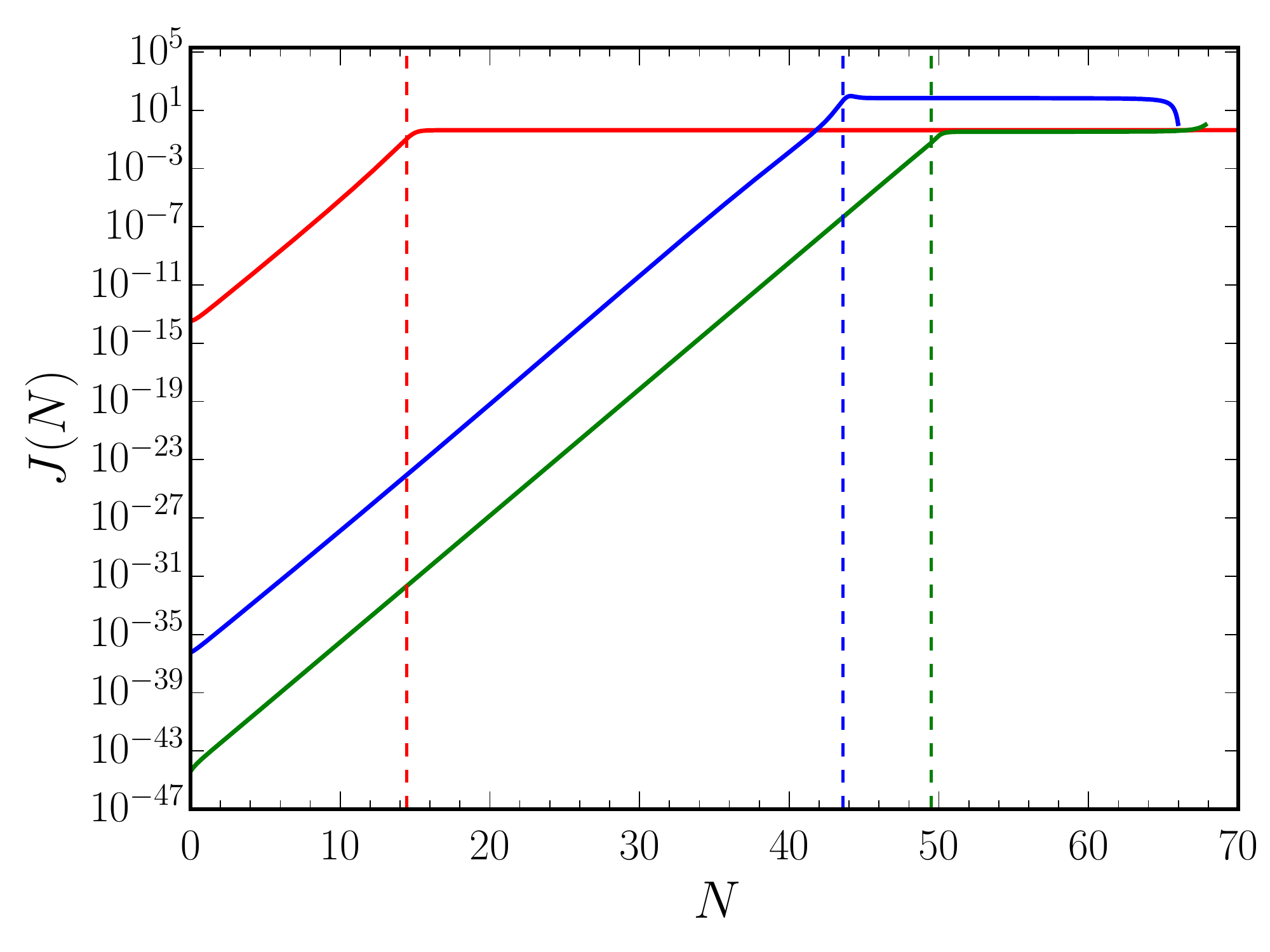}
\caption{The evolution of the non-minimal coupling function~$J$ [as given by 
Eqs.~\eqref{eq:Jcf-pi1} and~\eqref{eq:J-pbhm}] that we had considered in the 
models described by the potentials~\eqref{eq:pi1}, \eqref{eq:usr} and~\eqref{eq:pi2} 
has been plotted (in solid red, blue and green, respectively) as a function 
of the e-fold~$N$. 
The onset of the ultra slow roll phase corresponds to the time when the first 
slow roll parameter starts to decrease rapidly.
We have indicated the beginning of the ultra slow roll epoch (as dashed vertical 
lines of the corresponding color) in all these cases.
Recall that, we had constructed coupling functions $J(\phi)$ so that they behave
as~$a^2$ during the initial slow roll phase. 
For such choices of $J(\phi)$, the coupling function does not seem to change  
appreciably (until very close to the end of inflation) after ultra slow has 
set in.}
\label{fig:J-usr-pi2}
\end{figure}

Recall that, in these scenarios, we had constructed the coupling 
function $J(\phi)$ so that we obtain a scale invariant spectrum for the 
magnetic field on large scales [cf. Eqs.~\eqref{eq:J-pbhm}; also 
see Eq.~\eqref{eq:Jcf-pi1}].  
In order to achieve such a scale invariant spectrum, during the initial stage 
of slow roll inflation, let us assume that $J(\eta) \propto a^2$.
Note that, in these models, for our choices of the dependence of the coupling 
function on the field, we find that $J$ freezes when the epoch of ultra slow
roll sets in.
This is evident from Fig.~\ref{fig:J-usr-pi2} wherein we have plotted the 
evolution of the coupling function in the first and second models of
punctuated inflation [cf. Eqs.~\eqref{eq:pi1} and \eqref{eq:pi2}] as well
as in the model of ultra slow roll inflation [cf. Eq.~\eqref{eq:usr}].
Therefore, we can assume that, after a time, say, $\eta_1$, $J(\eta) \simeq 
\mathrm{constant}$.
In such a case, during the initial stage, the electromagnetic modes~$\cA_k$ 
can be easily obtained to be
\begin{equation}
\cA_k^\mathrm{I}(\eta)
=\f{1}{\sqrt{2\,k}}\,\l(1-\f{3\,i}{k\,\eta}-\f{3}{k^2\,\eta^2}\r)\,
\mathrm{e}^{-i\,k\,\eta}.\label{eq:cA1}
\end{equation}
It should be evident that, after $\eta_1$, the electromagnetic modes can be
written as
\begin{equation}
\cA_k^\mathrm{II}(\eta)
=\f{1}{\sqrt{2\,k}}\,\l(\alpha_k\,\mathrm{e}^{-i\,k\,\eta}
+\beta_k\,\mathrm{e}^{i\,k\,\eta}\r).\label{eq:cA-usr}
\end{equation}
The coefficients $\alpha_k$ and $\beta_k$ are to be determined by imposing 
the matching conditions on the modes at the transition at~$\eta_1$.

Since $J^{\prime}\simeq -2\,\eta_1^2/\eta^3$ prior to $\eta_1$ and $J^{\prime}
\simeq 0$ after, there is a discontinuity in $J'$ at $\eta_1$.
This leads to a Dirac delta function in the behavior of $J''/J$ at the 
transition at~$\eta_1$.
As a result, the modes in the two domains are related by the matching 
conditions 
\begin{subequations}\label{eq:mc-sr-usr}
\begin{eqnarray}
\cA_k^\mathrm{I}(\eta_1)&=&\cA_k^\mathrm{II}(\eta_1),\\
\cA_k^{\mathrm{II}\prime}(\eta)-\cA_k^{\mathrm{I}\prime}(\eta)
&=&\f{2}{\eta_1}\,\cA_k^\mathrm{I}(\eta_1).
\end{eqnarray}
\end{subequations}
These conditions lead to the following expressions for the coefficients 
$\alpha_k$ and $\beta_k$:
\begin{subequations}\label{eq:ab-1}
\begin{eqnarray}
\alpha_k &=& 1+\f{2\,i\,k_1}{k}-\f{3\,k_1^2}{2\,k^2},\\
\beta_k &=& \l(\f{i\,k_1}{k}
- \f{3\,k_1^2}{2\,k^2}\r)\,\mathrm{e}^{2\,i\,k/k_1},
\end{eqnarray}
where we have set $k_1=-1/\eta_1$, i.e. the wave number which leaves
the Hubble radius at the onset of the ultra slow roll epoch.
\end{subequations}
The power spectra of the magnetic and electric fields at late times 
[i.e. in the limit $(-k\,\ee) \ll 1$] can be evaluated to be 
\begin{subequations}
\begin{eqnarray}
\label{eq:ab_1}
\pb(k)&=&\f{\HI^4}{4\,\pi^2}\,\l(-k\,\ee\r)^4\,
\vert\alpha_k+\beta_k\vert^2,\\
\pe(k)&=&\f{\HI^4}{4\,\pi^2}\,\l(-k\,\ee\r)^4\,
\vert\alpha_k-\beta_k\vert^2.\label{eq:ab_2}
\end{eqnarray}
\end{subequations}
For large $k$ such that $k/k_1\gg 1$, we find that $\alpha_k\to 1$ 
and $\beta_k\to 0$ [cf. Eqs.~\eqref{eq:ab-1}].
Therefore, in such a limit, both the above power spectra behave as $k^4$,
which is what we observe numerically (see Figs.~\ref{fig:pbe-pi1} 
and~\ref{fig:pbe-usr-pi2}).
It can be shown that, in the limit $k/k_1\ll 1$, 
\begin{equation}
\vert\alpha_k+\beta_k\vert^2
=\f{9\,k_1^4}{k^4},\quad
\vert\alpha_k-\beta_k\vert^2
=\f{16\,k_1^2}{k^2},
\end{equation}
so that the above spectra reduce to the following forms:
\begin{subequations}
\begin{eqnarray}
\pb(k) & \simeq &\f{9\,\HI^4}{4\,\pi^2}\,\l[\f{a(\eta_1)}{a(\ee)}\r]^4,\\
\pe(k) & \simeq & \f{\HI^4}{4\,\pi^2}\,\l(\f{4\,k}{k_1}\r)^2\,
\l[\f{a(\eta_1)}{a(\ee)}\r]^4.
\end{eqnarray}
\end{subequations}
In other words, on the large scales, we obtain spectral shapes that are 
expected to occur when the coupling function behaves as $J \simeq a^2$
[cf. Eqs.~\eqref{eq:ne2}].
This should not come as a surprise since these modes leave during the 
initial slow roll regime.
However, note that the factor $[a(\eta_1)/a(\ee)]^4$ considerably suppresses
the amplitudes of the electromagnetic spectra on large scales.
In fact, the earlier the onset of the ultra slow roll regime, the larger is
the suppression.
It is for this reason that the electromagnetic spectra in the first punctuated
inflation model had substantially small amplitudes on large scales
(see Fig.~\ref{fig:pbe-pi1}).
%%%%%%%%%%%%%%%%%%%%%%%%%%%%%%%%%%%%%%%%%%%%%%%%%%%%%%%%%%%%%%%%%%%%%%%%%%%%%%%
%%%%%%%%%%%%%%%%%%%%%%%%%%%%%%%%%%%%%%%%%%%%%%%%%%%%%%%%%%%%%%%%%%%%%%%%%%%%%%%

Let us now examine the corresponding situation in the helical case.
In the case of the helical field, during the initial stage of slow roll
inflation, when $n=2$, the electromagnetic modes $\cA_k^{\sigma}$ are 
given by [cf. Eq.~\eqref{eq:cA-h}]
\begin{equation}
\cA_k^{\sigma\mathrm{I}}(\eta)
=\f{1}{\sqrt{2\,k}}\, \mathrm{e}^{-\pi\,\sigma\,\gamma}\,
W_{2\,i\,\sigma\,\gamma,\tfrac{5}{2}}(2\,i\,k\,\eta).
\end{equation}
Since the coupling function $J$ hardly evolves after the onset of ultra 
slow roll, the electromagnetic modes during the second stage, say, 
$\cA_k^{\sigma\mathrm{II}}$, can be expressed just as in Eq.~\eqref{eq:cA-usr} 
for the non-helical case.
Moreover, the matching conditions continue to be given by 
Eqs.~\eqref{eq:mc-sr-usr}.
However, we should clarify that the coefficients $\alpha_k$ and $\beta_k$
now depend on the polarization~$\sigma$.
The power spectra of the magnetic and electric fields at late times, i.e.
when $(-k\,\ee)\ll 1$, can be obtained to be 
\begin{subequations}\label{eq:pbe-sr-usr-h}
\begin{eqnarray}
\pb(k)&=&\f{\HI^4}{8\,\pi^2}\,\l(-k\,\ee\r)^4\,
\l(\vert\alpha_k^++\beta_k^+\vert^2
+\vert\alpha_k^-+\beta_k^-\vert^2\r),\nn\\
\\
\pe(k)&=&\f{\HI^4}{8\,\pi^2}\,\l(-k\,\ee\r)^4\,
\l(\vert\alpha_k^+-\beta_k^+\vert^2
+\vert\alpha_k^--\beta_k^-\vert^2\r).\nn\\
\end{eqnarray}
\end{subequations}

On matching the modes at $\eta_1$, we obtain the 
coefficients~$\alpha_k^{\sigma}$ and $\beta_k^{\sigma}$ to be
\begin{subequations}
\begin{eqnarray}
\alpha_k^\sigma 
&=&-\f{\mathrm{e}^{-i\,k/k_1}\, 
\mathrm{e}^{-\pi\,\sigma\,\gamma}}{2\, (k/k_1)}\,
\biggl[2\,(i +\sigma\,\gamma)\,W_{2\,i\,\sigma\,
\gamma,\tfrac{5}{2}}(-2\,i\,k/k_1)\nn\\
& & - i\, W_{1+2\,i\,\sigma\,\gamma,\tfrac{5}{2}}(-2\,i\,k/k_1)\biggr],\\
\beta_k^\sigma 
&=&-\f{\mathrm{e}^{i\,k/k_1}\, 
\mathrm{e}^{-\pi\,\sigma\,\gamma}}{2\, (k/k_1)}\,
\biggl[2\,\l(-i -\f{k}{k_1}-\sigma\,\gamma\r)\nn\\ 
& &\times\, W_{2\,i\,\sigma\,
\gamma,\tfrac{5}{2}}(-2\,i\,k/k_1)
+ i\, W_{1+2\,i\,\sigma\,\gamma,\tfrac{5}{2}}(-2\,i\,k/k_1)\biggr],\nn\\
\end{eqnarray}
\end{subequations}
where, as earlier, we have set $k_1=-1/\eta_1$.
In the limit $k/k_1 \gg 1$, we find that $\alpha_k^\sigma \to 1$ 
and $\beta_k^\sigma \to 0$, as in the non-helical case.
This suggests that the power spectra of both the electric and magnetic 
fields behave as $k^4$ in such a limit, which is indeed what we obtain 
numerically (see Figs.~\ref{fig:pbe-pi1} and~\ref{fig:pbe-usr-pi2}).
Whereas, in the limit $k/k_1\ll 1$, we find that~\cite{Mathematica} 
\begin{eqnarray}
\vert\alpha_k^\sigma+\beta_k^\sigma\vert^2
&=&\f{9\,\l(1-\mathrm{e}^{-4\,\pi\,\sigma\,\gamma}\r)}{4\,\pi\,
\sigma\,\gamma\,\l(1+5\,\gamma^2+4\,\gamma^4\r)}\,\l(\f{k}{k_1}\r)^{-4},\quad\\
\vert\alpha_k^\sigma-\beta_k^\sigma\vert^2
&=&\f{9\,\sigma\,\gamma^2\,\l(1-\mathrm{e}^{-4\,\pi\,\sigma\,\gamma}\r)}{4\,\pi\,
\gamma\, \l(1+5\,\gamma^2+4\,\gamma^4\r)}\,\l(\f{k}{k_1}\r)^{-4},
\end{eqnarray}
and hence the spectra~\eqref{eq:pbe-sr-usr-h} reduce to the following forms:
\begin{subequations}
\begin{eqnarray}
\pb(k) & \simeq &\f{9\,\HI^4}{4\,\pi^2}\,f(\gamma)\,
\l[\f{a(\eta_1)}{a(\ee)}\r]^4,\\
\pe(k) & \simeq & \f{9\,\HI^4}{4\,\pi^2}\,
f(\gamma)\,\gamma^2\, \l[\f{a(\eta_1)}{a(\ee)}\r]^4,
\end{eqnarray}
\end{subequations}
where, recall that, $f(\gamma)$ is given by Eq.~\eqref{eq:fg}.
Clearly, over large scales, the spectra of both the electric and magnetic
fields are scale invariant as is expected in the helical case when $J
\simeq a^2$ and the modes cross the Hubble radius during a regime of slow 
roll.
Moreover, note that, as in the non-helical case, the onset of the ultra 
slow roll epoch leads to a suppression in the amplitudes of the power 
spectra on large scales by the factor of $[a(\eta_1)/a(\ee)]^4$.

We have been able to understand the shape of the electromagnetic spectra 
arising in models involving an epoch of ultra slow roll inflation using 
analytical arguments.
Let us now compare the numerical results for the amplitudes of the spectra 
over large scales with the analytical estimates in both the non-helical and 
helical cases.
In the case of the ultra slow roll model described by the potential~\eqref{eq:usr},
we find that, when the pivot scale leaves the Hubble radius, the value of 
the Hubble parameter is $\HI=9.05\times10^{-6}\, \Mpl$. 
The epoch of ultra slow roll inflation can be said to begin when the first 
slow roll parameter~$\epsilon_1$ attains the maximum value (prior to the 
end of inflation) and begins to decrease rapidly thereafter.
We find that, in the model of our interest here, ultra slow roll sets in about 
$22.4$ e-folds before the end of inflation.
Also, the value of the wave number that equals $\sqrt{\vert J''/J\vert}$ at 
the onset of ultra slow roll inflation proves to be $k_1= 2.2 \times 10^{13}\,
\mathrm{Mpc}^{-1}$.
For these values, in the non-helical case, the analytical estimates we have 
obtained above lead to $\pb(k)\simeq 10^{-60}\, \Mpl^4$ and $\pe(k)\simeq 
10^{-89}\,\Mpl^4$ at the pivot scale. 
Numerically, we have obtained the corresponding values to be $\pb(k)\simeq 
10^{-63}\,\Mpl^4$ and $\pe(k) \simeq 10^{-84}\,\Mpl^4$.
In the helical case, for $\gamma=1$, the analytical estimates lead to $\pb(k)
= \pe(k) \simeq 10^{-57}\,\Mpl^4$ at the pivot scale. 
The corresponding numerical values turn out to be $\pb(k)=\pe(k) \simeq 
10^{-60}\,\Mpl^4$.

Similarly, in the case of the second model of punctuated inflation described
by the potential~\eqref{eq:pi2}, we find that the value of the Hubble parameter
at the time when the pivot scale exits the Hubble radius is $\HI =1.01 \times
10^{-5}\,\Mpl$.
Moreover, the onset of the ultra slow roll epoch occurs about $18.3$ e-folds
prior to the end of inflation, which implies that $k_1 \simeq 1.6\times10^{14}\, 
\mathrm{Mpc}^{-1}$. 
According to the analytical estimates, in the non-helical case, these values 
lead to $\pb(k)\simeq 10^{-53}\,\Mpl^4$ and $\pe(k)\simeq 10^{-84}\, \Mpl^4$ 
at the pivot scale. 
Numerically, we obtain the corresponding values to be $\pb(k)\simeq 10^{-50}\, 
\Mpl^4$ and $\pe(k) \simeq 10^{-83}\,\Mpl^4$.
In the case of the helical fields, when $\gamma=1$, the analytical estimates 
suggest that $\pb(k) =\pe(k) \simeq 10^{-50}\,\Mpl^4$ at the pivot scale,
while the corresponding numerical values turn out to be  $\pb(k) =\pe(k)\simeq 
10^{-47}\, \Mpl^4$.

While the analytical estimates broadly match the numerical results, there
arise differences of the order of $10^3$--$10^5$ in the values for the 
power spectra of the electromagnetic fields.
These differences can be attributed to the coarseness of the analytical 
modeling and the fact that $J$ evolves to a certain extent as one 
approaches the end of inflation.

%%%%%%%%%%%%%%%%%%%%%%%%%%%%%%%%%%%%%%%%%%%%%%%%%%%%%%%%%%%%%%%%%%%%%%%%%%%%%%%

\subsubsection{A closer look at the evolution of the modes at late times}

\begin{figure}
\centering
\includegraphics[width=8.50cm]{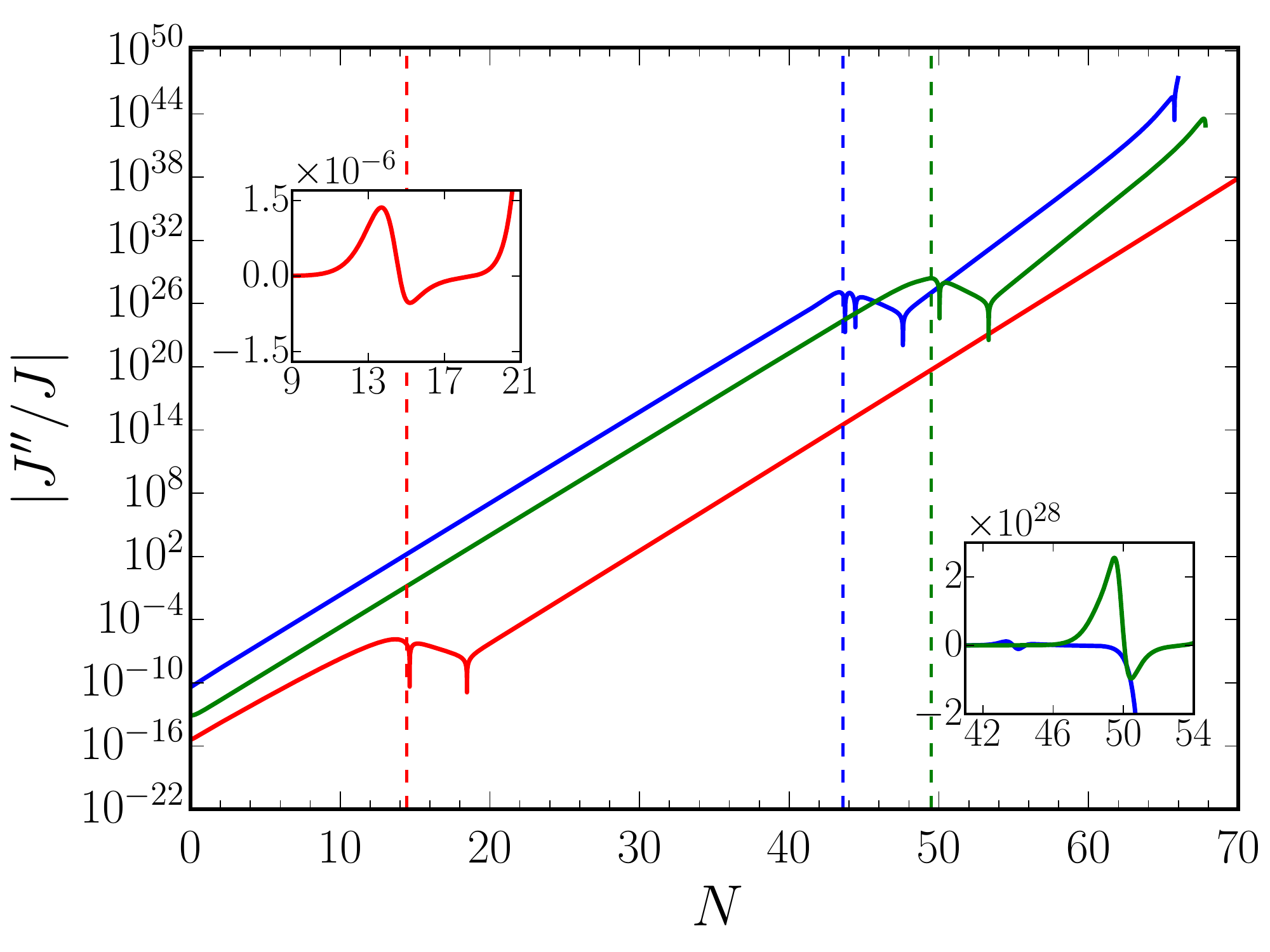}
\caption{The evolution of the quantity $J''/J$ corresponding to the three 
coupling functions we had illustrated in the previous figure has been plotted
as a function of e-fold~$N$ (with the same choice of colors). 
The insets highlight the behavior of the quantity around the onset of the 
epoch of ultra slow roll.
We find that $J''/J\propto \mathrm{e}^{2\,N}$ during the initial slow roll
phase, as expected.
It is clear $J''/J$ does not vanish once ultra slow roll inflation begins
(indicated by the vertical lines).
In fact, the quantity is almost a constant during the period of ultra slow 
roll and it actually grows (either as $\mathrm{e}^{2\,N}$ in the case of the 
first punctuated inflation model or as $\mathrm{e}^{5\,N/2}$ in the other 
two models) when the phase of ultra slow roll is complete and the first slow 
roll parameter begins to rise.
We should also mention the fact that $J''/J$ can turn negative during these
latter stages.}\label{fig:JppJ}
\end{figure}

In Fig.~\ref{fig:J-usr-pi2}, we had plotted the evolution of the non-minimal
coupling function in the ultra slow roll model and the two punctuated inflation 
models we have considered.
We had found that, once the epoch of ultra slow roll begins, the coupling
function~$J$ hardly evolves.
Based on such a behavior, we had assumed that $J'$ and $J''$ were zero and 
had arrived at the analytical form for the modes $\cA_k$ and, eventually, 
the power spectra of the electromagnetic fields.
While the coupling function $J$ is almost a constant, one can show that it 
is not correct to set $J'$ and $J''$ to zero in these scenarios.
In Fig.~\ref{fig:JppJ}, we have plotted the evolution of $\vert J''/J\vert$ 
in the three models.
%%%%%%%%%%%%%%%%%%%%%%%%%%%%%%%%%%%%%%%%%%%%%%%%%%%%%%%%%%%%%%%%%%%%%%%%%%%%%%%
%%%%%%%%%%%%%%%%%%%%%%%%%%%%%%%%%%%%%%%%%%%%%%%%%%%%%%%%%%%%%%%%%%%%%%%%%%%%%%%
It is clear from the figure that the quantity does not vanish once ultra slow 
begins, as we have assumed earlier. 
Therefore, it seems that we need to revise our previous discussion.

\begin{figure*}
\centering
\includegraphics[width=8.50cm]{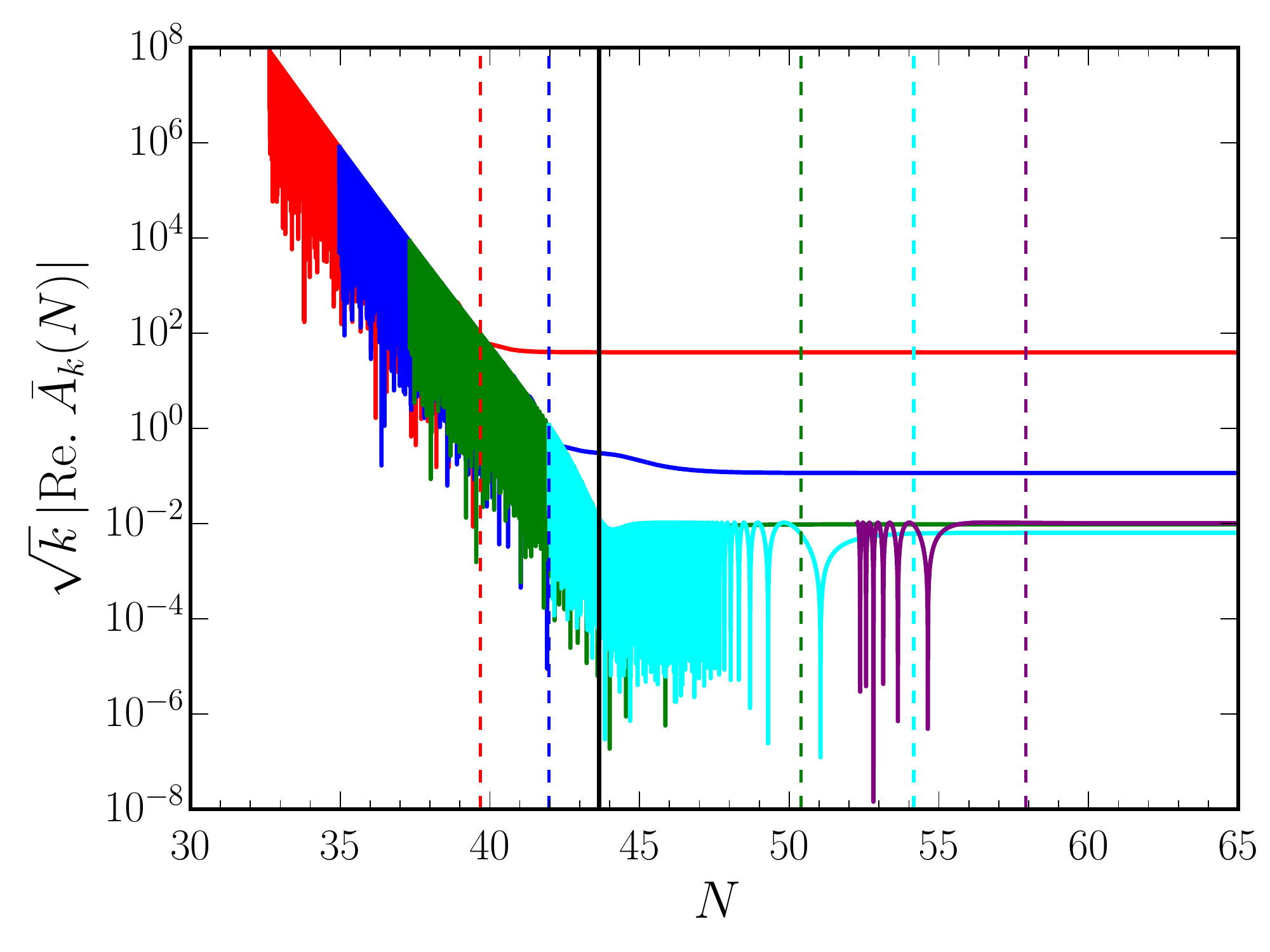}
\includegraphics[width=8.50cm]{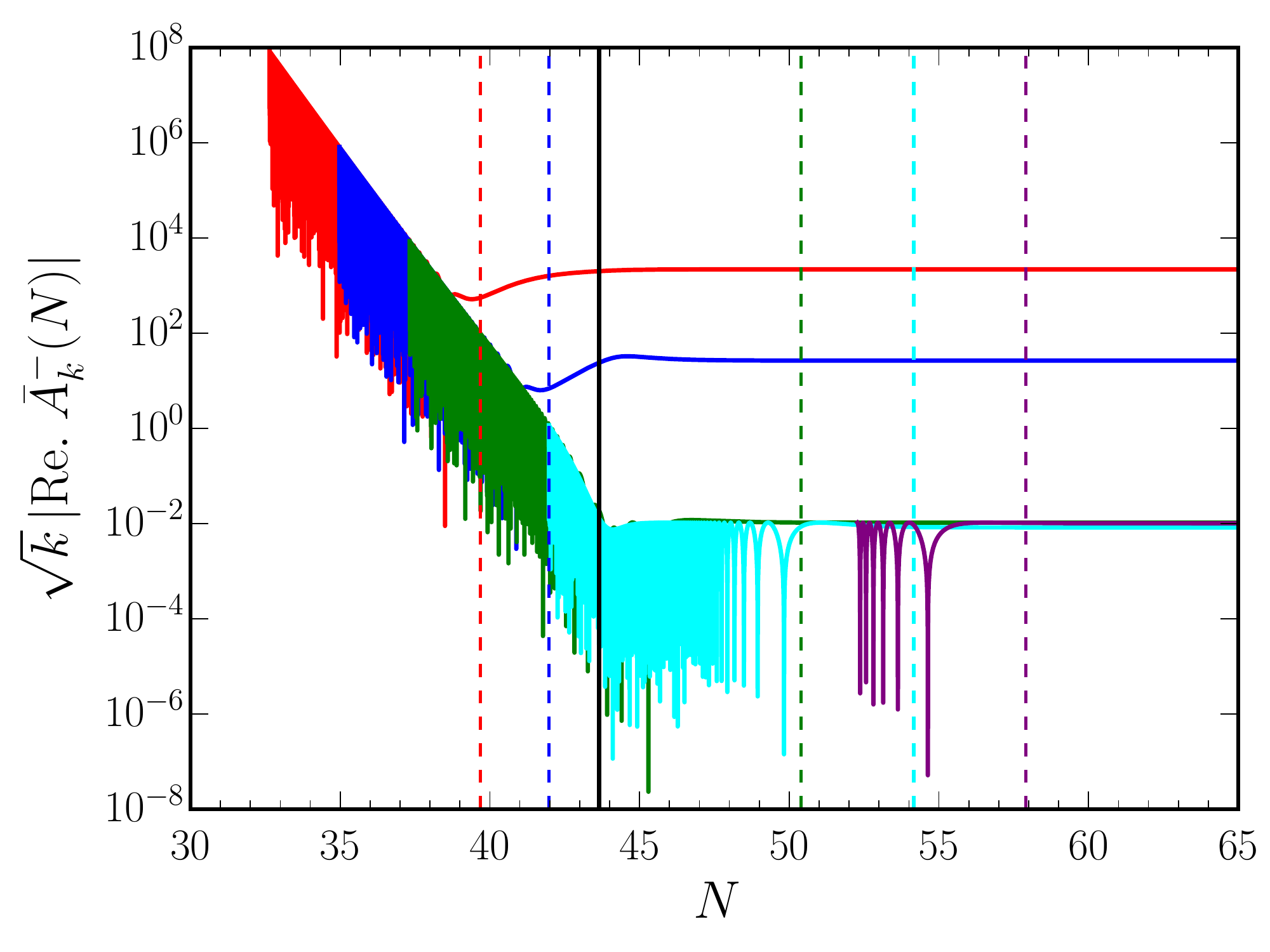}
\caption{The evolution of the electromagnetic modes in the case of the ultra 
slow roll inflation model~\eqref{eq:usr} has been plotted for the five choices 
of the wave numbers~$k=(10^{12},10^{13},10^{14}, 10^{16}, 10^{18})\;
\mathrm{Mpc}^{-1}$ (in red, blue, green, cyan and purple), respectively.
We have worked with the coupling function~\eqref{eq:J-usr} and have plotted the 
evolution of the dominant real part of the quantity $\sqrt{k}\,\vert\bar{A}_k\vert$ 
in the non-helical case (on the left) and the quantity $\sqrt{k}\,\vert \bar{A}_k^-\vert$ 
in the helical case (on the right).
We have also indicated the onset of the ultra slow roll epoch (as the solid vertical 
line in black) and the e-folds corresponding to the time~$\eta_k$, i.e. when $k^2
=\vert J^{\prime\prime}/J\vert$, for the different wave numbers (as dashed vertical 
lines, with the same choice of colors as the modes).
It is clear that the amplitude of the electromagnetic modes freeze at late times.
Importantly, we find that, for $k \gtrsim 10^{13}\,\mathrm{Mpc}^{-1}$,
the late time values of the quantities $\sqrt{k}\,\vert\bar{A}_k\vert$ and 
$\sqrt{k}\,\vert \bar{A}_k^-\vert$ are the same for the different wave numbers, 
which points to the $k^4$ behavior for the spectrum of the magnetic field over
small scales.}\label{fig:mode_USR2}
\end{figure*}

One can expect that, since $J$ as well as $J''/J$ behave as~$a^2$ during the 
initial slow roll phase, the power spectra over modes that leave the Hubble 
radius~---~to be precise, when $k =\sqrt{\vert J''/J\vert}$~---~will be scale 
invariant.
However, in the ultra slow roll and the second punctuated inflation models,
once the epoch of ultra slow roll comes to an end, $J''/J$ behaves as $a^{5/2}$
(as illustrated in Fig.~\ref{fig:JppJ}), while $J$ is a constant.
Let us now focus on large wave numbers in these models over which, numerically, 
we find that the power spectra of the magnetic as well as electric fields behave 
as~$k^4$.
In these cases, at suitably early times when $k\gg \sqrt{\vert J''/J\vert}$, 
the Fourier modes of the non-helical vector potential [governed by Eq.~\eqref{eq:de-cAk}]
can be written as
\begin{equation}
\cA_k^\mathrm{I}(\eta)=\f{1}{\sqrt{2\,k}}\,\mathrm{e}^{-i\,k\,\eta}.
\end{equation}
Also, since, $J$ is a constant, at late times when $k\ll \sqrt{\vert J''/J\vert}$,
we can express the non-helical electromagnetic modes as
\begin{equation}
\cA_k^\mathrm{II}(\eta)=\f{1}{\sqrt{2\,k}}\,\l[\alpha_k+ \beta_k\,\eta \r],
\label{eq:cA-a2}
\end{equation}
where the coefficients $\alpha_k$ and $\beta_k$ are to be determined by matching 
the above solutions and their derivatives at the time $\eta_k$ corresponding to 
$k=\sqrt{\vert J''/J\vert}$.
The coefficients $\alpha_k$ and $\beta_k$ can be easily obtained to be
\begin{equation}
\alpha_k=(1+i\,k\,\eta_k)\,\mathrm{e}^{-i\,k\,\eta_k},
\quad
\beta_k=-i\,k\,\eta_k\,\mathrm{e}^{-i\,k\,\eta_k},
\end{equation}
and hence, at late times, we have
\begin{equation}
\cA_k^\mathrm{II}(\eta)=\f{1}{\sqrt{2\,k}}\,
\l[1-i\,k\,(\eta-\eta_k)\r]\,\mathrm{e}^{-i\,k\,\eta_k}.
\end{equation}
Since $J$ is constant, this implies that the quantity $\sqrt{k}\, \bar{A}_k$
will have the same value at late times [i.e. when $(-k\,\ee) \ll 1$] for large
wave numbers provided $(k\,\eta_k)$ is small.
We shall see below that $(k\,\eta_k)$ is indeed small in the models of our
interest.
In Fig.~\ref{fig:mode_USR2}, we have plotted the evolution of the electromagnetic 
modes at late times in the case of the ultra slow roll inflation model~\eqref{eq:usr} 
for a range of wave numbers.
%%%%%%%%%%%%%%%%%%%%%%%%%%%%%%%%%%%%%%%%%%%%%%%%%%%%%%%%%%%%%%%%%%%%%%%%%%%%%%%
%%%%%%%%%%%%%%%%%%%%%%%%%%%%%%%%%%%%%%%%%%%%%%%%%%%%%%%%%%%%%%%%%%%%%%%%%%%%%%%
It is clear from the figure that, over large enough wave numbers for which 
$\eta_k$ occurs after the epoch of ultra slow roll, the quantity $\sqrt{k}\, 
\vert A_k\vert$ has the same amplitude at late times.
This, in turn, implies that the power spectrum of the magnetic field will 
behave as~$k^4$, which is what we obtain numerically.

Note that, because of the fact that the first slow roll parameter remains small 
until we approach close to the end of inflation, the de Sitter expression for 
the scale factor remains valid.
As a result, on using the above form for the electromagnetic modes, we obtain 
the spectra of the magnetic and electric fields in the limit $(-k\,\ee)\ll 1$
to be
\begin{subequations}
\begin{eqnarray}
\pb(k) &=& \f{\HI^4}{4\,\pi^2}\, (-k\,\ee)^4\, \l(1+k^2\,\eta_k^2\r),\\
\pe(k) &=& \f{\HI^4}{4\,\pi^2}\,(-k\,\ee)^4.
\end{eqnarray}
\end{subequations}
While $\pe(k)$ is independent of $\eta_k$ and evidently behaves as $k^4$ over
large wave numbers, we need to determine $\eta_k$ in order to understand the 
shape of $\pb(k)$.
Since $J''/J \propto a^{5/2}$ at late times, on using the behavior of the scale 
factor in de Sitter, based on dimensional grounds, we can write $J''/J
=(k_\mathrm{t}\, \eta^5)^{-1/2}$, where $k_\mathrm{t}$ is a wave number.
The quantity $k_\mathrm{t}$ needs to be determined from the numerical value of 
$J''/J$ at the end of the ultra slow roll phase.
Hence, the condition $k^2=J''/J=(k_\mathrm{t}\,\eta_k^5)^{-1/2}$ leads to
$k^2\,\eta_k^2=(k/k_\mathrm{t})^{2/5}$.
In the ultra slow roll and the second punctuated inflation models, we find that,
for our choices of the coupling functions, $k_\mathrm{t} \simeq 10^{23}\,
\mathrm{Mpc}^{-1}$, whereas the largest wave number of our interest is 
$k\simeq 10^{19}\, \mathrm{Mpc}^{-1}$.
These imply that $(k^2\,\eta_k^2) \lesssim 10^{-2}$. 
Therefore, we can expect $\pb(k)$ to behave as $k^4$ over the wave numbers
$10^{15}\,\mathrm{Mpc}^{-1} \lesssim k\lesssim 10^{19}\,\mathrm{Mpc}^{-1}$,
which is what we observe numerically.

In retrospect, it should be clear that  the approaches in the last two 
subsections yielded similar results for the behavior of the spectra at
large wave numbers because of the fact that the modes $\cA_k^\mathrm{II}$ 
as given by Eqs.~\eqref{eq:cA-usr} and~\eqref{eq:cA-a2} have the same 
amplitudes at late times.

%%%%%%%%%%%%%%%%%%%%%%%%%%%%%%%%%%%%%%%%%%%%%%%%%%%%%%%%%%%%%%%%%%%%%%%%%%%%%%%

\section{Can the features be ironed out?}\label{sec:iof}

It is now interesting to examine whether the features in the spectra of the
electromagnetic fields can be ironed out so that we arrive at nearly scale
invariant spectra for the magnetic field.
In this section, we shall discuss this possibility in the second Starobinsky 
model [cf. Eqs.~\eqref{eq:sm2} and~\eqref{eq:ssm}] that leads to features in 
the scalar power spectrum over the large scales.

\begin{figure*}
\centering
\includegraphics[width=8.50cm]{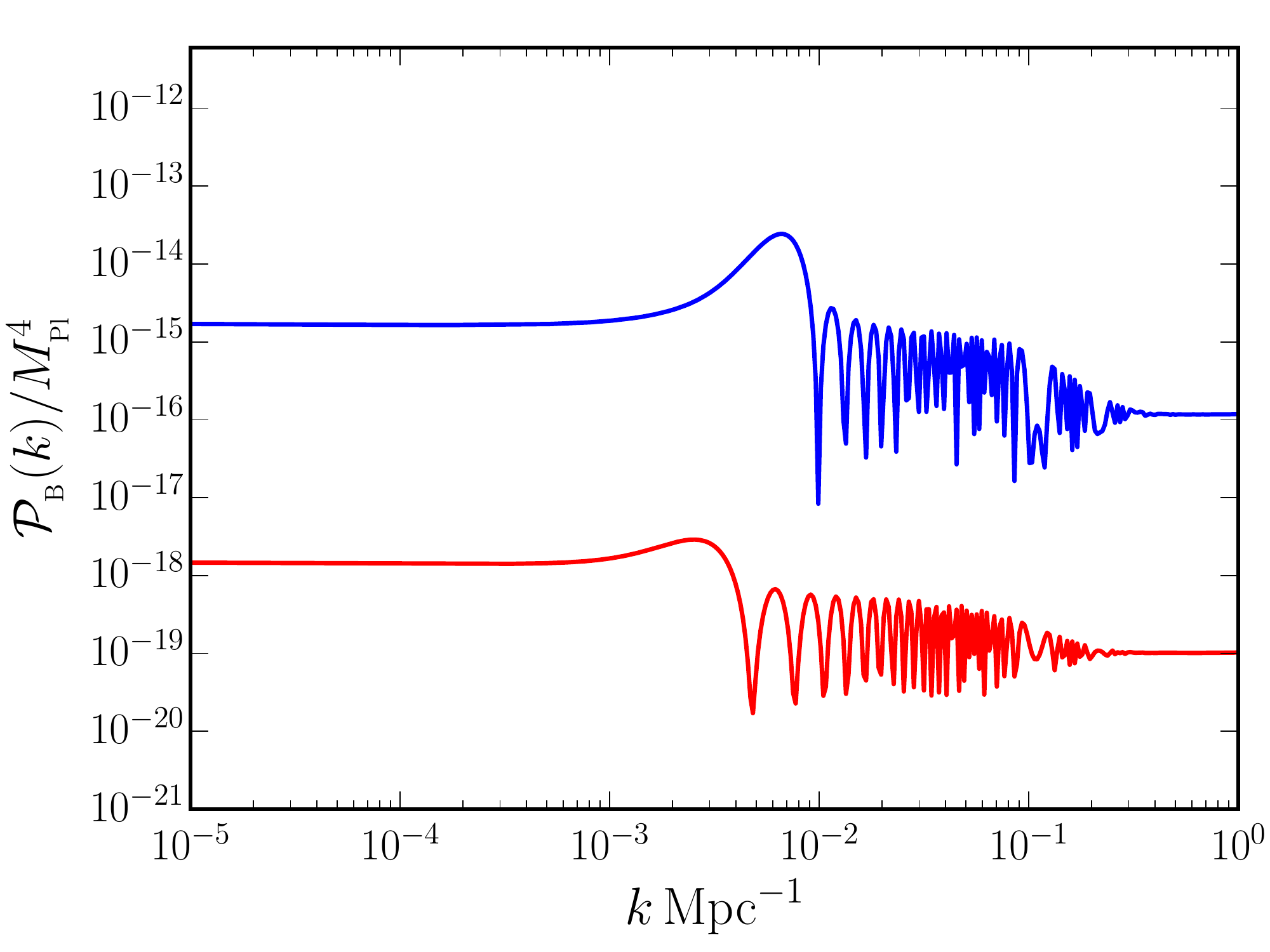}
\hskip 5pt
\includegraphics[width=8.50cm]{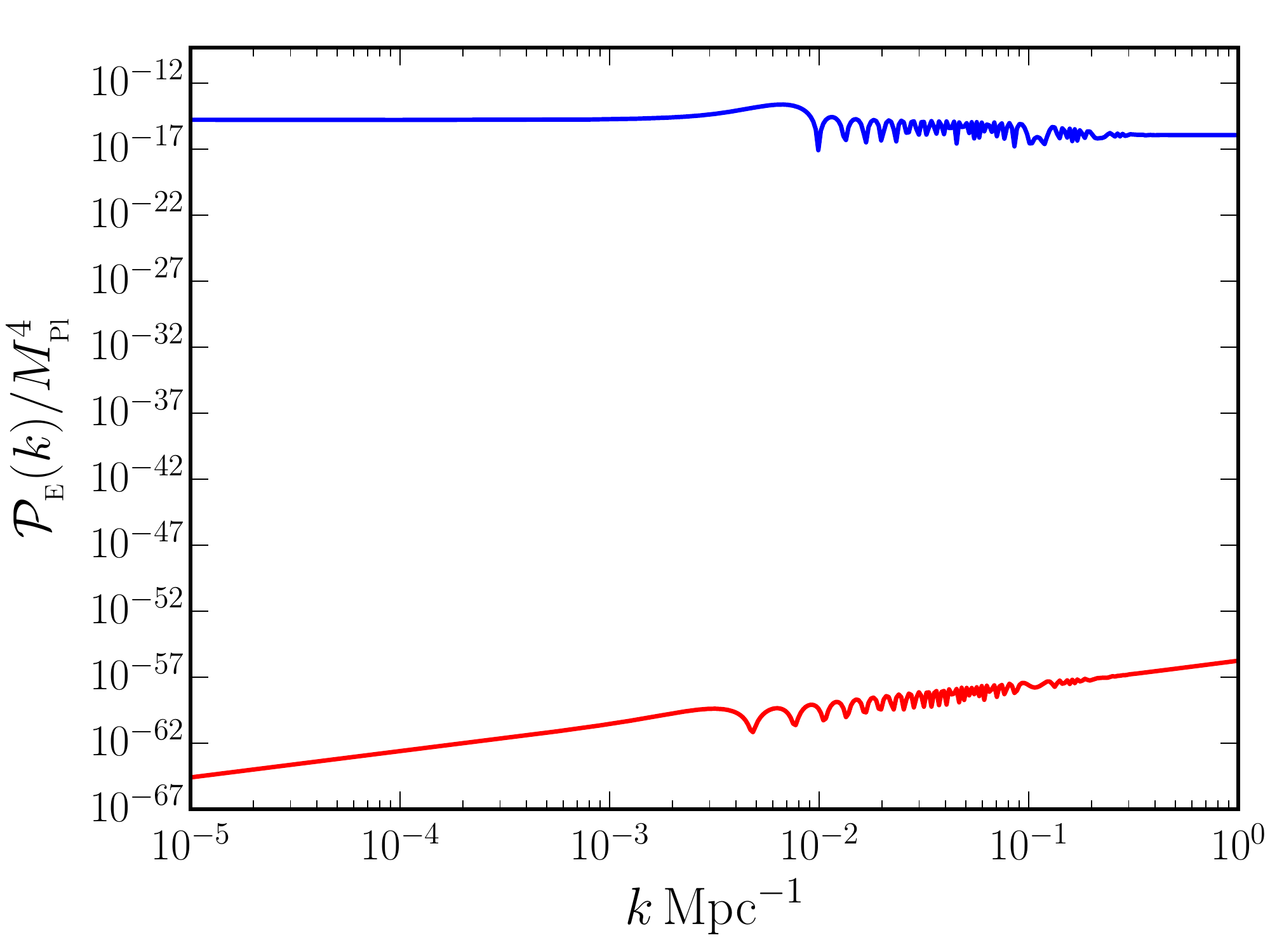} 
\caption{The spectra of the magnetic (on the left) and electric (on the right) 
fields arising for the choice of the coupling function~\eqref{eq:J-sm2-iof} 
in the second Starobinsky model~\eqref{eq:ssm} have been plotted for both the
non-helical (in red) and helical (in blue) cases. 
As before, we have set $n=2$ and $\gamma=1$ when computing the spectra.
Note that, with the new coupling function, the strong features have disappeared
and we are left with relatively smaller features that can be expected to be 
consistent with the current constraints.
Evidently, the burst of oscillations that remain in the spectra occurs because 
of the departure from slow roll as the field crosses the point~$\phi_0$.}
\label{fig:sm2-iof}
\end{figure*}

Earlier, we had arrived at the spectra of the magnetic field in this model 
assuming that the coupling function was given by either $J_+(\phi)$ or
$J_-(\phi)$ described by Eqs.~\eqref{eq:J-phipm}.
In order to remove the strong features that arise in the spectrum of the 
magnetic field, it seems reasonable to stitch together these two coupling 
functions in the following fashion:
\begin{eqnarray} 
J(\phi)&=& 
\f{J_1}{2\,J_{0+}}\,\l[1+\mathrm{tanh}\l(\f{\phi -\phi_0}{\Delta\phi_1}\r)\r]\,
J_+(\phi)\nn\\
& &+\,\f{J_1}{2\,J_{0-}}\,
\l[1-\mathrm{tanh}\l(\f{\phi-\phi_0}{\Delta\phi_1}\r)\r]\, J_-(\phi),\quad
\label{eq:J-sm2-iof}
\end{eqnarray}
where $J_1$ is constant which is determined by the condition that $J(\phi)$ 
reduces to unity at the end of inflation and $\Delta\phi_1$ is another 
constant which we shall choose suitably.
Note that, for a small enough $\Delta \phi_1$, the quantities within the 
square brackets (involving the hyperbolic tangent functions) in the above 
expression behave as step functions.  
It should then be evident that the above coupling function has been constructed 
in such a fashion that it is essentially described by $J_+(\phi)$ when $\phi>\phi_0$
and  $J_-(\phi)$ when $\phi < \phi_0$.
In Fig.~\ref{fig:sm2-iof}, we have plotted the resulting spectra for the 
magnetic as well as electric fields obtained numerically in the non-helical 
and helical cases.
%%%%%%%%%%%%%%%%%%%%%%%%%%%%%%%%%%%%%%%%%%%%%%%%%%%%%%%%%%%%%%%%%%%%%%%%%%%%%%%
%%%%%%%%%%%%%%%%%%%%%%%%%%%%%%%%%%%%%%%%%%%%%%%%%%%%%%%%%%%%%%%%%%%%%%%%%%%%%%%
As can be seen from the figure, there arise two nearly scale invariant regions 
in the power spectra of the magnetic field (and in the case of the helical 
electric field), with a burst of oscillations in between.
Clearly, the scale invariant parts correspond to the evolution of the field 
over the two linear parts of the potential and the oscillations arise as the 
deviations from slow roll occur when the field crosses $\phi_0$.
Thus, in a model involving a strong departure from slow roll, with a suitable 
choice of the coupling function, we have been able to arrive at electromagnetic 
spectra that do not lead to significant backreaction and can also be largely 
consistent with the current constraints.
However, we should stress the fact that it has been achieved only at the severe 
cost of an extremely fine tuned non-minimal coupling function.

%%%%%%%%%%%%%%%%%%%%%%%%%%%%%%%%%%%%%%%%%%%%%%%%%%%%%%%%%%%%%%%%%%%%%%%%%%%%%%%%

\section{Conclusions}\label{sec:c}

A nearly scale invariant primordial scalar power spectrum, as is generated 
in slow roll inflationary models, is remarkably consistent with the CMB 
data~\cite{Planck:2018jri,BICEPKeck:2021gln}.
However, it has been repeatedly noticed that certain features in the scalar 
power spectrum can improve the fit to the data.
Such features are often generated by considering potentials that induce 
departures from slow roll inflation~\cite{Contaldi:2003zv,Sinha:2005mn,
Powell:2006yg,Jain:2008dw,Jain:2009pm,Hazra:2010ve,Benetti:2013cja,
Hazra:2014jka,Hazra:2014goa,Chen:2016zuu,Chen:2016vvw,Ragavendra:2020old,
Antony:2021bgp}. 

Magnetic fields are generated during inflation by breaking the conformal 
invariance of the electromagnetic action.
In this work, we have investigated the effects of deviations from slow roll 
on the spectra of the electromagnetic fields generated during inflation.
Specifically, we have considered a class of inflationary models which 
allow transient deviations from slow roll and, as a result, generate 
localized features in the scalar power spectrum. 
When the electromagnetic fields are coupled to the scalar curvature, we found
that it proves to be challenging to obtain nearly scale invariant magnetic 
fields of the desired shapes and strengths even in slow roll inflation.
In contrast, this is easy to achieve when the electromagnetic field is coupled 
non-minimally to the inflaton, provided we work with model-dependent coupling 
functions.
Therefore, we focused on situations wherein the electromagnetic field is coupled 
to the inflaton and evaluated the spectra of non-helical as well as helical 
electromagnetic fields in non-trivial scenarios involving deviations from
slow roll.
We found that, when strong departures from slow roll arise, apart from 
generating features in the scalar power spectrum, quite generically, these 
deviations also led to features in the spectra of electromagnetic fields.
Moreover, in certain scenarios, it is also possible that the strengths of
the magnetic fields are considerably suppressed on large scales.
While it seems possible to remove the strong features in the spectra of the 
electromagnetic fields allowing us to arrive at nearly scale invariant spectra 
of required strengths, it is achieved at the terrible cost of extreme
fine-tuning.
In summary, if future observations confirm the presence of strong features 
in the primordial scalar power spectrum and, if the electromagnetic fields are 
to be generated by coupling them to the inflaton that is responsible for these
features, then there seems to arise a severe challenge in being able to 
produce magnetic fields of the desired shape and strength in single field 
models of inflation.
We are currently exploring possible ways of overcoming the challenge.

There are a couple of related points we wish to clarify before we conclude.
As we have stressed earlier, in this work, we have focused on a domain wherein 
backreaction due to the electromagnetic fields is negligible~\cite{Kanno:2009ei,
Markkanen:2017kmy}. 
Another interesting aspect of generating electromagnetic fields during inflation is 
that they can induce non-adiabatic pressure perturbations which can source the 
adiabatic scalar perturbations on super-Hubble scales (in this context, see, for
instance, Refs.~\cite{Bonvin:2011dt,Ferreira:2014hma,Markkanen:2017kmy}). 
This additional contribution can lead to distinguishable features in the CMB both 
at the level of the power spectrum as well as non-Gaussianities. 
However, for most of the models we have considered in this work, since the 
strength of generated magnetic fields over CMB scales is relatively weak, the effects 
arising from the induced curvature perturbations can be expected to be negligible.
Nevertheless, it seems important to investigate these effects more closely in 
non-trivial scenarios involving departures from slow roll inflation
We are also presently examining these issues.

%%%%%%%%%%%%%%%%%%%%%%%%%%%%%%%%%%%%%%%%%%%%%%%%%%%%%%%%%%%%%%%%%%%%%%%%%%%%%%%

\section*{Acknowledgments}

The authors wish to thank Kandaswamy Subramanian and Ramkishor Sharma for
clarifications concerning the behavior of the Fourier modes of the helical 
electromagnetic fields.
We also wish to thank H.~V.~Ragavendra and Shiv Sethi for discussions.
ST would like to thank the Indian Institute of Technology Madras, Chennai, 
India, for support through the Half-Time Research Assistantship.
DC's work is supported by the STFC grant ST/T000813/1.
LS and RKJ wish to acknowledge support from the Science and Engineering 
Research Board, Department of Science and Technology, Government of India, 
through the Core Research Grant~CRG/2018/002200.
RKJ also acknowledges financial support from the new faculty seed start-up 
grant of the Indian Institute of Science, Bengaluru, India.
RKJ also wishes to thank the Infosys Foundation, Bengaluru, India, for 
support through the Infosys Young Investigator Award. 

%%%%%%%%%%%%%%%%%%%%%%%%%%%%%%%%%%%%%%%%%%%%%%%%%%%%%%%%%%%%%%%%%%%%%%%%%%%%%%%

\appendix

\section{The electromagnetic spectral indices in slow roll inflation}\label{app:si}

In this appendix, we shall derive the spectral indices of the non-helical
magnetic and electric fields, viz. $\nb$ and $\ne$, in the slow roll 
approximation.

Given the form $J=[a(\eta)/a(\ee)]^n$ for the non-minimal coupling function 
[cf. Eq.~\eqref{eq:J}], one finds that
\begin{equation}
\f{J''}{J}
=\mathcal{H}^2\,\l(n^2+n-n\,\epsilon_1\r),
\end{equation}
where $\epsilon_1=-\dot{H}/H^2$ is the first slow roll parameter, and we 
should emphasize that this relation is exact.
In the slow roll approximation, one can express the conformal Hubble parameter
as~\cite{Mukhanov:1990me,Martin:2003bt,Martin:2004um,Bassett:2005xm,Baumann:2008bn,
Sriramkumar:2009kg,Kinney:2009vz,Baumann:2009ds,Sriramkumar:2012mik,Linde:2014nna,
Martin:2015dha}
\begin{equation}
\mathcal{H}=\f{a'}{a}\simeq -\f{1}{(1-\epsilon_1)\,\eta}.
\label{eq:cH-sr}
\end{equation}
so that, at the first order in the slow roll parameter~$\epsilon_1$, we have
\begin{equation}
\f{J''}{J}
\simeq\f{1}{\eta^2}\,\l[n^2+n+(2\,n^2+n)\,\epsilon_1\r].
\end{equation}

In such a case, the solution to Eq.~\eqref{eq:de-cAk} that satisfies 
the Bunch-Davies initial conditions is given by
\begin{equation}
\cA_k(\eta) 
= \sqrt{-\f{\pi\,\eta}{4}}\,
\mathrm{e}^{i\,[\nu+(1/2)]\,\pi/2}\,
H^{(1)}_{\nu}(-k\,\eta),\label{eq:nhs-sr}
\end{equation}
where, as we had mentioned earlier, $H_\nu^{(1)}(z)$ is the Hankel 
function of the first kind.
For $\epsilon_1\ll 1$, at the first order in the slow roll parameter, 
the index $\nu$ is given by
\begin{equation}
\nu \simeq \l(n+\f{1}{2}\r)+n\,\epsilon_1.
\end{equation}
Note that, when $\epsilon_1=0$, the above solution reduces to the de
Sitter solution~\eqref{eq:nhs}, as required.
Since we are eventually interested in the case $n=2$, for convenience,
we shall assume that $\nu>1$.
In such a case, we find that the power spectra of the magnetic and 
electric fields evaluated at late times can be expressed as
\begin{equation}
\pb(k)\propto k^{5-2\,\nu},\quad
\pe(k)\propto k^{7-2\,\nu}, \label{eq:SR-PB_PE}
\end{equation}
which correspond to the spectral indices of 
\begin{equation}
\nb = 4-2\,n\,(1+\epsilon_1),\quad
\ne = 6-2\,n\,(1+\epsilon_1).
\end{equation}
For $n=2$, these correspond to $\nb=-4\,\epsilon_1$ and
$\ne=2-4\,\epsilon_1$.

Since $0 < \epsilon_1 \ll 1$, the above results imply that, for $n=2$, 
in the non-helical case, the spectrum of the magnetic field should be 
red in slow roll inflation.
However, on closer inspection of Fig.~\ref{fig:pbe-srm}, we find that 
the spectrum of the magnetic field is red in the case of the quadratic
potential~\eqref{eq:qp}, but is mildly blue in the cases of the small 
field model~\eqref{eq:sfm} and the first Starobinsky model~\eqref{eq:sm1},
which lead to slow roll inflation.
This can be attributed to the fact that the coupling functions~\eqref{eq:J-qp},
\eqref{eq:J-sfm} and~\eqref{eq:J-sm1} do not exactly mimic the coupling
function $J=[a(\eta)/a(\ee)]^n$.
In the case of the quadratic potential, for the choice of the coupling 
function~\eqref{eq:J-qp}, we find that the quantity $J^{\prime\prime}/J$
can be expressed as
\begin{eqnarray}
\f{J^{\prime\prime}}{J}
&=&a^2\,H^2\, 
\biggl[\f{n^2\,H^2}{m^2}\,(3\,\epsilon_1-\epsilon_1^2) - n\,\epsilon_1 \nn\\
& &+\,\f{n\,H}{m}\,\l(3\,\epsilon_1-\epsilon_1^2\r)^{1/2}\,
\l(1-\epsilon_1+\f{\epsilon_2}{2}\r)\biggr].
\end{eqnarray}
We should mention that no approximations have been made in arriving at this
expression.
It does not seem possible to express the quantity $J^{\prime\prime}/J$ purely
in terms of the slow roll parameters.
For $n=2$, if we make use of the expression~\eqref{eq:cH-sr} for the conformal
Hubble parameter~$\mathcal{H}$, we obtain that
\begin{eqnarray}
\f{J^{\prime\prime}}{J}
&=&\f{1}{\eta^2} \biggl\{\f{1}{(1-\epsilon_1)^2}\, 
\biggl[\f{4\,H^2}{m^2}\,(3\,\epsilon_1-\epsilon_1^2) - 2\,\epsilon_1 \nn\\
& &+\,\f{2\,H}{m}\,\l(3\,\epsilon_1-\epsilon_1^2\r)^{1/2}\,
\l(1-\epsilon_1+\f{\epsilon_2}{2}\r)\biggr]\biggr\}.\quad
\end{eqnarray}
We should clarify that, while the quantity within the square brackets in this
expression is an exact one, the conformal Hubble parameter has been evaluated
in the slow roll approximation. 
Clearly, in such a case, the solution to the electromagnetic vector potential 
can be written in terms of the Hankel function as in Eq.~\eqref{eq:nhs-sr}. 
The index $\nu$ can be determined by equating the quantity within the curly 
brackets in the above expression for~$J^{\prime\prime}/J$ to $\nu^2-(1/4)$.
At the time when the pivot scale leaves the Hubble radius, for the choice
of the parameters we have worked with, we find that $\nu=2.513$. 
Since $2\,\nu>5$, the spectrum of the magnetic field exhibits a red tilt for 
our choice of the coupling function in the case of the quadratic potential  
[cf. Eq.~\eqref{eq:SR-PB_PE}].

We find that, in general, the quantity $J^{\prime\prime}/J$ can be expressed as 
\begin{equation}
\f{J^{\prime\prime}}{J} = a^2\,H^2\, \mub^2(N),
\end{equation}
where $\mub(N)$ is given by
\begin{equation}
\mub^2(N)=\f{J_{NN}}{J}+(1-\epsilon_1)\,\f{J_N}{J},    
\end{equation}
with $J_N=\d J/\d N$ and $J_{NN}=\d^2 J/\d^2 N$.
If we make use of the conformal Hubble parameter in the slow roll approximation
[cf. Eq.~\eqref{eq:cH-sr}], then, we can write 
\begin{equation}
\f{J^{\prime\prime}}{J} = \f{1}{\eta^2}\,\f{\mub^2(N)}{\l(1-\epsilon_1\r)^2},
\end{equation}
which implies that $\nu^2-(1/4)=\mub^2/(1-\epsilon_1)^2$, with $\mub$ and
$\epsilon_1$ evaluated, say, when the pivot scale leaves the Hubble radius.
Note that, one obtains a strictly scale invariant spectrum for the magnetic
field when $\mub^2/(1-\epsilon_1)^2=6$, which corresponds to $2\,\nu=5$.
For our choice of the coupling function, in the case of the quadratic potential, 
at the time the pivot scale leaves the Hubble radius, we find that 
$\mub^2/(1-\epsilon_1)^2 = 6.068$, which leads to $\nu =2.513$ that we 
mentioned above.
In the cases of the small field and the first Starobinsky models, for the
choices of the coupling functions~\eqref{eq:J-sfm} and~\eqref{eq:J-sm1}, 
we find that, when the pivot scale exits the Hubble radius, 
$\mub^2/(1-\epsilon_1)^2= 5.935$ and $5.939$ which correspond 
to $\nu=2.487$ and $2.488$, respectively.
Since, $2\,\nu < 5$, we obtain magnetic field spectra with blue tilts in these 
two cases.

%%%%%%%%%%%%%%%%%%%%%%%%%%%%%%%%%%%%%%%%%%%%%%%%%%%%%%%%%%%%%%%%%%%%%%%%%%%%%%%
\bibliographystyle{apsrev4-2}
\bibliography{references} 
%%%%%%%%%%%%%%%%%%%%%%%%%%%%%%%%%%%%%%%%%%%%%%%%%%%%%%%%%%%%%%%%%%%%%%%%%%%%%%%
\end{document}